\documentclass[
    article, 
    superscriptaddress,
    amsmath,
    amssymb,
    aps,
    prd,
    showkeys
]{revtex4-2}

\usepackage{graphicx}
\usepackage{bm}
\usepackage{mathtools}
\usepackage{array}
\usepackage{booktabs}
\usepackage{multirow}
\usepackage{hyperref}

\usepackage{tabularray}
\NewTblrTheme{prdwidetable}{
    \DefTblrTemplate{caption-tag}{prd}{\MakeUppercase{\tablename}~\thetable}
    \SetTblrTemplate{caption-tag}{prd}
    \DefTblrTemplate{caption-sep}{prd}{.\enspace}
    \SetTblrTemplate{caption-sep}{prd}
}

\usepackage{placeins}
\usepackage{physics}

\newcommand{\RNumb}[1]{\textup{\uppercase\expandafter{\romannumeral#1}}}
\let\rnumb\RNumb

\makeatletter
\newcommand{\pmp}{\mathrel{\mathchoice
  {\pmp@make\scriptstyle{1.3em}{-0.10ex}{0.30ex}}%
  {\pmp@make\scriptstyle{1.3em}{-0.10ex}{0.30ex}}%
  {\pmp@make\scriptscriptstyle{1.0em}{-0.15ex}{0.15ex}}%
  {\pmp@make\scriptscriptstyle{0.8em}{-0.15ex}{0.15ex}}%
}}
\newcommand{\pmp@make}[4]{%
  \vcenter{\offinterlineskip
    \hbox to #2{\hfil\m@th$#1(+)$\hfil}%
    \kern #3%
    \hbox to #2{\hfil\m@th$#1\raisebox{-#4}{$-$}$\hfil}%
  }%
}
\makeatother

\begin{document}

\title{Doubly heavy open--flavor tetraquarks \\ within the relativistic quark model}

\author{Vladimir O. Galkin}
\email{vgalkin@frccsc.ru}
\affiliation{Federal Research Center ``Computer Science and Control'', Russian Academy of Sciences}

\author{Elena M. Savchenko}
\email{savchenko.em16@physics.msu.ru}
\affiliation{Federal Research Center ``Computer Science and Control'', Russian Academy of Sciences}

\date{\today}

\begin{abstract}
Mass spectra of the ground, orbitally and radially excited states of doubly heavy open--flavor tetraquarks are calculated within the relativistic quark model based on the quasipotential approach and quantum chromodynamics. Tetraquarks are considered as bound states of a heavy diquark $\mathrm{QQ^{\prime}}$ ($\mathrm{Q,Q^{\prime} = c,b}$) and a light antidiquark $\mathrm{\overline{qq^{\prime}}}$ ($\mathrm{q,q^{\prime} = u,d,s}$). The diquark--antidiquark interaction quasipotential includes both spin--independent and spin--dependent relativistic contributions. The finite sizes of the diquark and antidiquark are taken into account via the form factors of their interaction with gluons. Masses of the ground ($1S$) and excited ($1P$, $2S$, $1D$, $2P$, $3S$) states are calculated including the mixing of states with the same observable quantum numbers $J^{P}$ but different spins $S$. The obtained masses are compared with the lowest allowed thresholds for strong fall--apart decays into a pair of heavy--light mesons in $S$--wave. Most of the states lie significantly above these thresholds. However, several states, predominantly orbital excitations with large $J$, lie below or slightly above the corresponding thresholds and, thus, could be promising candidates for experimental observation. The number of such near--threshold states increases for tetraquarks containing bottom quarks. The predicted mass of the ground state of the $\mathrm{cc\overline{ud}}$ tetraquark with $J^{P}=1^{+}$ is consistent with the experimentally observed $\mathrm{T}_{\mathrm{cc}}(3875)^{+}$ state.
\end{abstract}


\maketitle

\section{Introduction}
\label{sec:introduction}

\par
The possible existence of hadrons with a content of valence quarks and antiquarks different from the ordinary quark--antiquark pair for mesons and three quarks for baryons has been considered since the early days of the quark model~\cite{GellMann1964,Zweig1964}. However, absence of convincing experimental evidence for such exotic states significantly reduced interest in their investigation for several decades. The situation has changed since the early 2000s when numerous candidates for exotic multiquark hadrons have finally been discovered. The most important experimental results are the following.
\begin{itemize}
	\item The first exotic state discovered was $X(3872)$ (now $\chi_{\mathrm{c}1}(3872)$) observed in 2003 by the Belle Collaboration~\cite{X3872:Belle:2003}, with estimated minimal quark content $\mathrm{cu\overline{cu}}$.
	
	\item The first identified explicitly exotic charged state was $Z_{\mathrm{c}}(4430)$ (now $\mathrm{T}_{\mathrm{c\overline{c}}1}(4430)$) observed in 2008 by the Belle Collaboration~\cite{Zc4430:Belle:2008} and in 2014 confirmed by the LHCb Collaboration~\cite{Zc4430:LHCb:2014}, with estimated minimal quark content $\mathrm{cu\overline{cd}}$, $\mathrm{cd\overline{cu}}$.
	
	\item The first discovered pentaquark states were a family of hidden--charm pentaquarks $\mathrm{P}_{\mathrm{c\overline{c}}}^{+}$ observed between 2015--2019 by the LHCb Collaboration~\cite{uudcc:LHCb:2015,uudcc:LHCb:2019}, with estimated minimal quark content $\mathrm{uudc\overline{c}}$.
	
	\item The first detected fully heavy exotic states were a family of fully charmed tetraquarks $X(6600)$, $X(6900)$, $X(7100)$ (now $\mathrm{T}_{\mathrm{cc\overline{cc}}}$) observed between 2020--2026 by the LHCb, ATLAS and CMS Collaborations~\cite{cccc:LHCb:2020,cccc:ATLAS:2023,cccc:ATLAS:2026,cccc:CMS:2026}, with estimated minimal quark content $\mathrm{cc\overline{cc}}$.
\end{itemize}
\noindent
More recently, tetraquark states with open heavy flavors (charm in particular) have also been observed by the LHCb Collaboration: 
\begin{itemize}
	\item $\mathrm{T}_{\mathrm{cs}}$~\cite{csud:LHCb:2020A} ($\mathrm{cs\overline{ud}}$), 
	
	\item $\mathrm{T}_{\mathrm{c\overline{s}}}$~\cite{cuds;cdus:LHCb:2022} ($\mathrm{cu\overline{ds}}$, $\mathrm{cd\overline{us}}$), 
	
	\item $\mathrm{T}_{\mathrm{cc}}$~\cite{ccud:LHCb:2021} ($\mathrm{cc\overline{ud}}$).
\end{itemize}
\noindent
Detailed reviews of the experimental and theoretical progress in exotic hadron spectroscopy can be found in Refs.~\cite{Review2020,Review2023,Review2024LHCb}.
\par
Despite the considerable progress, there is still no consensus on the internal structure of many observed exotic hadrons. Several theoretical descriptions have been proposed for four--quark states. The most widely used are the tetraquark picture, in which the tetraquark is considered as a compact bound state of a diquark and an antidiquark~\cite{Jaffe1977,Diq2005,Tetra2006}, and the molecular model, describing it as a loosely bound system of two mesons~\cite{Molecule1991,Molecule1994,Molecule2018}. Other interpretations, such as hadroquarkonium~\cite{HadroQ2008} and kinematic threshold effects~\cite{Cusp2015}, have also been discussed. In addition, the properties of exotic hadrons are actively investigated within different QCD approaches, including QCD sum rules~\cite{Melikhov2019} and lattice QCD~\cite{Lattice2023}. Yet distinguishing between these descriptions remains a complicated theoretical and experimental task.
\par
Heavy tetraquarks are of particular interest, since the presence of heavy quarks can increase the binding energy of a multiquark system and, thus, the probability of its stability against strong fall--apart decays into a pair of mesons~\cite{Badalian1987,tetrons2018}. In this respect, doubly heavy tetraquarks with open charm and/or bottom having the quark composition $\mathrm{QQ^{\prime}\overline{qq^{\prime}}}$ ($\mathrm{Q,Q^{\prime}=c,b}$, $\mathrm{q,q^{\prime}=u,d,s}$) are especially interesting. These states are explicitly exotic, since they contain two heavy quarks without the corresponding heavy antiquarks. Their possible existence and stability have been investigated in various theoretical approaches, including relativistic quark models~\cite{Tetra2007,Savch2021Uni}, hadronic molecular models~\cite{QQqq.molecule.2021}, potential models~\cite{Pot2018} and lattice QCD~\cite{bbud.lattice.2019,ccud.lattice.2023}. In particular, some doubly heavy tetraquarks are predicted to lie below the thresholds for strong fall--apart decays into two heavy--light mesons.
\par
An important experimental confirmation of the existence of doubly heavy tetraquark states was obtained by the LHCb Collaboration in 2021~\cite{ccud:LHCb:2021}: a narrow state $\mathrm{T}_{\mathrm{cc}}(3875)^{+}$ with the minimal quark composition $\mathrm{cc\overline{ud}}$ was observed in the $D^{0}D^{0}\pi^{+}$ invariant mass spectrum, very close to the $D^{0}D^{*+}$ threshold~\cite{ccud:LHCb:2022std}. Its small decay width and proximity to the meson pair threshold make this state particularly interesting for theoretical studies. The observation of $\mathrm{T}_{\mathrm{cc}}(3875)^{+}$ also stimulates further searches for other doubly heavy tetraquarks, especially those containing bottom quarks, which are expected to have better chances of being stable against strong decays~\cite{Parkhomenko2018,bbud.lattice.2019}.
\par
In this paper we continue our investigation of heavy tetraquarks within the relativistic quark model based on the quasipotential approach and QCD. We employ the diquark--antidiquark picture, which has been successfully used in our previous studies of various tetraquark systems \cite{Tetra2006,Tetra2007,Savch2020PhysRevD,Savch2021Uni,Savch2022Sym,Savch2024EPJA}. Here we focus on doubly heavy open--flavor tetraquarks, particularly composed of a heavy diquark and a light antidiquark $\mathrm{QQ^{\prime}\overline{qq^{\prime}}}$ ($\mathrm{Q,Q^{\prime}=c,b}$, $\mathrm{q,q^{\prime}=u,d,s}$). We calculate the masses of their ground, orbitally and radially excited states and compare the obtained results with the lowest allowed thresholds for strong fall--apart decays into meson pairs. The calculated spectra are also compared with other theoretical predictions and available experimental data.
\par
The paper is organized as follows. In Sec.~\ref{sec:RDiDiM} we describe the relativistic diquark--antidiquark model. The calculated mass spectra are presented in Sec.~\ref{sec:numres}. In Sec.~\ref{sec:thresholdanalysis} we compare these masses with the thresholds for strong fall--apart decays, while Sec.~\ref{sec:othermodels} is devoted to comparisons with other theoretical approaches. The experimental data are discussed in Sec.~\ref{sec:experiment}, followed by our conclusions in Sec.~\ref{sec:conclusion}.

\section{Relativistic diquark--antidiquark model}
\label{sec:RDiDiM}

\subsection{Diquark--antidiquark picture}
\label{subsec:DiDi}

\par
We consider doubly heavy open--flavor tetraquarks with the quark composition $\mathrm{QQ^{\prime}\overline{qq^{\prime}}}$, where $\mathrm{Q,Q^{\prime}=c,b}$ are heavy quarks and $\mathrm{q,q^{\prime}=u,d,s}$ are light quarks. Within the diquark--antidiquark picture these states consist of a heavy diquark $\mathrm{QQ^{\prime}}$ and a light antidiquark $\mathrm{\overline{qq^{\prime}}}$. Three possible heavy diquark flavor compositions are considered: $\mathrm{cc}$, $\mathrm{cb}$ and $\mathrm{bb}$. The light antidiquark can consist of two nonstrange antiquarks $\mathrm{\overline{qq^{\prime}}}$, one nonstrange and one strange antiquark $\mathrm{\overline{qs}}$, or two strange antiquarks $\mathrm{\overline{ss}}$. Thus, the investigated tetraquarks have two different open heavy flavors or two units of the same open heavy flavor without the corresponding heavy antiquarks. They also have strangeness equal to zero, one, or even two, depending on the flavor content of light antidiquark.
\par
The diquark--antidiquark picture significantly simplifies the calculation of tetraquark properties: instead of solving the relativistic four--body problem for two quarks and two antiquarks at once, we consider two consecutive relativistic two--body problems. At the first stage, the masses and wave functions of the diquarks and antidiquarks are determined from the corresponding quark--quark and antiquark--antiquark interactions. At the second stage, these results are used to calculate the masses and wave functions of tetraquarks treated as diquark--antidiquark bound states. This procedure allows us to employ the same relativistic quasipotential formalism for both stages of the calculation.
\par
An important ingredient of our approach is that the diquark and antidiquark are considered as spatially extended objects. We assume that they interact with each other as a whole. Therefore, we do not explicitly consider the cross--interactions between individual quarks of the diquark and antiquarks of the antidiquark. Instead, their effective interaction is described by the diquark--antidiquark quasipotential. The internal structure of the diquark and antidiquark is taken into account via their interaction with gluons, described by the corresponding form factors calculated from the overlap integrals of their wave functions.
\par
Since quarks are fermions, the total diquark wave function must be antisymmetric. It can be represented as a product of the spatial, color, flavor and spin parts:
\begin{equation}\label{eq:Psid}
\Psi_{d} = \psi_{\mathrm{space}} \times \psi_{\mathrm{color}} \times \psi_{\mathrm{flavor}} \times \psi_{\mathrm{spin}}.
\end{equation}
\noindent
The same applies to antidiquark states.
\par
Based on the physical considerations we impose the following restrictions on the diquark and antidiquark wave functions.
\begin{itemize}
	\item First, we consider spatial part of the diquarks wave function. We limit ourselves with diquarks and antidiquarks in their ground states with zero internal orbital angular momentum. Consequently, their spatial wave functions are symmetric under the interchange of the constituents
	\begin{equation}\label{eq:Psid.space}
	\begin{gathered}
	L_{d}=0 \quad\Longrightarrow\quad P_{d}=(-1)^{L_{d}}=+1
	\\
	\Downarrow
	\\
	\psi_{\mathrm{space}}=\psi_{\mathrm{sym}}.
	\end{gathered}
	\end{equation}
	\noindent
	The ground state antidiquark also has positive intrinsic parity. This assumption is motivated by the fact that the ground state diquark and antidiquark are the most spatially compact and, thus, are less likely to overlap with each other and subsequently rearrange in a pair of mesons.
	
	\item Second, we consider the possible color configurations of the diquark and antidiquark. Since each quark and antiquark transforms as a triplet $3_{\mathrm{c}}$ and antitriplet $\overline{3}_{\mathrm{c}}$, respectively, under the $SU(3)_{\mathrm{c}}$ group, the product of two quark color representations is decomposed as
	\begin{equation}\label{eq:di.color}
	3_{\mathrm{c}} \otimes 3_{\mathrm{c}} = \overline{3}_{\mathrm{c}} \oplus 6_{\mathrm{c}},
	\end{equation}
	\noindent
	where the color--antitriplet is antisymmetric and the color--sextet is symmetric under the interchange of the quarks. Similarly, for two antiquarks we have antisymmetric color--triplet and symmetric color--antisextet
	\begin{equation}\label{eq:adi.color}
	\overline{3}_{\mathrm{c}} \otimes \overline{3}_{\mathrm{c}} = 3_{\mathrm{c}} \oplus \overline{6}_{\mathrm{c}}.
	\end{equation}
	\noindent
	The tetraquark is identified with the observable color singlet state. It can be produced either by the color--antitriplet diquark with color--triplet antidiquark or by the color--sextet diquark with color--antisextet antidiquark
	\begin{equation}\label{eq:tetra.color}
	\begin{aligned}
	\overline{3}_{\mathrm{c}} \otimes  3_{\mathrm{c}} &= 8_{\mathrm{c}} \oplus 1_{\mathrm{c}}
	\\
	6_{\mathrm{c}} \otimes \overline{6}_{\mathrm{c}} &= 27_{\mathrm{c}} \oplus 8_{\mathrm{c}} \oplus 1_{\mathrm{c}}.
	\end{aligned}
	\end{equation}
	\noindent
	The choice between these color configurations is determined by the character of the quark--quark interaction. In the one--gluon exchange approximation, its color dependence is proportional to the expectation value of the product of the $SU(3)_{\mathrm{c}}$ generators
	\begin{equation}\label{eq:colorfactors}
    \left\langle T_{1} \cdot T_{2} \right\rangle =
    \begin{cases}
        -\frac{2}{3}, & \overline{3}_{\mathrm{c}}, \\[3pt]
        +\frac{1}{3}, & 6_{\mathrm{c}}.
    \end{cases}
	\end{equation}
	\noindent
	Thus, the one--gluon exchange interaction is attractive for quarks in the color--antitriplet state and repulsive for quarks in the color--sextet state. Similarly, the one--gluon exchange interaction is attractive for antiquarks in the color--triplet state and repulsive for antiquarks in the color--antisextet state. We restrict our consideration to color--antitriplet diquarks and color--triplet antidiquarks. Therefore, color wave functions of the diquark and antidiquark are antisymmetric
	\begin{equation}\label{eq:Psid.color}
	\psi_{\mathrm{color}}=\psi_{\mathrm{asym}}.
	\end{equation}
	\noindent
	Note that the color--sextet diquark with color--antisextet antidiquark channel can be stabilized if cross--interactions between individual quarks of the diquark and antiquarks of the antidiquark are considered, but it goes beyond the scope of the current study.
\end{itemize}
\par
Since the spatial wave function of the ground state diquark is symmetric and its color wave function is considered to be antisymmetric, in order for the whole diquark wave function to be antisymmetric the remaining spin--flavor part of the wave function must be symmetric
\begin{equation}\label{eq:Psid.spinflavor}
\psi_{\mathrm{flavor}} \times \psi_{\mathrm{spin}} = \psi_{\mathrm{sym}}.
\end{equation}
\noindent
This condition can be satisfied in two different ways: either both the flavor and spin wave functions are symmetric, or both are antisymmetric. These possibilities correspond to two types of diquarks:
\begin{equation}\label{eq:Psid.spin.flavor}
\begin{aligned}
&\begin{cases}
	\psi_{\mathrm{flavor}} = \psi_{\mathrm{sym}} \\
	\psi_{\mathrm{spin}}=\psi_{\mathrm{sym}}
\end{cases} && &\Longrightarrow \quad S_{\mathrm{d}}=1 \quad (\mathrm{A}),
\\
&\begin{cases}
	\psi_{\mathrm{flavor}} = \psi_{\mathrm{antisym}} \\
	\psi_{\mathrm{spin}}=\psi_{\mathrm{antisym}}
\end{cases} && &\Longrightarrow \quad S_{\mathrm{d}}=0 \quad (\mathrm{S}).
\end{aligned}
\end{equation}
\noindent
Here $A$ denotes the axialvector diquark with spin $S_{\mathrm{d}}=1$ and $S$ denotes the scalar diquark with spin $S_{\mathrm{d}}=0$. Since only the internal ground states are considered, these diquarks have quantum numbers $J_{\mathrm{d}}^{P}=1^{+}$ and $J_{\mathrm{d}}^{P}=0^{+}$, respectively. The corresponding scalar and axialvector antidiquarks are defined analogously.
\par
The allowed spin configurations depend on the flavor composition of the diquark. For two quarks of the same flavor the flavor wave function can only be symmetric (otherwise it vanishes), and therefore only the axialvector diquark is allowed. In particular, the heavy $\mathrm{cc}$, $\mathrm{bb}$ and light $\mathrm{\overline{uu}}$, $\mathrm{\overline{dd}}$, $\mathrm{\overline{ss}}$ (anti)diquarks necessarily have spin $S_{\mathrm{d}}=1$. For a diquark consisting of two different flavors, both symmetric and antisymmetric flavor combinations are possible. Consequently, the heavy $\mathrm{cb}$ and light $\mathrm{\overline{ud}}$, $\mathrm{\overline{us}}$, $\mathrm{\overline{ds}}$ (anti)diquarks can be either scalar or axialvector. As a result, the possible spin configurations of doubly heavy tetraquarks are determined by the allowed combinations of their constituent diquarks and antidiquarks. We denote these configurations by $\mathrm{A\overline{A}}$, $\mathrm{A\overline{S}}$, $\mathrm{S\overline{A}}$ and $\mathrm{S\overline{S}}$, where the first symbol refers to the heavy diquark and the second to the light antidiquark. The number of allowed configurations depends on the tetraquark flavor composition.
\par
The total spin $\mathbf{S}$ of the tetraquark is obtained by coupling the spins of the diquark $\mathbf{S}_{d}$ and antidiquark $\mathbf{S}_{\overline{d}}$
\begin{equation}\label{eq:tetra.S}
\mathbf{S} = \mathbf{S}_{d} + \mathbf{S}_{\overline{d}}.
\end{equation}
\noindent
Similarly, the total angular momentum $\mathbf{J}$ of the tetraquark is obtained by coupling the spin of the tetraquark $\mathbf{S}$ and the orbital angular momentum of the relative diquark--antidiquark motion $\mathbf{L}$
\begin{equation}\label{eq:tetra.J}
\mathbf{J} = \mathbf{L} + \mathbf{S}.
\end{equation}
\noindent
Since both constituents are in their internal ground states, the tetraquark parity is determined by their relative orbital angular momentum
\begin{equation} \label{eq:tetra.P}
P=(-1)^{L}.
\end{equation}
\noindent
The investigated states are characterized by the quantum numbers $J^{P}$ and are denoted by the spectroscopic notation $n^{2S+1}\mathrm{L}_{J}$. We calculate the masses of the ground states as well as their orbital and radial excitations.

\subsection{Relativistic quasipotential approach}
\label{subsec:rel.quasi.app}

\par
The description of these tetraquarks requires a consistent account of relativistic effects. The relative velocities of heavy quarks in hadrons are not always sufficiently small to neglect relativistic corrections~\cite{Bodwin1995}, while the light quarks and antiquarks are intrinsically relativistic constituents~\cite{Godfrey1985}. Therefore, both the relativistic kinematics and the relativistic dynamics must be included in the calculation. For this purpose, we employ the relativistic quark model developed in Refs.~\cite{Param1992,Param1995}, which is based on the quasipotential approach~\cite{Logunov1963,Faustov1973,Faustov1985eng} and has been extensively applied to the spectroscopy of ordinary mesons~\cite{Meson1998,Meson2003,Meson2005}, baryons~\cite{Baryo1997,Baryo2002,Baryo2005} and exotic tetraquark states~\cite{Tetra2006,Tetra2007}.
\par
The quasipotential approach provides a three--dimensional description of a relativistic two--body bound state. It is based on a reduction of the four--dimensional Bethe--Salpeter equation by eliminating the dependence on the relative time of the constituents~\cite{Salpeter1951,Logunov1963,Faustov1970quasi.eng}. The resulting equation can be brought to a Schr\"odinger--type form with relativistic kinematics and an effective quasipotential describing the interaction between the constituents~\cite{Faustov1985eng}. This quasipotential is constructed with the help of the off--mass--shell scattering amplitude projected onto positive--energy states and contains both spin--independent and spin--dependent relativistic contributions. In the present model, the effective interaction includes short--range one--gluon exchange and long--range confinement with a mixture of vector and scalar Lorentz structures~\cite{Param1992,Meson2003,Tetra2006}.
\par
Within this framework, the masses of the heavy diquarks and light antidiquarks are first determined from their relativistic quark--quark quasipotentials. Their masses and wave functions together with the parameters of the diquark--gluon interaction form factors are then used as input for the tetraquark calculations. The masses of the tetraquarks are found by solving the corresponding relativistic quasipotential equation for the diquark--antidiquark bound states.
\par
The masses and wave functions of diquarks and tetraquarks are obtained as solutions of the relativistic Schr\"odinger--type quasipotential equation~\cite{Tetra2007,Savch2020PhysRevD,Savch2024EPJA}
\begin{widetext}
\begin{equation}\label{eq:Schr}
\Big( \frac{b^{2}(M)}{2\mu_{\mathrm{R}}(M)} - \frac{\mathbf{p}^{2}}{2\mu_{\mathrm{R}}(M)} \Big) \Psi_{\mathrm{d,T}}(\mathbf{p}) = \int\frac{d^{3}q}{(2\pi)^{3}}V(\mathbf{p},\mathbf{q};M)\Psi_{\mathrm{d,T}}(\mathbf{q}),
\end{equation}
\end{widetext}
where
\begin{itemize}
	\item $\mathrm{d}$ and $\mathrm{T}$ denote diquark and tetraquark bound states, respectively;
	
	\item $\Psi_{\mathrm{d,T}}$ is the wave function of a diquark or a tetraquark, respectively;
	
	\item $\mathbf{p}$ is the relative momentum;
	
	\item $M$ is the mass of the bound state;
	
	\item $\mu_{\mathrm{R}}(M)$ is the relativistic reduced mass given by
	\begin{equation}\label{eq:muR}
	\mu_{\mathrm{R}}(M) = \frac{E_{1}(M)E_{2}(M)}{E_{1}(M)+E_{2}(M)} = \frac{M^{4}-(m_{1}^{2}-m_{2}^{2})^{2}}{4M^{3}},
	\end{equation}
	\noindent
	here
	\begin{itemize}
		\item $m_{1,2}$ denotes the constituent masses,
		
		\item $E_{1,2}(M)$ are energies of constituents on--mass--shell
		\begin{equation}\label{eq:Edi}
		E_{1,2}(M)=\frac{M^{2}+m_{1,2}^{2}-m_{2,1}^{2}}{2M};
		\end{equation}
	\end{itemize}

	\item $b^{2}(M)$ is the on--mass--shell relative momentum in the center--of--mass system squared given by
	\begin{equation}\label{eq:b2}
	b^{2}(M)=\frac{[M^{2}-(m_{1}+m_{2})^{2}][M^{2}-(m_{1}-m_{2})^{2}]}{4M^{2}};
	\end{equation}
	
	\item $V(\mathbf{p},\mathbf{q};M)$ is the interaction quasipotential of constituents (quarks or diquark and antidiquark).
\end{itemize} 
\par
The quark--quark quasipotential is constructed using one--gluon exchange and a mixture of vector and scalar confining interactions. For a quark pair in the color--antitriplet state, the interaction is one half of the corresponding quark--antiquark interaction in the color--singlet state. Its explicit form and the calculation of the diquark properties were discussed in detail in Refs.~\cite{Tetra2006,Tetra2007,Savch2022Sym}. We use the resulting masses and wave functions as input for the tetraquark calculation.
\par
For the diquark--antidiquark system, we employ the same short-- and long--distance interaction mechanisms, taking into account the integer spins and finite spatial sizes of both constituents. In momentum space the quasipotential consists of the one--gluon exchange contribution and the vector--scalar confinement contribution \cite{Tetra2006,Tetra2007,Savch2022Sym,Savch2024EPJA}:
\begin{widetext}
\begin{equation}\label{eq:Vdidi}
\begin{aligned}
V_{\mathrm{d\overline{d}}}(\mathbf{p},\mathbf{q};M) = &\underbrace{\frac{\bra{d(P)} J_{\mu} \ket{d(Q)}}{2\sqrt{E_{\mathrm{d}}}\sqrt{E_{\mathrm{d}}}} \frac{4}{3}\alpha_{\mathrm{s}}D^{\mu\nu}(\mathbf{k}) \frac{\bra{\overline{d}(P^{\prime})} J_{\nu} \ket{\overline{d}(Q^{\prime})}}{2\sqrt{E_{\mathrm{\overline{d}}}}\sqrt{E_{\mathrm{\overline{d}}}}}}_{\text{one--gluon exchange potential}}
\\
+ &\underbrace{\Psi_{\mathrm{d}}^{*}(P)\Psi_{\mathrm{\overline{d}}}^{*}(P^{\prime}) \Big( J_{\mathrm{d};\mu}J_{\mathrm{\overline{d}}}^{\mu}V_{\mathrm{conf}}^{\mathrm{V}}(\mathbf{k})+V_{\mathrm{conf}}^{\mathrm{S}}(\mathbf{k}) \Big) \Psi_{\mathrm{d}}(Q)\Psi_{\mathrm{\overline{d}}}(Q^{\prime})}_{\text{confinement potential}}.
\end{aligned}
\end{equation}
\end{widetext}
Here 
\begin{itemize}
	\item $\mathrm{d}$ and $\mathrm{\overline{d}}$ denote the diquark and antidiquark, respectively (further on prime denotes everything related to the antidiquark);
	
	\item $Q^{(\prime)} = (E_{\mathrm{d}(\mathrm{\overline{d}})}, \pmp \mathbf{q})$, $P^{(\prime)} = (E_{\mathrm{d}(\mathrm{\overline{d}})}, \pmp \mathbf{p})$ are diquark and antidiquark initial and final four--momenta, respectively;
	
	\item $\mathbf{k} = \mathbf{p}-\mathbf{q}$ is the transferred three--momentum.
	
	\item One--gluon exchange potential dominant at short distances contains the following ingredients
	
	\begin{itemize}
		\item $E_{\mathrm{d},\mathrm{\overline{d}}}$ are the on--shell diquark and antidiquark energies, respectively, defined by Eq.~\eqref{eq:Edi} 
		\begin{equation}\label{eq:EdM}
		E_{\mathrm{d},\mathrm{\overline{d}}}(M_{\mathrm{T}})=\frac{M_{\mathrm{T}}^{2}+M_{\mathrm{d},\mathrm{\overline{d}}}^{2} - M_{\mathrm{\overline{d}},\mathrm{d}}^{2}}{2M_{\mathrm{T}}},
		\end{equation}
		\noindent
		where 
		\begin{itemize}
			\item $M_{\mathrm{d},\mathrm{\overline{d}}}$ are the diquark and antidiquark masses, respectively,
			
			\item $M_{\mathrm{T}}$ is the tetraquark mass;
		\end{itemize}
		
		\item $\alpha_{\mathrm{s}}$ is the running QCD coupling constant with freezing~\cite{alphaS1995,alphaS2004} given by
		\begin{equation}\label{eq:alphaS}
		\alpha_{\mathrm{s}}(\mu^{2}) = \frac{4\pi}{\big( 11 - \frac{2}{3}n_{\mathrm{f}} \big) \ \ln \Big( \frac{\mu^{2} + M_{\mathrm{BG}}^{2}}{\Lambda^{2}} \Big)},
		\end{equation}
		\noindent
		where
		\begin{itemize}
			\item $n_{\mathrm{f}}$ is the number of open quark flavors
			\begin{equation}\label{eq:nf}
			n_{\mathrm{f}} = 
			\begin{cases}
				4, &\mathrm{Q} = \mathrm{Q}^{\prime} = \mathrm{c} \\
				5, &\mathrm{Q},\mathrm{Q}^{\prime} = \mathrm{c},\mathrm{b},
			\end{cases}
			\end{equation}
			
			\item $\mu$ is energy scale taken as twice the reduced constituents mass
			\begin{equation}\label{eq:mu}
			\mu = 2\frac{M_{\mathrm{d}}M_{\mathrm{\overline{d}}}}{M_{\mathrm{d}} + M_{\mathrm{\overline{d}}}},
			\end{equation}
			
			\item $M_{\mathrm{BG}} = 0.95$ GeV is the background mass,
			
			\item $\Lambda = 414$ MeV is the parameter of the running coupling constant obtained from the analysis of meson mass spectra;
		\end{itemize}
		
		\item $D^{\mu\nu}(\mathbf{k})$ is the gluon propagator in the Coulomb gauge
		\begin{equation}\label{eq:Dmunu}
		\begin{cases}
			D^{00}(\mathbf{k}) = -\frac{4\pi}{\mathbf{k}^{2}}, \\
			D^{ij}(\mathbf{k}) = -\frac{4\pi}{k^{2}} \Big( \delta^{ij} - \frac{k^{i}k^{j}}{\mathbf{k}^{2}} \Big), \\
			D^{0i}(\mathbf{k}) = D^{i0}(\mathbf{k}) = 0;
		\end{cases}
		\end{equation}
		
		\item $\bra{d(P)} J_{\mu} \ket{d(Q)}$ and $\bra{\overline{d}(P^{\prime})} J_{\nu} \ket{\overline{d}(Q^{\prime})}$ are the elastic diquark--gluon and antidiquark--gluon vertices, respectively, via which the internal structure of the diquark and antidiquark is taken into account.
	\end{itemize}
	
	\item Confinement potential dominant at large distances contains the following ingredients
	
	\begin{itemize}
		\item $\Psi_{\mathrm{d}}(P)$ is the diquark wave function given by
		\begin{equation}\label{eq:Psid.axial}
		\Psi_{\mathrm{d}}(P) = 
		\begin{cases}
			1, &\text{scalar} \\
			\epsilon_{\mathrm{d}}(P), &\text{axialvector}
		\end{cases} \text{diquark},
		\end{equation}
		\noindent
		where $\epsilon_{\mathrm{d}}(P)$ is the polarization vector for the axialvector diquark
		\begin{equation}\label{eq:epsilon}
		\begin{gathered}
		\epsilon_{\mathrm{d}}(p) = \Big( \frac{(\boldsymbol{\epsilon_{\mathrm{d}}} \cdot \mathbf{p})}{M_{\mathrm{d}}}, \ \boldsymbol{\epsilon_{\mathrm{d}}} + \frac{(\boldsymbol{\epsilon_{\mathrm{d}}} \cdot \mathbf{p}) \mathbf{p}}{M_{\mathrm{d}} \big( M_{\mathrm{d}} + E_{\mathrm{d}}(\mathbf{p}) \big)} \Big),
		\\
		\epsilon_{\mathrm{d}}^{\mu}(p)p_{\mu} = 0,
		\end{gathered}
		\end{equation}
		where $E_{\mathrm{d}}(\mathbf{p})$ is the diquark energy
		\begin{equation}\label{eq:Edp}
		E_{\mathrm{d}}(\mathbf{p}) = \sqrt{M_{\mathrm{d}}^{2} + \mathbf{p}^{2}};
		\end{equation}
	
		\item $J_{\mathrm{d};\mu}$ is the effective long--range vector interaction vertex of the diquark given by
		\begin{equation}\label{eq:J}
		J_{\mathrm{d};\mu} = 
		\begin{cases}
			\frac{(P+Q)_{\mu}}{2\sqrt{E_{\mathrm{d}}(\mathbf{p})}\sqrt{E_{\mathrm{d}}(\mathbf{q})}}, &\text{scalar} \\
			-\frac{(P+Q)_{\mu}}{2\sqrt{E_{\mathrm{d}}(\mathbf{p})}\sqrt{E_{\mathrm{d}}(\mathbf{q})}} + \frac{i\mu_{\mathrm{d}}}{2M_{\mathrm{d}}} \Sigma_{\mu}^{\nu} \tilde{k}_{\nu}, &\text{axialvector}
		\end{cases} \text{diquark},
		\end{equation}
		\noindent
		where 
		\begin{itemize}
			\item $\tilde{k}_{\nu} = (0, \mathbf{k})$,
			
			\item $\Sigma_{\mu}^{\nu}$ is the antisymmetric tensor
			\begin{equation}
			(\Sigma_{\rho\sigma})_{\mu}^{\nu} = -i (g_{\mu\rho}\delta_{\sigma}^{\nu} - g_{\mu\sigma}\delta_{\rho}^{\nu}),
			\end{equation}
			
			\item $\mu_{\mathrm{d}}$ is the total chromomagnetic moment of the diquark;
		\end{itemize}

		\item $V_{\mathrm{conf}}^{\mathrm{V,S}}(\mathbf{k})$ are vector and scalar confining potentials having the following nonrelativistic limits in configuration space consistent with the lattice calculations
		\begin{equation}\label{eq:Vconf}
		\begin{aligned}
		V_{\mathrm{conf}}^{\mathrm{V}}(r) &= (1-\varepsilon)(Ar+B),\\
		V_{\mathrm{conf}}^{\mathrm{S}}(r) &= \varepsilon(Ar+B),\\
		V_{\mathrm{conf}}(r) &= V_{\mathrm{conf}}^{\mathrm{V}}(r)+V_{\mathrm{conf}}^{\mathrm{S}}(r) = Ar+B,
		\end{aligned}
		\end{equation}
		\noindent
		where $\varepsilon$ specifies the mixture of the vector and scalar Lorentz structures.
	\end{itemize}
	
\end{itemize}
\par
\begin{figure*}
    \centering
    \begin{minipage}{0.48\textwidth}
    \centering
    \includegraphics[width=0.75\linewidth]{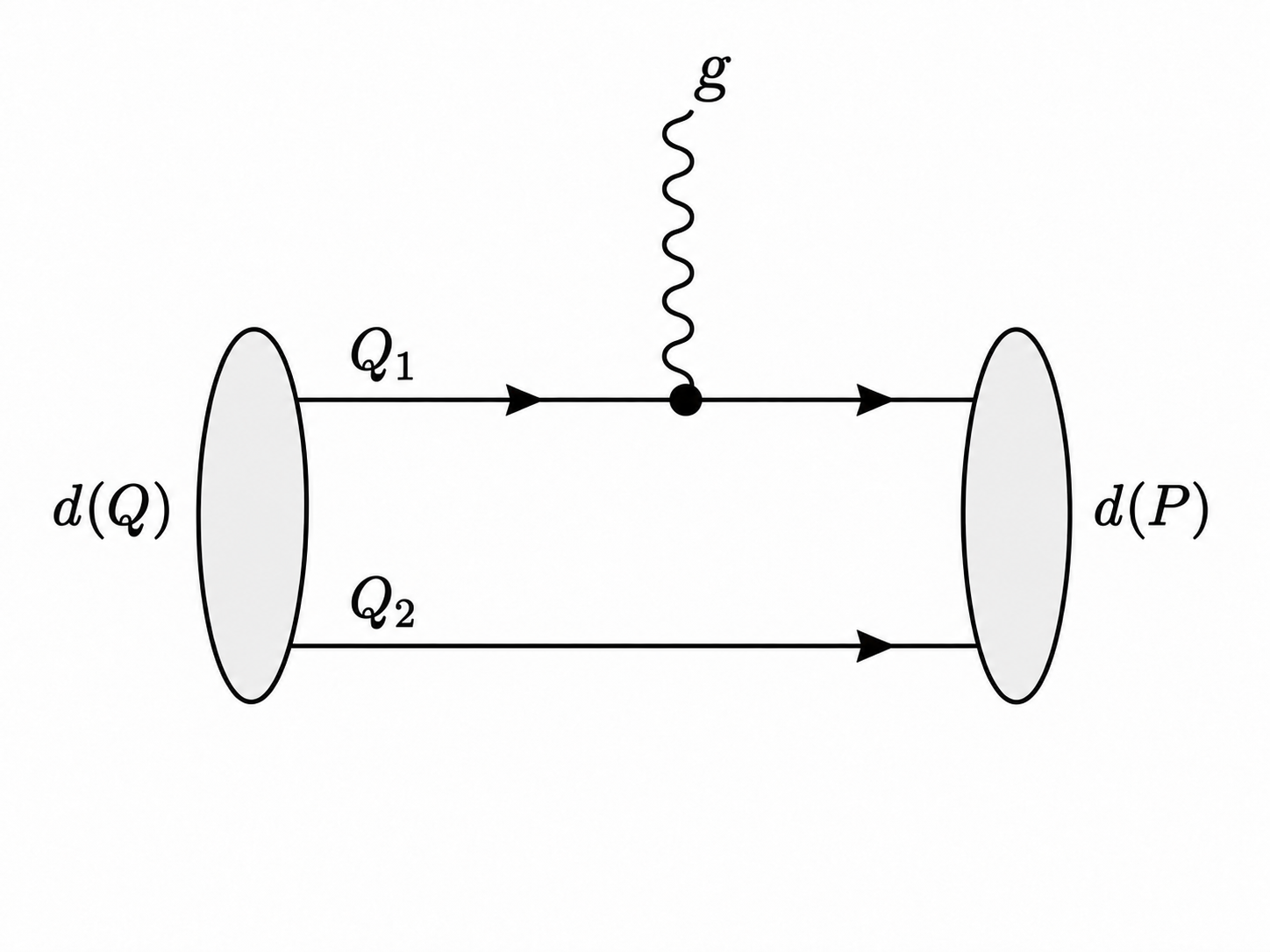}
    \par\vspace{3pt}
    \small (a) First quark interaction with a gluon
    \end{minipage}
    \hfill
    \begin{minipage}{0.48\textwidth}
    \centering
    \includegraphics[width=0.75\linewidth]{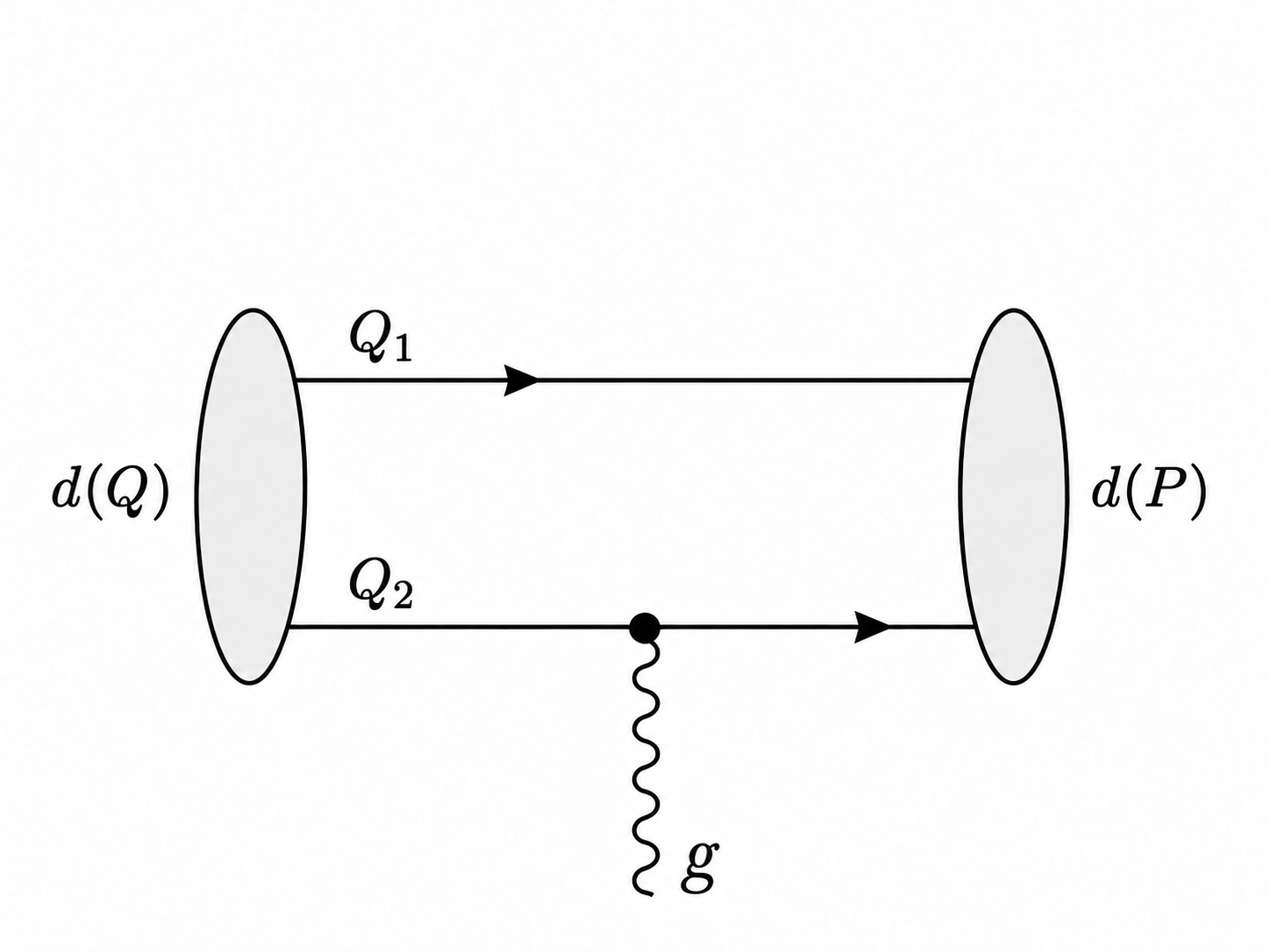}
    \par\vspace{3pt}
    \small (b) Second quark interaction with a gluon
    \end{minipage}
    \caption{Feynman diagrams of quark--gluon interactions within a diquark}
    \label{fig:Feynman_OGE}
\end{figure*}

The finite size of the diquark is taken into account through the elastic matrix elements of the quark current $J_{\mu}$ between the initial and final diquark states. These matrix elements define the diquark--gluon vertex and can be expressed in terms of the corresponding wave functions as~\cite{Baryo2002,Baryo2005,Savch2022Sym}
\begin{equation}\label{eq:dicurr}
\bra{d(P)} J_{\mu} \ket{d(Q)} = \int\frac{d^{3}p\,d^{3}q}{(2\pi)^{6}} \ \overline{\Psi}_{\mathrm{d}}^{P}(\mathbf{p}) \Gamma_{\mu}(\mathbf{p},\mathbf{q}) \Psi_{\mathrm{d}}^{Q}(\mathbf{q}),
\end{equation}
where $\Psi_{\mathrm{d}}^{Q,P}(\mathbf{q},\mathbf{p})$ are diquark wave functions, transformed from the rest frame to the corresponding moving frames. In the leading approximation, the gluon couples to either of the two constituent quarks while the other quark acts as a spectator. The vertex function $\Gamma_{\mu}(\mathbf{p},\mathbf{q})$ is therefore given by
\begin{widetext}
\begin{equation}\label{eq:Gamma}
\Gamma_{\mu}(\mathbf{p},\mathbf{q}) \approx \Gamma_{\mu}^{(1)}(\mathbf{p},\mathbf{q}) = \overline{u}_{1}(p_{1})\gamma_{\mu}u_{1}(q_{1})(2\pi)^{3}\delta^{(3)}(\mathbf{p}_{2}-\mathbf{q}_{2}) + \overline{u}_{2}(p_{2})\gamma_{\mu}u_{2}(q_{2})(2\pi)^{3}\delta^{(3)}(\mathbf{p}_{1}-\mathbf{q}_{1}),
\end{equation}
\end{widetext}
where $q_{i}$ and $p_{i}$ are the initial and final four--momenta of the constituent quarks, respectively. Thus, the two contributions describe the interaction of the gluon with the first or the second quark in the diquark (see Fig.~\ref{fig:Feynman_OGE}).

\par
The elastic current matrix elements can be decomposed into independent Lorentz structures multiplied by invariant form factors. For the scalar diquark ($\mathrm{S}$), only one form factor $h_{+}(k^{2})$ is required, while the axialvector diquark ($\mathrm{A}$) is described by three form factors $h_{1,2,3}(k^{2})$ \cite{Baryo2002,Savch2022Sym}:
\begin{widetext}
\begin{align}
\bra{S(P)} J_{\mu} \ket{S(Q)} = &+h_{+}(k^{2})(P+Q)_{\mu},\label{eq:scalar_current}
\\
\bra{A(P)} J_{\mu} \ket{A(Q)} = &-h_{1}(k^{2}) \big( \epsilon_{\mathrm{d}}^{*}(P)\cdot\epsilon_{\mathrm{d}}(Q)\big) (P+Q)_{\mu} \notag
\\
 &+h_{2}(k^{2})\Big( \big( \epsilon_{\mathrm{d}}^{*}(P) \cdot Q \big) \epsilon_{\mathrm{d};\mu}(Q) + \big( \epsilon_{\mathrm{d}}(Q) \cdot P \big) \epsilon_{\mathrm{d};\mu}^{*}(P) \Big) \notag
\\
 &+h_{3}(k^{2}) \frac{1}{M_{\mathrm{A}}^{2}} \big( \epsilon_{\mathrm{d}}^{*}(P) \cdot Q \big) \big( \epsilon_{\mathrm{d}}(Q) \cdot P \big) (P+Q)_{\mu},
\label{eq:axial_current}
\end{align}
\end{widetext}
\noindent
where 
\begin{itemize}
	\item $k=P-Q$ is the four--momentum transfer; 
	
	\item $M_{\mathrm{A}}$ is the axialvector diquark mass;
	
	\item $\epsilon_{\mathrm{d}}(Q)$ and $\epsilon_{\mathrm{d}}^{*}(P)$ are diquarks initial and final polarization vectors, respectively. 
\end{itemize}
\noindent
The quantities $h_{+}$ and $h_{1,2,3}$ are therefore the coefficients of the Lorentz decomposition of the elastic current.
\par
\begin{figure*}
    \centering
    \begin{minipage}{0.48\textwidth}
    \centering
    \includegraphics[width=0.99\linewidth]{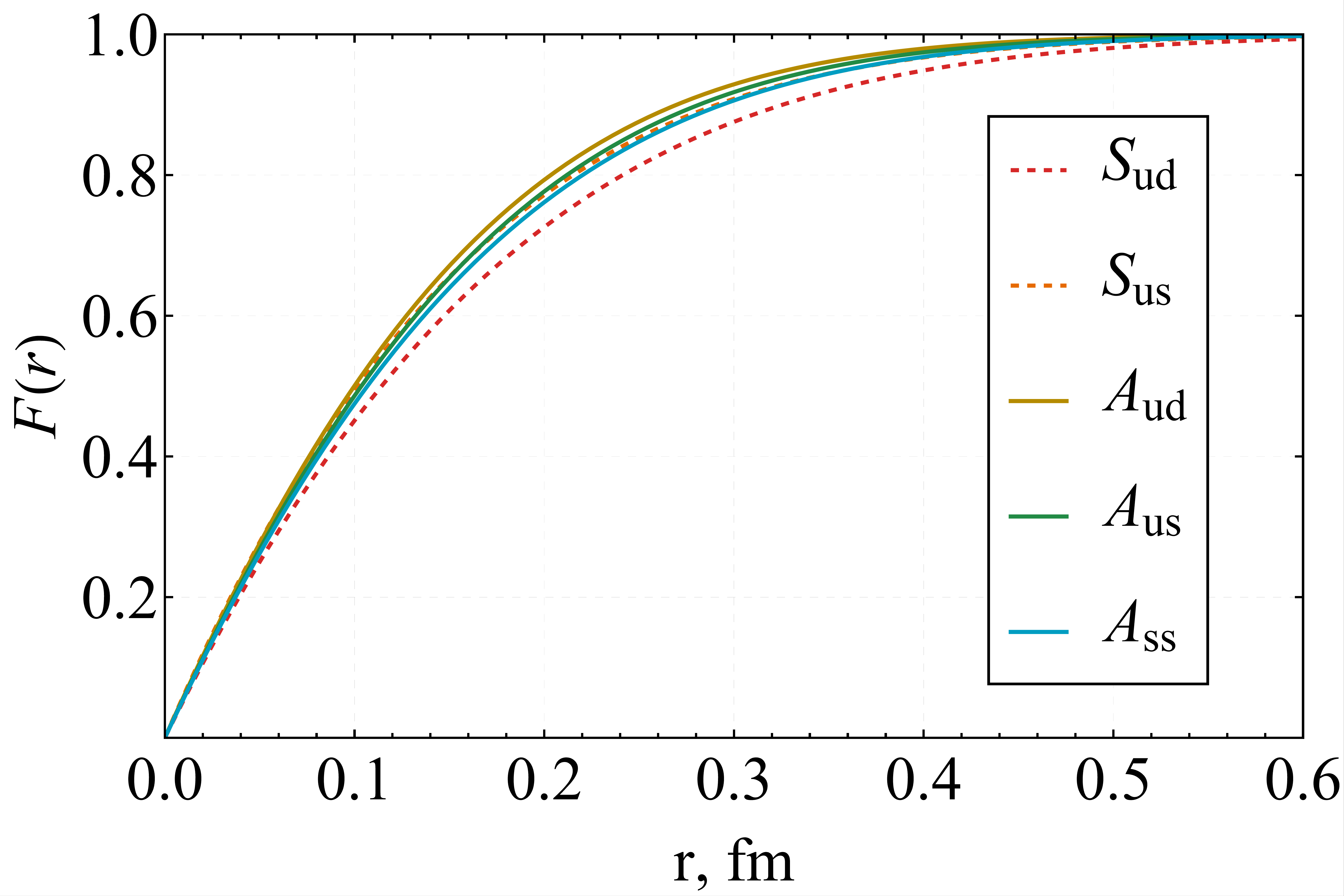}
    \par\vspace{3pt}
    \small (a) light diquarks
    \end{minipage}
    \hfill
    \begin{minipage}{0.48\textwidth}
    \centering
    \includegraphics[width=0.99\linewidth]{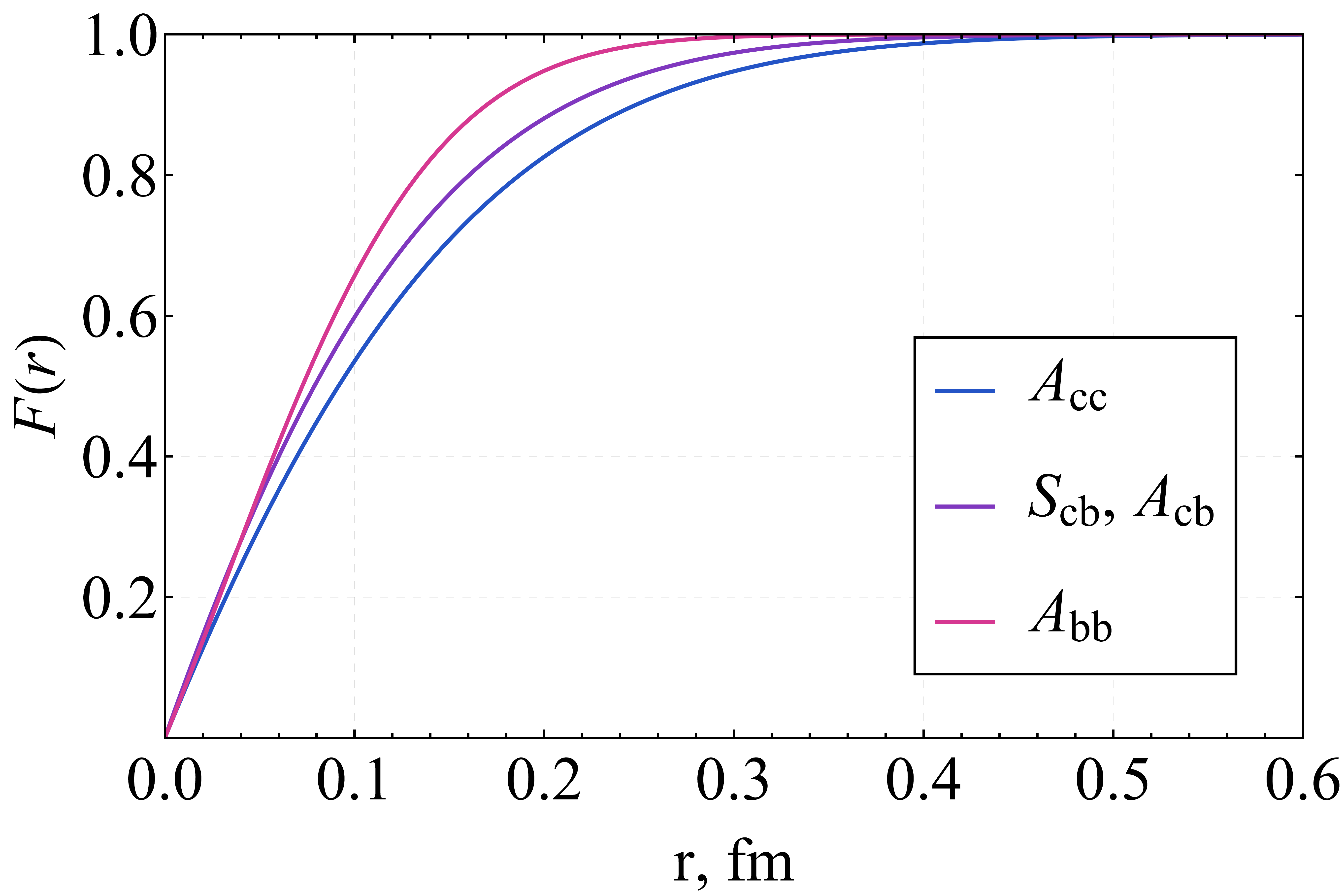}
    \par\vspace{3pt}
    \small (b) heavy diquarks
    \end{minipage}
    \caption{Form factors of heavy and light diquarks.}
    \label{fig:formFactor}
\end{figure*}

In the adopted relativistic quark model, evaluating these matrix elements in terms of the diquark wave functions yields \cite{Baryo2005,Tetra2007,Savch2022Sym}
\begin{equation}\label{eq:hFk}
\begin{aligned}
h_{+}(k^{2})=h_{1}(k^{2})=h_{2}(k^{2}) &= F(\mathbf{k}^{2}),
\\
h_{3}(k^{2}) &= 0,
\end{aligned}
\end{equation}
\noindent
where the function $F(\mathbf{k}^{2})$ is obtained from the overlap integral of the relativistic diquark wave functions~\cite{Savch2022Sym}. In the configuration space, the finite size form factor $F(r)$ is defined through the Fourier transform of $F(\mathbf{k}^{2})/\mathbf{k}^{2}$, multiplied by $r$. The resulting function is approximated with high accuracy by~\cite{Baryo2002,Tetra2007}
\begin{equation}\label{eq:F}
F(r) = 1 - e^{-\xi r-\zeta r^{2}},
\end{equation}
\noindent
where $\xi$ and $\zeta$ depend on the flavor composition and spin of the diquark. For the diquark--antidiquark interaction, the convolution of the corresponding form factors can be accurately approximated by the product $F_{1}(r)F_{2}(r)$ in the Coulomb--like term~\cite{Savch2022Sym}
\begin{equation}\label{eq:Coul}
V_{\mathrm{Coul}}(r) = -\frac{4}{3}\alpha_{\mathrm{s}}\frac{F_{1}(r)F_{2}(r)}{r}.
\end{equation}
\noindent
Since
\begin{equation}\label{eq:Frto0}
F_{i}(r) \mathrel{\underset{r \to 0}{\scalebox{2.5}[1]{$\rightarrow$}}} \xi_{i}r + O(r^{2}),
\end{equation}
\noindent
the modified Coulomb interaction behaves as 
\begin{equation}\label{eq:Vrto0}
V_{\mathrm{Coul}}(r) \mathrel{\underset{r \to 0}{\scalebox{2.5}[1]{$\sim$}}} -\frac{4}{3}\alpha_{\mathrm{s}}\xi_{1}\xi_{2}r
\end{equation}
\noindent
and has no $\dfrac{1}{r}$ singularity. At large distances
\begin{equation}\label{eq:Frtoinfty}
F_{i}(r) \mathrel{\underset{r \to \infty}{\scalebox{2.5}[1]{$\rightarrow$}}} 1
\end{equation}
\noindent
and the standard Coulomb--like interaction is recovered (see Fig.~\ref{fig:formFactor}). The form factors are not introduced into the confining potential, since the diquark behaves effectively as a pointlike color source on the corresponding long--distance scale.

\par
The quasipotential of the diquark--antidiquark interaction in configuration space is obtained by replacing the constituent energies $E_{i}(\mathbf{p})$ (Eq.~\eqref{eq:Edp}) with their on--mass--shell values $E_{i}(M)$ (Eq.~\eqref{eq:EdM}) in order to construct a local interaction. Here $M_{1,2}$ and $E_{1,2}(M)$ are the masses and on--mass--shell energies of the diquark and antidiquark. The resulting quasipotential contains all spin--independent and spin--dependent relativistic contributions considered in the model and has the form \cite{Tetra2007,Savch2022Sym,Savch2024EPJA}:
\begin{widetext}
\begin{subequations}\label{eq:Vr}
\begin{align}
V(r;M) = &\Bigg\{ V_{\mathrm{Coul}}(r) + V_{\mathrm{conf}}(r) + \frac{1}{E_{1}E_{2}} \bigg( \mathbf{p} \Big( V_{\mathrm{Coul}}(r) + V_{\mathrm{conf}}^{\mathrm{V}}(r) \Big) \mathbf{p} - \frac{1}{4}\Delta V_{\mathrm{conf}}^{\mathrm{V}}(r) + V_{\mathrm{Coul}}^{\prime}(r) \frac{\mathbf{L}^{2}}{2r} \bigg) \Bigg\}_{a} \label{eq:Vra}
\\[2pt]
+ &\Bigg\{ \frac{1}{2} \bigg( \frac{1}{E_{1}(E_{1}+M_{1})} + \frac{1}{E_{2}(E_{2}+M_{2})} \bigg)\frac{V_{\mathrm{Coul}}^{\prime}(r)}{r} - \frac{1}{2} \bigg( \frac{1}{M_{1}(E_{1}+M_{1})} + \frac{1}{M_{2}(E_{2}+M_{2})} \bigg) \frac{V_{\mathrm{conf}}^{\prime}(r)}{r} \notag
\\[2pt]
&\quad + \frac{\mu_{\mathrm{d}}}{4} \bigg( \frac{1}{M_{1}^{2}} + \frac{1}{M_{2}^{2}} \bigg) \frac{V_{\mathrm{conf}}^{\mathrm{V}\prime}(r)}{r} + \frac{1}{E_{1}E_{2}} \bigg( V_{\mathrm{Coul}}^{\prime}(r) + \frac{\mu_{\mathrm{d}}}{4} \Big( \frac{E_{1}}{M_{1}} + \frac{E_{2}}{M_{2}} \Big) V_{\mathrm{conf}}^{\mathrm{V}\prime}(r) \bigg) \frac{1}{r} \Bigg\}_{b} \bigg( \mathbf{L}\cdot(\mathbf{S}_{1}+\mathbf{S}_{2}) \bigg) \label{eq:Vrb}
\\[2pt]
+ &\Bigg\{ \frac{1}{2} \bigg( \frac{1}{E_{1}(E_{1}+M_{1})} - \frac{1}{E_{2}(E_{2}+M_{2})} \bigg) \frac{V_{\mathrm{Coul}}^{\prime}(r)}{r} - \frac{1}{2} \bigg( \frac{1}{M_{1}(E_{1}+M_{1})} - \frac{1}{M_{2}(E_{2}+M_{2})} \bigg) \frac{V_{\mathrm{conf}}^{\prime}(r)}{r} \notag
\\[2pt]
&\quad + \frac{\mu_{\mathrm{d}}}{4} \bigg( \frac{1}{M_{1}^{2}}-\frac{1}{M_{2}^{2}} \bigg) \frac{V_{\mathrm{conf}}^{\mathrm{V}\prime}(r)}{r} + \frac{\mu_{\mathrm{d}}}{4} \frac{1}{E_{1}E_{2}} \bigg( \frac{E_{1}}{M_{1}} - \frac{E_{2}}{M_{2}} \bigg) \frac{V_{\mathrm{conf}}^{\mathrm{V}\prime}(r)}{r} \Bigg\}_{c} \bigg( \mathbf{L}\cdot(\mathbf{S}_{1}-\mathbf{S}_{2}) \bigg) \label{eq:Vrc}
\\[2pt]
+ &\Bigg\{ \frac{1}{3E_{1}E_{2}} \bigg( \frac{1}{r} V_{\mathrm{Coul}}^{\prime}(r) - V_{\mathrm{Coul}}^{\prime\prime}(r) + \frac{\mu_{\mathrm{d}}^{2}}{4} \frac{E_{1}E_{2}}{M_{1}M_{2}} \Big( \frac{1}{r} V_{\mathrm{conf}}^{\mathrm{V}\prime}(r) - V_{\mathrm{conf}}^{\mathrm{V}\prime\prime}(r) \Big) \bigg) \Bigg\}_{d} \bigg( \frac{3}{r^{2}} \Big( \mathbf{S}_{1}\cdot\mathbf{r}\Big) \Big( \mathbf{S}_{2}\cdot\mathbf{r}\Big) - \Big( \mathbf{S}_{1}\cdot\mathbf{S}_{2} \Big) \bigg) \label{eq:Vrd}
\\[2pt]
+ &\Bigg\{ \frac{2}{3E_{1}E_{2}} \bigg( \Delta V_{\mathrm{Coul}}(r) + \frac{\mu_{\mathrm{d}}^{2}}{4} \frac{E_{1}E_{2}}{M_{1}M_{2}} \Delta V_{\mathrm{conf}}^{\mathrm{V}}(r) \bigg) \Bigg\}_{e} \bigg( \mathbf{S}_{1}\cdot\mathbf{S}_{2} \bigg). \label{eq:Vre}
\end{align}
\end{subequations}
\end{widetext}
\noindent
Here 
\begin{itemize}
	\item $\mathbf{p}$ is the relative momentum;
	
	\item $M_{1,2}$ are masses of diquark and antidiquark;
	
	\item $E_{1,2}$ are on--mass--shell energies of diquark and antidiquark (Eq.~\eqref{eq:EdM});
	
	\item $\mathbf{L}$ is the relative orbital angular momentum of the diquark--antidiquark system;
	
	\item $\mu_{\mathrm{d}}$ is the total chromomagnetic moment of the axialvector diquark chosen to be zero to eliminate the long--range chromomagnetic interactions;
	
	\item $\mathbf{S}_{1,2}$ are spins of diquark and antidiquark.
\end{itemize}
\noindent
The potential includes the spin--independent interaction (term~\eqref{eq:Vra}), the symmetric (term~\eqref{eq:Vrb}) and antisymmetric (term~\eqref{eq:Vrc}) spin--orbit interactions, the tensor interaction (term~\eqref{eq:Vrd}), and the spin--spin interaction (term~\eqref{eq:Vre}).

\subsection{Mixing of states}
\label{subsec:mixing}

\par
For fixed orbital angular momentum $L$, total spin $S$, and total angular momentum $J$, we evaluate the matrix elements of the operators entering Eq.~\eqref{eq:Vr} in the $\ket{L \ S \ J}$ basis, where 
\begin{equation}\label{eq:LSJ}
\begin{aligned}
\mathbf{J} &= \mathbf{L} + \mathbf{S},
\\
\mathbf{S} &= \mathbf{S}_{1} + \mathbf{S}_{2},
\\
\mathbf{K}^{2}\ket{L \ S \ J} &= K(K+1)\ket{L \ S \ J} \quad \mathbf{K} = \mathbf{J},\mathbf{L},\mathbf{S}_{(1,2)}
\end{aligned}
\end{equation}
\begin{itemize}
	\item The orbital term $\mathbf{L}^{2}$ (in term~\eqref{eq:Vra}) is diagonal
	\begin{equation}\label{eq:LLmatrel}
	\bra{L \ S \ J} \mathbf{L}^{2} \ket{L^{\prime} \ S^{\prime} \ J^{\prime}} = L(L+1) \cdot \delta_{L,L^{\prime}} \delta_{S,S^{\prime}} \delta_{J,J^{\prime}}.
	\end{equation}
	
	\item The spin--orbit operators $\mathrm{LS}_{\pm} \equiv \mathbf{L} \cdot \big( \mathbf{S}_{1} \pm \mathbf{S}_{2} \big)$ matrix elements (in terms~\eqref{eq:Vrb},\eqref{eq:Vrc}) can be expressed through the Wigner $6j$--symbols~\cite{MatrixEl2004,Savch2024EPJA}
	\begin{widetext}
	\begin{equation}\label{eq:LSmatrel}
	\begin{aligned}
	\bra{L \ S \ J} \mathrm{LS}_{\pm} \ket{L^{\prime} \ S^{\prime} \ J^{\prime}} = &(-1)^{L+S+J+S_{1}+S_{2}+1} \sqrt{L(L+1)(2L+1)(2S+1)(2S^{\prime}+1)}
	\begin{Bmatrix}
	L & S & J \\
	S^{\prime} & L & 1
	\end{Bmatrix}
	\\
	\cdot \Big( &(-1)^{S^{\prime}} \sqrt{S_{1}(S_{1}+1)(2S_{1}+1)}
	\begin{Bmatrix}
	S_{1} & S_{1} & 1 \\
	S^{\prime} & S & S_{2}
	\end{Bmatrix}
	\\
	\pm &(-1)^{S} \sqrt{S_{2}(S_{2}+1)(2S_{2}+1)}
	\begin{Bmatrix}
	S_{2} & S_{2} & 1 \\
	S^{\prime} & S & S_{1}
	\end{Bmatrix}
	\Big) \cdot \delta_{L,L^{\prime}} \delta_{J,J^{\prime}}.
	\end{aligned}
	\end{equation}
	\end{widetext}
	\noindent
	The symmetric combination can also be evaluated directly using $\mathbf{L} \cdot \big( \mathbf{S}_{1} + \mathbf{S}_{2} \big) = \mathbf{L} \cdot \mathbf{S}$:
	\begin{widetext}
	\begin{equation}\label{eq:LSplusmatrel}
	\bra{L \ S \ J} \mathrm{LS}_{+} \ket{L^{\prime} \ S^{\prime} \ J^{\prime}} = \frac{1}{2} \Big( J(J+1)-L(L+1)-S(S+1) \Big) \cdot \delta_{L,L^{\prime}} \delta_{S,S^{\prime}} \delta_{J,J^{\prime}}.
	\end{equation}
	\end{widetext}
	\noindent
	Thus, the symmetric spin--orbit interaction is diagonal in $L$, $S$, and $J$, whereas the antisymmetric one can mix states with different total spins $S$ at fixed $L$ and $J$.
	
	\item The tensor interaction (in term~\eqref{eq:Vrd}) is governed by the operator
	\begin{equation}\label{eq:T}
	\mathrm{T} \equiv \frac{3}{r^{2}} \big( \mathbf{S}_{1} \cdot \mathbf{r} \big) \big( \mathbf{S}_{2} \cdot \mathbf{r} \big) - \big( \mathbf{S}_{1} \cdot \mathbf{S}_{2} \big).
	\end{equation}
	\noindent
	Treating its orbital and spin parts as irreducible tensors of rank two, its matrix elements can be expressed through the Wigner $3j$--, $6j$--, and $9j$--symbols \cite{MatrixEl2004,Savch2024EPJA}:
	\begin{widetext}
	\begin{equation}\label{eq:Tmatrel}
	\begin{aligned}
	\bra{L \ S \ J} \mathrm{T} \ket{L^{\prime} \ S^{\prime} \ J^{\prime}} = (-1)^{S+J} &\sqrt{30(2L+1)(2L^{\prime}+1)(2S+1)(2S^{\prime}+1)}
	\\
	\cdot &\sqrt{S_{1}(S_{1}+1)(2S_{1}+1)} \sqrt{S_{2}(S_{2}+1)(2S_{2}+1)}
	\\
	\cdot 
	&\begin{pmatrix}
	L^{\prime} & 2 & L \\
	0 & 0 & 0
	\end{pmatrix}
	\begin{Bmatrix}
	J & S^{\prime} & L^{\prime} \\
	2 & L & S
	\end{Bmatrix}
	\begin{Bmatrix}
	S_{1} & S_{2} & S \\
	S_{1} & S_{2} & S^{\prime} \\
	1 & 1 & 2
	\end{Bmatrix} \cdot \delta_{J,J^{\prime}}.
	\end{aligned}
	\end{equation}
	\end{widetext}
	\noindent
	The tensor matrix element is diagonal in $J$ but can be non-diagonal in $L$ and $S$. In the present calculations the tetraquark states are classified by a fixed orbital excitation $L$, and the corresponding matrix elements within each $L$ sector are used.
	
	\item The spin--spin operator matrix element (in term~\eqref{eq:Vre}) can be obtained the same way as the $\mathrm{LS}_{+}$ term, which gives
	\begin{widetext}
	\begin{equation}\label{eq:SSmatrel}
	\bra{L \ S \ J} \mathbf{S}_{1} \cdot \mathbf{S}_{2} \ket{L^{\prime} \ S^{\prime} \ J^{\prime}} = \frac{1}{2}\Big( S(S+1) - S_{1}(S_{1}+1) - S_{2}(S_{2}+1) \Big) \cdot \delta_{L,L^{\prime}} \delta_{S,S^{\prime}} \delta_{J,J^{\prime}}.
	\end{equation}
	\end{widetext}
\end{itemize}
\par
The non-diagonal spin--orbit~\eqref{eq:LSmatrel} and tensor~\eqref{eq:Tmatrel} matrix elements lead to mixing within the same excitation $nL$ between states with the same $J^{P}$, but different total spins $S$. This is relevant in particular for the $\mathrm{A\overline{A}}$ configurations, where $S=0,1,2$. For the $\mathrm{A\overline{S}}$ and $\mathrm{S\overline{A}}$ configurations the total spin is $S=1$, while for the $\mathrm{S\overline{S}}$ configuration it is $S=0$. Thus, within a fixed diquark--antidiquark configuration these states do not require mixing between different values of $S$.
\par
We account for the mixing by constructing the mass matrix $\mathcal{M}_{L,J}$ in the basis of allowed spin states. Diagonal matrix elements are $M_{L,J}(S,S)$ and nondiagonal matrix elements are 
\begin{equation}\label{eq:DeltaM}
\Delta M_{L,J}(S,S^{\prime}) \equiv (M_{L,J}(S,S^{\prime}))_{\text{full}} - (M_{L,J}(S,S^{\prime}))_{S\text{--indep.}},
\end{equation}
\noindent
where 
\begin{itemize}
	\item $(M_{L,J}(S,S^{\prime}))_{\text{full}}$ is calculated with the full Eq.~\eqref{eq:Vr} quasipotential,
	
	\item $(M_{L,J}(S,S^{\prime}))_{S\text{--indep.}}$ is calculated with only the spin--independent part of Eq.~\eqref{eq:Vr} (term~\eqref{eq:Vra}). 
\end{itemize}
\noindent
Masses of the mixed states are obtained as its eigenvalues,
\begin{equation}\label{eq:mixEigen}
M_{L,J}^{(i)}=\lambda_{i}(\mathcal{M}_{L,J}).
\end{equation}
\noindent
For axialvector diquark--antidiquark configurations, the mixing matrices have the following structures:
\begin{equation}\label{eq:mixMatrix}
\begin{aligned}
\mathcal{M}_{L,J}^{(3)} &=
\begin{pmatrix}
M_{L,J}(0,0) & \Delta M_{L,J}(0,1) & \Delta M_{L,J}(0,2) \\
\Delta M_{L,J}(1,0) & M_{L,J}(1,1) & \Delta M_{L,J}(1,2) \\
\Delta M_{L,J}(2,0) & \Delta M_{L,J}(2,1) & M_{L,J}(2,2)
\end{pmatrix},
\\
\mathcal{M}_{L,J}^{(2)} &=
\begin{pmatrix}
M_{L,J}(1,1) & \Delta M_{L,J}(1,2) \\
\Delta M_{L,J}(2,1) & M_{L,J}(2,2)
\end{pmatrix}.
\end{aligned}
\end{equation}
\noindent
The $3\times 3$ matrices arise for the orbitally excited states with $J=L$, e.g., $P$--wave states with $J=1$ and the $D$--wave states with $J=2$. The $2\times 2$ matrices correspond to the $P$--wave states with $J=2$ and the $D$--wave states with $J=1,3$. For states with the maximal and minimal value of $J$ a single value of $S$ is allowed, therefore, no diagonalization is needed. In the considered doubly heavy tetraquarks, the diquark and antidiquark masses are different, so the antisymmetric spin--orbit interaction may give a significant contribution to the mixing. The spin--dependent matrix elements are evaluated using Eqs.~\eqref{eq:LSmatrel}--\eqref{eq:SSmatrel}.

\section{Numerical results}
\label{sec:numres}

\subsection{Numerical task}
\label{subsec:num}

\par
The quasipotential equation~\eqref{eq:Schr} is solved numerically in configuration space. Its main feature is the nonlinear dependence on the unknown bound--state mass $M$, which enters the relativistic reduced mass $\mu_{\mathrm{R}}(M)$~\eqref{eq:muR}, the squared on--mass--shell relative momentum $b^{2}(M)$~\eqref{eq:b2}, the constituent energies $E_{1,2}(M)$~\eqref{eq:Edi}, and the interaction quasipotential itself~\eqref{eq:Vr}. We therefore determine the mass and wave function self--consistently using the method of successive approximations~\cite{Faustov1990eng}.
\par
The quantity $\mathcal{E}(M)$, which plays the role of the eigenvalue of the relativistic Schr\"odinger--type quasipotential equation~\eqref{eq:Schr}, is introduced as
\begin{widetext}
\begin{equation}\label{eq:E.eigen}
\mathcal{E}(M) = \frac{b^{2}(M)}{2\mu_{\mathrm{R}}(M)} = \frac{M\big[M^{2}-(M_{1}+M_{2})^{2}\big]\big[M^{2}-(M_{1}-M_{2})^{2}\big]}{2\big[M^{4}-(M_{1}^{2}-M_{2}^{2})^{2}\big]},
\end{equation}
\end{widetext}
\noindent
where $M_{1,2}$ are the masses of the constituents. Then the relativistic Schr\"odinger--type quasipotential equation~\eqref{eq:Schr} can be rewritten and treated as an eigenvalue problem: 
\begin{widetext}
\begin{equation}\label{eq:eigenSchr}
\frac{\mathbf{p}^{2}}{2\mu_{\mathrm{R}}(M)} \Psi_{\mathrm{d,T}}(\mathbf{p}) + \int\frac{d^{3}q}{(2\pi)^{3}} V(\mathbf{p},\mathbf{q};M) \Psi_{\mathrm{d,T}}(\mathbf{q}) = \mathcal{E}(M) \Psi_{\mathrm{d,T}}(\mathbf{p}).
\end{equation}
\end{widetext}
\noindent
The full wave function in the $LSJ$ basis can be expressed as
\begin{widetext}
\begin{equation}\label{eq:PsinLSJ}
\Psi_{nLSJM_{J}}(\mathbf{r}) = R_{nL}(r) \sum_{M_{L},M_{S}} \bra{L \ M_{L} \ S \ M_{S}} \ket{J \ M_{J}} Y_{LM_{L}}(\theta,\varphi) \chi_{SM_{S}},
\end{equation}
\end{widetext}
\noindent
where 
\begin{itemize}
	\item $n=n_{r}+1$ is the radial excitation number;
	
	\item $M_{K} = -K, \ldots, +K$, $K=L,S,J$ are projections of corresponding moments;
	
	\item $R_{nL}(r)Y_{LM_{L}}(\theta,\varphi)$ is the spatial part of the wave function of a bound state, expanded in terms of spherical harmonics $Y_{LM_{L}}(\theta,\varphi)$
	
	\item $\bra{L \ M_{L} \ S \ M_{S}} \ket{J \ M_{J}}$ are the Clebsch--Gordan coefficients;
	
	\item $\chi_{SM_{S}}$ is the spin wave function;
\end{itemize}
\noindent
Introducing the reduced radial wave function $y_{n_{r}L}(r) = rR_{nL}(r)$, we obtain the radial eigenvalue equation~\cite{WolframSchr1985}
\begin{widetext}
\begin{equation}\label{eq:radial}
\Bigg( \frac{1}{2\mu_{\mathrm{R}}(M)} \bigg( -\frac{d^{2}}{dr^{2}} + \frac{L(L+1)}{r^{2}} \bigg) + V_{LSJ}(r;M) \Bigg) y_{n_{r}L}(r) = \mathcal{E}(M) y_{n_{r}L}(r).
\end{equation}
\end{widetext}
\noindent
The effective radial quasipotential $V_{LSJ}(r;M)$ is obtained from Eq.~\eqref{eq:Vr} by evaluating the matrix elements of the orbital, spin--orbit, tensor, and spin--spin operators using Eqs.~\eqref{eq:LLmatrel}, \eqref{eq:LSmatrel}, \eqref{eq:Tmatrel}, and~\eqref{eq:SSmatrel}. The momentum--dependent terms are retained as differential operators in configuration space.
\par
For a fixed set of quantum numbers $\{ n_{r},L,S,J \}$, the radial equation~\eqref{eq:radial} is solved by the shooting method in Wolfram Mathematica~\cite{WolframSchr1999}. The eigenvalue $\mathcal{E}$ is selected so that the radial wave function is regular at the origin, has the prescribed number of nodes, and decreases at large distances. The corresponding mass is reconstructed numerically from Eq.~\eqref{eq:E.eigen}. The initial reduced mass is chosen in the nonrelativistic limit,
\begin{equation}\label{eq:muR0}
\mu_{\mathrm{R}}^{(0)} = \frac{M_{1}M_{2}}{M_{1}+M_{2}}.
\end{equation}
\noindent
The reduced mass $\mu_{\mathrm{R}}^{(i)}$, eigenvalue $\mathcal{E}^{(i)}$, bound--state mass $M^{(i)}$, constituent energies $E_{1,2}(M^{(i)})$, and quasipotential $V_{LSJ}(r;M^{(i)})$ are then recalculated iteratively
\begin{widetext}
\begin{equation}\label{eq:iter}
\begin{aligned}
\mu_{\mathrm{R}}^{(i)} \xrightarrow{\text{numerical solution of Eq.~\eqref{eq:radial}}} \mathcal{E}^{(i)} \xrightarrow{\text{from Eq.~\eqref{eq:E.eigen}}} &M^{(i)} \xrightarrow{\text{from Eq.~\eqref{eq:muR}}} \mu_{\mathrm{R}}(M^{(i)}) \equiv \mu_{\mathrm{R}}^{(i+1)},
\\
&M^{(i)} \xrightarrow{\text{from Eq.~\eqref{eq:Edi}}} E_{1,2}(M^{(i)}) \xrightarrow{\text{Eqs.~\eqref{eq:Vr},\eqref{eq:LLmatrel}-\eqref{eq:SSmatrel}}} V_{LSJ}(r;M^{(i)}).
\end{aligned}
\end{equation}
\end{widetext}
\noindent
The iterations are continued until the required precision ($\varepsilon_{M}$, $\varepsilon_{\mu_{\mathrm{R}}}$) is reached
\begin{equation}\label{eq:converge}
\begin{cases}
\big| M^{(i+1)} - M^{(i)} \big| < \varepsilon_{M}, \\
\big| \mu_{\mathrm{R}}^{(i+1)} - \mu_{\mathrm{R}}^{(i)} \big| < \varepsilon_{\mu_{\mathrm{R}}}.
\end{cases}
\end{equation}
\par
For orbitally excited states with the same $J^{P}$ but different total spins $S$, the nondiagonal spin--orbit and tensor matrix elements lead to mixing. As described in Sec.~\ref{subsec:mixing}, the corresponding mass matrices~\eqref{eq:mixMatrix} are constructed using the spin--dependent contributions~\eqref{eq:DeltaM}, and the mixed--state masses are obtained as their eigenvalues~\eqref{eq:mixEigen}. For unmixed states, a single radial equation is sufficient.
\par
All free model parameters were fixed previously from the spectroscopy of ordinary mesons and baryons and are not readjusted for tetraquarks~\cite{Param1986,Param1990,Param1992,Param1995}. The parameters of the interaction are given in Table~\ref{tab:param}. The masses and form factor parameters of the constituent diquarks and antidiquarks are taken from previous calculations~\cite{Baryo2002,Baryo2005,Tetra2006,Tetra2007,Savch2021Uni} and listed in Table~\ref{tab:di.mff}.

\begin{table}
\caption{Parameters of the running QCD coupling and the interaction potential~\cite{Param1986,Param1990,Param1992,Param1995}.}
\label{tab:param}
\begin{ruledtabular}
\begin{tabular}{cccccc}
$\Lambda$, $\mathrm{MeV}$ & $M_{\mathrm{BG}}$ $\mathrm{GeV}$ & $A$, $\mathrm{GeV}^{2}$ & $B$, $\mathrm{GeV}$ & $\varepsilon$ & $\mu_{\mathrm{d}}$
\\
\hline
$414$ & $0.95$ & $0.18$ & $-0.30$ & $-1$ & $0$
\\
\end{tabular}
\end{ruledtabular}
\end{table}

\begin{table}
\caption{Masses $M_{\mathrm{d}}$ and form factor parameters $\xi$ and $\zeta$ (Eq.~\eqref{eq:F}) of the light and heavy diquarks and antidiquarks employed in the tetraquark mass calculations~\cite{Baryo2002,Tetra2007,Savch2021Uni}.}
\label{tab:di.mff}
\begin{ruledtabular}
\begin{tabular}{lcccc}
Quark composition & Type & $M_{\mathrm{d}}$, MeV & $\xi$, GeV & $\zeta$, GeV$^{2}$
\\
\hline
$\mathrm{ud}$ & $\mathrm{S}$ & $710$ & $1.09$ & $0.185$
\\
$\mathrm{ud}$ & $\mathrm{A}$ & $909$ & $1.185$ & $0.365$
\\
$\mathrm{us}$ & $\mathrm{S}$ & $948$ & $1.23$ & $0.225$
\\
$\mathrm{us}$ & $\mathrm{A}$ & $1069$ & $1.15$ & $0.325$
\\
$\mathrm{ss}$ & $\mathrm{A}$ & $1203$ & $1.13$ & $0.280$
\\
\hline
$\mathrm{cc}$ & $\mathrm{A}$ & $3226$ & $1.30$ & $0.42$
\\
$\mathrm{cb}$ & $\mathrm{S}$ & $6519$ & $1.50$ & $0.59$
\\
$\mathrm{cb}$ & $\mathrm{A}$ & $6526$ & $1.50$ & $0.59$
\\
$\mathrm{bb}$ & $\mathrm{A}$ & $9778$ & $1.30$ & $1.60$
\\
\end{tabular}
\end{ruledtabular}
\end{table}

\subsection{Mass spectra}
\label{subsec:res}

\par
The calculated spectra of doubly heavy tetraquarks $\mathrm{QQ^{\prime}\overline{qq^{\prime}}}$, where $\mathrm{Q,Q^{\prime}=c,b}$ and $\mathrm{q,q^{\prime}=u,d,s}$, are presented in Tables~\ref{tab:res.cc}--\ref{tab:res.bb}. We consider the ground states $1S$, orbital excitations $1P,1D$, and radial excitations $2S,2P,3S$. The constituent diquarks and antidiquarks remain in their internal ground states, the listed excitations refer exclusively to their relative motion. The tables specify the constituent spin configuration, the quantum numbers $n_{r}$, $L$, $S$, and $J^{P}$, and the calculated masses in MeV.
\par
The theoretical uncertainties arise from both the relativistic quark model and the diquark--antidiquark approximation. Comparisons with conventional hadron spectra suggest model uncertainties of a few MeV~\cite{ReggeBaryo2011,ReggeMeson2011}, whereas the uncertainty associated with the diquark--antidiquark picture is difficult to quantify. Based on previous calculations within the same framework~\cite{Tetra2006,Tetra2007,Tetra2008}, we estimate the overall uncertainty to be approximately $20$--$50$~MeV.

\begin{table*}
\caption{Calculated masses (in MeV) of the ground, radial, and orbital excitations of doubly charmed tetraquarks.}
\label{tab:res.cc}
\centering
\begin{ruledtabular}
\begin{tabular}{ccccccccc}
$\mathrm{d\overline{d^{\prime}}}$ & $nL$ & $n_{r}$ & $L$ & $S$ & $J^{P}$ & $M_{\mathrm{cc\overline{qq^{\prime}}}}$ & $M_{\mathrm{cc\overline{sq}}}$ & $M_{\mathrm{cc\overline{ss}}}$
\\
\hline
\multirow[c]{32}{*}{$\mathrm{A\overline{A}}$} & \multirow[c]{3}{*}{$1S$} & \multirow[c]{3}{*}{$0$} & \multirow[c]{3}{*}{$0$} & $0$ & $0^{+}$ & $3983$ & $4161$ & $4308$
\\
\cline{5-9}
 &  &  &  & $1$ & $1^{+}$ & $4017$ & $4188$ & $4330$
\\
\cline{5-9}
 &  &  &  & $2$ & $2^{+}$ & $4074$ & $4233$ & $4368$
\\
\cline{2-9}
 & \multirow[c]{7}{*}{$1P$} & \multirow[c]{7}{*}{$0$} & \multirow[c]{7}{*}{$1$} & $1$ & $0^{-}$ & $4456$ & $4600$ & $4724$
\\
\cline{5-9}
 &  &  &  & $0$ & \multirow[c]{3}{*}{$1^{-}$} & $4423$ & $4575$ & $4702$
\\
 &  &  &  & $1$ &  & $4471$ & $4612$ & $4733$
\\
 &  &  &  & $2$ &  & $4500$ & $4635$ & $4753$
\\
\cline{5-9}
 &  &  &  & $1$ & \multirow[c]{2}{*}{$2^{-}$} & $4443$ & $4591$ & $4716$
\\
 &  &  &  & $2$ &  & $4500$ & $4636$ & $4754$
\\
\cline{5-9}
 &  &  &  & $2$ & $3^{-}$ & $4473$ & $4616$ & $4738$
\\
\cline{2-9}
 & \multirow[c]{3}{*}{$2S$} & \multirow[c]{3}{*}{$1$} & \multirow[c]{3}{*}{$0$} & $0$ & $0^{+}$ & $4760$ & $4897$ & $5013$
\\
\cline{5-9}
 &  &  &  & $1$ & $1^{+}$ & $4764$ & $4899$ & $5015$
\\
\cline{5-9}
 &  &  &  & $2$ & $2^{+}$ & $4768$ & $4901$ & $5016$
\\
\cline{2-9}
 & \multirow[c]{9}{*}{$1D$} & \multirow[c]{9}{*}{$0$} & \multirow[c]{9}{*}{$2$} & $2$ & $0^{+}$ & $4910$ & $5025$ & $5128$
\\
\cline{5-9}
 &  &  &  & $1$ & \multirow[c]{2}{*}{$1^{+}$} & $4878$ & $5002$ & $5109$
\\
 &  &  &  & $2$ &  & $4921$ & $5033$ & $5135$
\\
\cline{5-9}
 &  &  &  & $0$ & \multirow[c]{3}{*}{$2^{+}$} & $4834$ & $4969$ & $5083$
\\
 &  &  &  & $1$ &  & $4892$ & $5013$ & $5119$
\\
 &  &  &  & $2$ &  & $4941$ & $5049$ & $5148$
\\
\cline{5-9}
 &  &  &  & $1$ & \multirow[c]{2}{*}{$3^{+}$} & $4846$ & $4980$ & $5093$
\\
 &  &  &  & $2$ &  & $4912$ & $5029$ & $5133$
\\
\cline{5-9}
 &  &  &  & $2$ & $4^{+}$ & $4863$ & $4994$ & $5104$
\\
\cline{2-9}
 & \multirow[c]{7}{*}{$2P$} & \multirow[c]{7}{*}{$1$} & \multirow[c]{7}{*}{$1$} & $1$ & $0^{-}$ & $5150$ & $5266$ & $5368$
\\
\cline{5-9}
 &  &  &  & $0$ & \multirow[c]{3}{*}{$1^{-}$} & $5084$ & $5215$ & $5325$
\\
 &  &  &  & $1$ &  & $5149$ & $5265$ & $5366$
\\
 &  &  &  & $2$ &  & $5188$ & $5297$ & $5395$
\\
\cline{5-9}
 &  &  &  & $1$ & \multirow[c]{2}{*}{$2^{-}$} & $5095$ & $5223$ & $5332$
\\
 &  &  &  & $2$ &  & $5162$ & $5276$ & $5376$
\\
\cline{5-9}
 &  &  &  & $2$ & $3^{-}$ & $5112$ & $5237$ & $5343$
\\
\cline{2-9}
 & \multirow[c]{3}{*}{$3S$} & \multirow[c]{3}{*}{$2$} & \multirow[c]{3}{*}{$0$} & $0$ & $0^{+}$ & $5456$ & $5567$ & $5663$
\\
\cline{5-9}
 &  &  &  & $1$ & $1^{+}$ & $5454$ & $5565$ & $5660$
\\
\cline{5-9}
 &  &  &  & $2$ & $2^{+}$ & $5445$ & $5556$ & $5653$
\\
\hline
\multirow[c]{12}{*}{$\mathrm{A\overline{S}}$} & $1S$ & $0$ & $0$ & \multirow[c]{12}{*}{$1$} & $1^{+}$ & $3869$ & $4093$ & \multirow[c]{12}{*}{}
\\
\cline{2-4}
\cline{6-8}
 & \multirow[c]{3}{*}{$1P$} & \multirow[c]{3}{*}{$0$} & \multirow[c]{3}{*}{$1$} &  & $0^{-}$ & $4291$ & $4493$ & 
\\
\cline{6-8}
 &  &  &  &  & $1^{-}$ & $4299$ & $4500$ & 
\\
\cline{6-8}
 &  &  &  &  & $2^{-}$ & $4316$ & $4514$ & 
\\
\cline{2-4}
\cline{6-8}
 & $2S$ & $1$ & $0$ &  & $1^{+}$ & $4622$ & $4801$ & 
\\
\cline{2-4}
\cline{6-8}
 & \multirow[c]{3}{*}{$1D$} & \multirow[c]{3}{*}{$0$} & \multirow[c]{3}{*}{$2$} &  & $1^{+}$ & $4721$ & $4900$ & 
\\
\cline{6-8}
 &  &  &  &  & $2^{+}$ & $4733$ & $4909$ & 
\\
\cline{6-8}
 &  &  &  &  & $3^{+}$ & $4750$ & $4923$ & 
\\
\cline{2-4}
\cline{6-8}
 & \multirow[c]{3}{*}{$2P$} & \multirow[c]{3}{*}{$1$} & \multirow[c]{3}{*}{$1$} &  & $0^{-}$ & $5000$ & $5162$ & 
\\
\cline{6-8}
 &  &  &  &  & $1^{-}$ & $5000$ & $5162$ & 
\\
\cline{6-8}
 &  &  &  &  & $2^{-}$ & $5000$ & $5162$ & 
\\
\cline{2-4}
\cline{6-8}
 & $3S$ & $2$ & $0$ &  & $1^{+}$ & $5332$ & $5479$ & 
\\
\end{tabular}
\end{ruledtabular}
\end{table*}

\begin{table*}
\caption{Calculated masses (in MeV) of the ground, radial, and orbital excitations of charmed--bottom tetraquarks in the $\mathrm{A\overline{A}}$ configuration.}
\label{tab:res.cb.AA}
\centering
\begin{ruledtabular}
\begin{tabular}{ccccccccc}
$\mathrm{d\overline{d^{\prime}}}$ & $nL$ & $n_{r}$ & $L$ & $S$ & $J^{P}$ & $M_{\mathrm{cb\overline{qq^{\prime}}}}$ & $M_{\mathrm{cb\overline{sq}}}$ & $M_{\mathrm{cb\overline{ss}}}$
\\
\hline
\multirow[c]{32}{*}{$\mathrm{A\overline{A}}$} & \multirow[c]{3}{*}{$1S$} & \multirow[c]{3}{*}{$0$} & \multirow[c]{3}{*}{$0$} & $0$ & $0^{+}$ & $7337$ & $7499$ & $7634$
\\
\cline{5-9}
 &  &  &  & $1$ & $1^{+}$ & $7353$ & $7511$ & $7645$
\\
\cline{5-9}
 &  &  &  & $2$ & $2^{+}$ & $7381$ & $7535$ & $7665$
\\
\cline{2-9}
 & \multirow[c]{7}{*}{$1P$} & \multirow[c]{7}{*}{$0$} & \multirow[c]{7}{*}{$1$} & $1$ & $0^{-}$ & $7722$ & $7863$ & $7983$
\\
\cline{5-9}
 &  &  &  & $0$ & \multirow[c]{3}{*}{$1^{-}$} & $7691$ & $7837$ & $7961$
\\
 &  &  &  & $1$ &  & $7731$ & $7870$ & $7989$
\\
 &  &  &  & $2$ &  & $7756$ & $7889$ & $8006$
\\
\cline{5-9}
 &  &  &  & $1$ & \multirow[c]{2}{*}{$2^{-}$} & $7702$ & $7847$ & $7969$
\\
 &  &  &  & $2$ &  & $7748$ & $7884$ & $8002$
\\
\cline{5-9}
 &  &  &  & $2$ & $3^{-}$ & $7718$ & $7861$ & $7982$
\\
\cline{2-9}
 & \multirow[c]{3}{*}{$2S$} & \multirow[c]{3}{*}{$1$} & \multirow[c]{3}{*}{$0$} & $0$ & $0^{+}$ & $7964$ & $8101$ & $8217$
\\
\cline{5-9}
 &  &  &  & $1$ & $1^{+}$ & $7967$ & $8103$ & $8218$
\\
\cline{5-9}
 &  &  &  & $2$ & $2^{+}$ & $7972$ & $8106$ & $8221$
\\
\cline{2-9}
 & \multirow[c]{9}{*}{$1D$} & \multirow[c]{9}{*}{$0$} & \multirow[c]{9}{*}{$2$} & $2$ & $0^{+}$ & $8100$ & $8216$ & $8319$
\\
\cline{5-9}
 &  &  &  & $1$ & \multirow[c]{2}{*}{$1^{+}$} & $8066$ & $8190$ & $8298$
\\
 &  &  &  & $2$ &  & $8107$ & $8222$ & $8324$
\\
\cline{5-9}
 &  &  &  & $0$ & \multirow[c]{3}{*}{$2^{+}$} & $8016$ & $8153$ & $8268$
\\
 &  &  &  & $1$ &  & $8075$ & $8198$ & $8305$
\\
 &  &  &  & $2$ &  & $8121$ & $8233$ & $8334$
\\
\cline{5-9}
 &  &  &  & $1$ & \multirow[c]{2}{*}{$3^{+}$} & $8024$ & $8160$ & $8274$
\\
 &  &  &  & $2$ &  & $8087$ & $8208$ & $8314$
\\
\cline{5-9}
 &  &  &  & $2$ & $4^{+}$ & $8034$ & $8168$ & $8282$
\\
\cline{2-9}
 & \multirow[c]{7}{*}{$2P$} & \multirow[c]{7}{*}{$1$} & \multirow[c]{7}{*}{$1$} & $1$ & $0^{-}$ & $8281$ & $8401$ & $8505$
\\
\cline{5-9}
 &  &  &  & $0$ & \multirow[c]{3}{*}{$1^{-}$} & $8225$ & $8356$ & $8468$
\\
 &  &  &  & $1$ &  & $8284$ & $8402$ & $8506$
\\
 &  &  &  & $2$ &  & $8314$ & $8426$ & $8526$
\\
\cline{5-9}
 &  &  &  & $1$ & \multirow[c]{2}{*}{$2^{-}$} & $8232$ & $8362$ & $8473$
\\
 &  &  &  & $2$ &  & $8292$ & $8409$ & $8512$
\\
\cline{5-9}
 &  &  &  & $2$ & $3^{-}$ & $8244$ & $8372$ & $8481$
\\
\cline{2-9}
 & \multirow[c]{3}{*}{$3S$} & \multirow[c]{3}{*}{$2$} & \multirow[c]{3}{*}{$0$} & $0$ & $0^{+}$ & $8517$ & $8633$ & $8733$
\\
\cline{5-9}
 &  &  &  & $1$ & $1^{+}$ & $8516$ & $8632$ & $8732$
\\
\cline{5-9}
 &  &  &  & $2$ & $2^{+}$ & $8513$ & $8629$ & $8729$
\\
\end{tabular}
\end{ruledtabular}
\end{table*}

\begin{table*}
\caption{Calculated masses (in MeV) of the ground, radial, and orbital excitations of charmed--bottom tetraquarks in the $\mathrm{A\overline{S}}$, $\mathrm{S\overline{A}}$, and $\mathrm{S\overline{S}}$ configurations.}
\label{tab:res.cb.other}
\centering
\begin{ruledtabular}
\begin{tabular}{ccccccccc}
$\mathrm{d\overline{d^{\prime}}}$ & $nL$ & $n_{r}$ & $L$ & $S$ & $J^{P}$ & $M_{\mathrm{cb\overline{qq^{\prime}}}}$ & $M_{\mathrm{cb\overline{sq}}}$ & $M_{\mathrm{cb\overline{ss}}}$
\\
\hline
\multirow[c]{12}{*}{$\mathrm{A\overline{S}}$} & $1S$ & $0$ & $0$ & \multirow[c]{24}{*}{$1$} & $1^{+}$ & $7198$ & $7410$ & \multirow[c]{12}{*}{}
\\
\cline{2-4}
\cline{6-8}
 & \multirow[c]{3}{*}{$1P$} & \multirow[c]{3}{*}{$0$} & \multirow[c]{3}{*}{$1$} &  & $0^{-}$ & $7552$ & $7751$ & 
\\
\cline{6-8}
 &  &  &  &  & $1^{-}$ & $7557$ & $7756$ & 
\\
\cline{6-8}
 &  &  &  &  & $2^{-}$ & $7568$ & $7765$ & 
\\
\cline{2-4}
\cline{6-8}
 & $2S$ & $1$ & $0$ &  & $1^{+}$ & $7822$ & $8004$ & 
\\
\cline{2-4}
\cline{6-8}
 & \multirow[c]{3}{*}{$1D$} & \multirow[c]{3}{*}{$0$} & \multirow[c]{3}{*}{$2$} &  & $1^{+}$ & $7903$ & $8084$ & 
\\
\cline{6-8}
 &  &  &  &  & $2^{+}$ & $7910$ & $8090$ & 
\\
\cline{6-8}
 &  &  &  &  & $3^{+}$ & $7921$ & $8100$ & 
\\
\cline{2-4}
\cline{6-8}
 & \multirow[c]{3}{*}{$2P$} & \multirow[c]{3}{*}{$1$} & \multirow[c]{3}{*}{$1$} &  & $0^{-}$ & $8123$ & $8292$ & 
\\
\cline{6-8}
 &  &  &  &  & $1^{-}$ & $8125$ & $8294$ & 
\\
\cline{6-8}
 &  &  &  &  & $2^{-}$ & $8129$ & $8297$ & 
\\
\cline{2-4}
\cline{6-8}
 & $3S$ & $2$ & $0$ &  & $1^{+}$ & $8389$ & $8545$ & 
\\
\cline{1-4}
\cline{6-9}
\multirow[c]{12}{*}{$\mathrm{S\overline{A}}$} & $1S$ & $0$ & $0$ &  & $1^{+}$ & $7361$ & $7516$ & $7649$
\\
\cline{2-4}
\cline{6-9}
 & \multirow[c]{3}{*}{$1P$} & \multirow[c]{3}{*}{$0$} & \multirow[c]{3}{*}{$1$} &  & $0^{-}$ & $7749$ & $7883$ & $7999$
\\
\cline{6-9}
 &  &  &  &  & $1^{-}$ & $7733$ & $7870$ & $7989$
\\
\cline{6-9}
 &  &  &  &  & $2^{-}$ & $7700$ & $7845$ & $7967$
\\
\cline{2-4}
\cline{6-9}
 & $2S$ & $1$ & $0$ &  & $1^{+}$ & $7963$ & $8098$ & $8213$
\\
\cline{2-4}
\cline{6-9}
 & \multirow[c]{3}{*}{$1D$} & \multirow[c]{3}{*}{$0$} & \multirow[c]{3}{*}{$2$} &  & $1^{+}$ & $8107$ & $8220$ & $8322$
\\
\cline{6-9}
 &  &  &  &  & $2^{+}$ & $8072$ & $8194$ & $8301$
\\
\cline{6-9}
 &  &  &  &  & $3^{+}$ & $8020$ & $8155$ & $8269$
\\
\cline{2-4}
\cline{6-9}
 & \multirow[c]{3}{*}{$2P$} & \multirow[c]{3}{*}{$1$} & \multirow[c]{3}{*}{$1$} &  & $0^{-}$ & $8306$ & $8419$ & $8518$
\\
\cline{6-9}
 &  &  &  &  & $1^{-}$ & $8281$ & $8399$ & $8502$
\\
\cline{6-9}
 &  &  &  &  & $2^{-}$ & $8229$ & $8359$ & $8469$
\\
\cline{2-4}
\cline{6-9}
 & $3S$ & $2$ & $0$ &  & $1^{+}$ & $8508$ & $8624$ & $8724$
\\
\hline
\multirow[c]{6}{*}{$\mathrm{S\overline{S}}$} & $1S$ & $0$ & $0$ & \multirow[c]{6}{*}{$0$} & $0^{+}$ & $7191$ & $7403$ & \multirow[c]{6}{*}{}
\\
\cline{2-4}
\cline{6-8}
 & $1P$ & $0$ & $1$ &  & $1^{-}$ & $7556$ & $7753$ & 
\\
\cline{2-4}
\cline{6-8}
 & $2S$ & $1$ & $0$ &  & $0^{+}$ & $7815$ & $7998$ & 
\\
\cline{2-4}
\cline{6-8}
 & $1D$ & $0$ & $2$ &  & $2^{+}$ & $7907$ & $8086$ & 
\\
\cline{2-4}
\cline{6-8}
 & $2P$ & $1$ & $1$ &  & $1^{-}$ & $8120$ & $8288$ & 
\\
\cline{2-4}
\cline{6-8}
 & $3S$ & $2$ & $0$ &  & $0^{+}$ & $8383$ & $8538$ & 
\\
\end{tabular}
\end{ruledtabular}
\end{table*}

\begin{table*}
\caption{Calculated masses (in MeV) of the ground, radial, and orbital excitations of the doubly bottom tetraquarks.}
\label{tab:res.bb}
\centering
\begin{ruledtabular}
\begin{tabular}{ccccccccc}
$\mathrm{d\overline{d^{\prime}}}$ & $nL$ & $n_{r}$ & $L$ & $S$ & $J^{P}$ & $M_{\mathrm{bb\overline{qq^{\prime}}}}$ & $M_{\mathrm{bb\overline{sq}}}$ & $M_{\mathrm{bb\overline{ss}}}$
\\
\hline
\multirow[c]{32}{*}{$\mathrm{A\overline{A}}$} & \multirow[c]{3}{*}{$1S$} & \multirow[c]{3}{*}{$0$} & \multirow[c]{3}{*}{$0$} & $0$ & $0^{+}$ & $10606$ & $10763$ & $10896$
\\
\cline{5-9}
 &  &  &  & $1$ & $1^{+}$ & $10617$ & $10772$ & $10904$
\\
\cline{5-9}
 &  &  &  & $2$ & $2^{+}$ & $10638$ & $10790$ & $10919$
\\
\cline{2-9}
 & \multirow[c]{7}{*}{$1P$} & \multirow[c]{7}{*}{$0$} & \multirow[c]{7}{*}{$1$} & $1$ & $0^{-}$ & $10969$ & $11108$ & $11228$
\\
\cline{5-9}
 &  &  &  & $0$ & \multirow[c]{3}{*}{$1^{-}$} & $10940$ & $11084$ & $11207$
\\
 &  &  &  & $1$ &  & $10976$ & $11113$ & $11232$
\\
 &  &  &  & $2$ &  & $10998$ & $11131$ & $11247$
\\
\cline{5-9}
 &  &  &  & $1$ & \multirow[c]{2}{*}{$2^{-}$} & $10947$ & $11091$ & $11212$
\\
 &  &  &  & $2$ &  & $10989$ & $11124$ & $11241$
\\
\cline{5-9}
 &  &  &  & $2$ & $3^{-}$ & $10958$ & $11100$ & $11221$
\\
\cline{2-9}
 & \multirow[c]{3}{*}{$2S$} & \multirow[c]{3}{*}{$1$} & \multirow[c]{3}{*}{$0$} & $0$ & $0^{+}$ & $11188$ & $11324$ & $11440$
\\
\cline{5-9}
 &  &  &  & $1$ & $1^{+}$ & $11191$ & $11326$ & $11442$
\\
\cline{5-9}
 &  &  &  & $2$ & $2^{+}$ & $11196$ & $11330$ & $11444$
\\
\cline{2-9}
 & \multirow[c]{9}{*}{$1D$} & \multirow[c]{9}{*}{$0$} & \multirow[c]{9}{*}{$2$} & $2$ & $0^{+}$ & $11325$ & $11440$ & $11543$
\\
\cline{5-9}
 &  &  &  & $1$ & \multirow[c]{2}{*}{$1^{+}$} & $11290$ & $11414$ & $11521$
\\
 &  &  &  & $2$ &  & $11330$ & $11444$ & $11546$
\\
\cline{5-9}
 &  &  &  & $0$ & \multirow[c]{3}{*}{$2^{+}$} & $11238$ & $11374$ & $11489$
\\
 &  &  &  & $1$ &  & $11296$ & $11419$ & $11526$
\\
 &  &  &  & $2$ &  & $11340$ & $11453$ & $11554$
\\
\cline{5-9}
 &  &  &  & $1$ & \multirow[c]{2}{*}{$3^{+}$} & $11243$ & $11379$ & $11493$
\\
 &  &  &  & $2$ &  & $11305$ & $11426$ & $11532$
\\
\cline{5-9}
 &  &  &  & $2$ & $4^{+}$ & $11250$ & $11385$ & $11499$
\\
\cline{2-9}
 & \multirow[c]{7}{*}{$2P$} & \multirow[c]{7}{*}{$1$} & \multirow[c]{7}{*}{$1$} & $1$ & $0^{-}$ & $11481$ & $11601$ & $11706$
\\
\cline{5-9}
 &  &  &  & $0$ & \multirow[c]{3}{*}{$1^{-}$} & $11430$ & $11561$ & $11672$
\\
 &  &  &  & $1$ &  & $11484$ & $11603$ & $11707$
\\
 &  &  &  & $2$ &  & $11511$ & $11625$ & $11725$
\\
\cline{5-9}
 &  &  &  & $1$ & \multirow[c]{2}{*}{$2^{-}$} & $11436$ & $11566$ & $11676$
\\
 &  &  &  & $2$ &  & $11490$ & $11608$ & $11712$
\\
\cline{5-9}
 &  &  &  & $2$ & $3^{-}$ & $11445$ & $11573$ & $11683$
\\
\cline{2-9}
 & \multirow[c]{3}{*}{$3S$} & \multirow[c]{3}{*}{$2$} & \multirow[c]{3}{*}{$0$} & $0$ & $0^{+}$ & $11691$ & $11808$ & $11909$
\\
\cline{5-9}
 &  &  &  & $1$ & $1^{+}$ & $11691$ & $11808$ & $11909$
\\
\cline{5-9}
 &  &  &  & $2$ & $2^{+}$ & $11690$ & $11807$ & $11908$
\\
\hline
\multirow[c]{12}{*}{$\mathrm{A\overline{S}}$} & $1S$ & $0$ & $0$ & \multirow[c]{12}{*}{$1$} & $1^{+}$ & $10461$ & $10670$ & \multirow[c]{12}{*}{}
\\
\cline{2-4}
\cline{6-8}
 & \multirow[c]{3}{*}{$1P$} & \multirow[c]{3}{*}{$0$} & \multirow[c]{3}{*}{$1$} &  & $0^{-}$ & $10799$ & $10996$ & 
\\
\cline{6-8}
 &  &  &  &  & $1^{-}$ & $10803$ & $11000$ & 
\\
\cline{6-8}
 &  &  &  &  & $2^{-}$ & $10812$ & $11007$ & 
\\
\cline{2-4}
\cline{6-8}
 & $2S$ & $1$ & $0$ &  & $1^{+}$ & $11045$ & $11228$ & 
\\
\cline{2-4}
\cline{6-8}
 & \multirow[c]{3}{*}{$1D$} & \multirow[c]{3}{*}{$0$} & \multirow[c]{3}{*}{$2$} &  & $1^{+}$ & $11125$ & $11306$ & 
\\
\cline{6-8}
 &  &  &  &  & $2^{+}$ & $11131$ & $11310$ & 
\\
\cline{6-8}
 &  &  &  &  & $3^{+}$ & $11138$ & $11317$ & 
\\
\cline{2-4}
\cline{6-8}
 & \multirow[c]{3}{*}{$2P$} & \multirow[c]{3}{*}{$1$} & \multirow[c]{3}{*}{$1$} &  & $0^{-}$ & $11323$ & $11493$ & 
\\
\cline{6-8}
 &  &  &  &  & $1^{-}$ & $11325$ & $11495$ & 
\\
\cline{6-8}
 &  &  &  &  & $2^{-}$ & $11329$ & $11498$ & 
\\
\cline{2-4}
\cline{6-8}
 & $3S$ & $2$ & $0$ &  & $1^{+}$ & $11563$ & $11721$ & 
\\
\end{tabular}
\end{ruledtabular}
\end{table*}

\section{Threshold analysis}
\label{sec:thresholdanalysis}

\par
The calculated mass spectra presented in Sec.~\ref{sec:numres} contain a large number of doubly heavy tetraquark states with different flavor compositions and quantum numbers. To identify tetraquark states which are the most promising for the experimental detection, we compare the calculated masses with the thresholds for strong fall--apart decays into pairs of heavy--light mesons. This approach was also employed in our previous studies of heavy tetraquarks~\cite{Savch2020PhysRevD,Savch2021Uni,Savch2022Sym,Savch2024EPJA}.
\par
The strong fall--apart decay proceeds through a rearrangement of the constituent quarks and antiquarks into two color--singlet mesons:
\begin{equation}\label{eq:fallapart}
\mathrm{QQ^{\prime}\overline{qq^{\prime}}} \longrightarrow
\left[
\begin{array}{ll}
	(\mathrm{Q\overline{q}})+(\mathrm{Q^{\prime}\overline{q^{\prime}}}), \\
    (\mathrm{Q\overline{q^{\prime}}})+(\mathrm{Q^{\prime}\overline{q}}).
\end{array}
\right.
\end{equation}
\noindent
The final mesons belong to the $D_{(\mathrm{s})}$ or $B_{(\mathrm{s})}$ families, including their excited states. The allowed decay channels are determined by flavor, total angular momentum, and spatial parity conservation (charge conjugation parity does not impose an additional selection rule for the considered open--flavor tetraquarks).
\par
Let $J_{\mathrm{A}}^{P_{\mathrm{A}}}$ and $J_{\mathrm{B}}^{P_{\mathrm{B}}}$ denote the spin--parities of the final mesons, and $L_{\mathrm{rel}}$ their relative orbital angular momentum. The spatial parity of the meson pair is
\begin{equation}\label{eq:Pparity}
P_{\mathrm{AB}} = P_{\mathrm{A}}P_{\mathrm{B}}(-1)^{L_{\mathrm{rel}}}.
\end{equation}
\noindent
For an $S$--wave final state ($L_{\mathrm{rel}}=0$), the conservation of total angular momentum $J$ and parity $P$ of the initial tetraquark state requires
\begin{equation}\label{eq:JP}
\begin{gathered}
J = |J_{\mathrm{A}}-J_{\mathrm{B}}|, \ldots, J_{\mathrm{A}}+J_{\mathrm{B}}, \\
P = P_{\mathrm{A}}P_{\mathrm{B}}.
\end{gathered}
\end{equation}
\noindent
Here, $L_{\mathrm{rel}}$ should not be confused with the orbital angular momentum $L$ of the initial diquark--antidiquark system. Thus, orbitally excited tetraquarks can decay into meson pairs in $S$--wave if the conservation laws are satisfied. The lowest meson--pair threshold with the required flavor composition is therefore not necessarily allowed for a given $J^{P}$, and excited heavy--light mesons must also be considered.
\par
Near the decay threshold, higher partial waves are suppressed by the centrifugal barrier. For a regular reduced decay amplitude, the partial width behaves as
\begin{equation}\label{eq:partialwidth}
\Gamma_{\mathrm{AB}}^{(L_{\mathrm{rel}})} \propto |\mathbf{k}_{\mathrm{AB}}| \cdot \Big( \frac{|\mathbf{k}_{\mathrm{AB}}|}{\Lambda_{\mathrm{had}}} \Big)^{2L_{\mathrm{rel}}},
\end{equation}
\noindent
where $\mathbf{k}_{\mathrm{AB}}$ is the relative momentum of the final mesons and $\Lambda_{\mathrm{had}}$ is a characteristic hadronic momentum scale. For $|\mathbf{k}_{\mathrm{AB}}| \ll \Lambda_{\mathrm{had}}$, higher partial waves acquire an additional suppression factor. We therefore restrict the present threshold analysis to the $S$--wave strong fall--apart channels, following Refs.~\cite{Savch2022Sym,Savch2024EPJA}.
\par
For each meson pair $\mathrm{AB}$, the strong fall--apart decay threshold and the corresponding mass difference are defined as
\begin{equation}\label{eq:delta}
\begin{aligned}
M_{\mathrm{thr}}^{\mathrm{AB}} &= M_{\mathrm{A}}+M_{\mathrm{B}}, \\
\Delta_{\mathrm{AB}} &= M_{\mathrm{T}}-M_{\mathrm{thr}}^{\mathrm{AB}},
\end{aligned}
\end{equation}
\noindent
where $M_{\mathrm{T}}$ is the calculated tetraquark mass and $M_{\mathrm{A,B}}$ are the meson masses. A channel is kinematically open for $\Delta_{\mathrm{AB}}>0$ and closed for $\Delta_{\mathrm{AB}}<0$. The relative momentum $\mathbf{k}_{\mathrm{AB}}$ in an open channel is given by
\begin{equation}\label{eq:kAB}
\begin{aligned}
|\mathbf{k}_{\mathrm{AB}}|^{2} &= \frac{\big[M_{\mathrm{T}}^{2}-(M_{\mathrm{A}}+M_{\mathrm{B}})^{2}\big]\big[M_{\mathrm{T}}^{2}-(M_{\mathrm{A}}-M_{\mathrm{B}})^{2}\big]}{4M_{\mathrm{T}}^{2}} \\
&\mathrel{\underset{\Delta_{\mathrm{AB}}\to 0^{+}}{\scalebox{4.5}[1]{$\sim$}}} 2\mu_{\mathrm{AB}}\Delta_{\mathrm{AB}}, \qquad \mu_{\mathrm{AB}} = \frac{M_{\mathrm{A}}M_{\mathrm{B}}}{M_{\mathrm{A}}+M_{\mathrm{B}}}.
\end{aligned}
\end{equation}
\noindent
A small positive $\Delta_{\mathrm{AB}}$ limits the phase space of the corresponding decay channel and may suppress its partial width. However, the total width also depends on other open channels and on the transition amplitudes, which are not evaluated here.
\par
For each tetraquark state, we consider the lowest $S$--wave strong fall--apart threshold allowed by its flavor composition and $J^{P}$. The meson masses are taken from the Particle Data Group~\cite{PDG2026}. Separate charge channels are retained to account for the mass differences between isospin partners. For the nonstrange tetraquarks, we distinguish the $\mathrm{\overline{uu}}$, $\mathrm{\overline{ud}}$, and $\mathrm{\overline{dd}}$ compositions where allowed by the antidiquark configuration, while for the singly strange states we consider $\mathrm{\overline{us}}$ and $\mathrm{\overline{ds}}$.
\par
The comparison is restricted to the $1S$, $1P$, and $1D$ states. The corresponding radial excitations have the same allowed $S$--wave decay channels but considerably higher masses. Tables~\ref{tab:comp.cc}--\ref{tab:comp.bb.sqss} list the states lying below or slightly above their lowest allowed $S$--wave thresholds ($\Delta\leq 60$~MeV).
\par
The calculated tetraquark masses have estimated theoretical uncertainties of order $20$--$50$~MeV. For channels involving unstable excited mesons, the thresholds are evaluated using their nominal masses, without accounting for their finite widths. These limitations should be taken into account when interpreting the relative positions of the predicted states and nearby thresholds.

\begin{table*}
\caption{Doubly charmed tetraquarks ($\mathrm{cc\overline{qq^{\prime}}}$, $\mathrm{cc\overline{sq}}$, $\mathrm{cc\overline{ss}}$) that lie below or slightly above the lowest allowed strong fall--apart decay thresholds into a pair of mesons with $\Delta \le 60$ MeV.}
\label{tab:comp.cc}
\centering
\begin{ruledtabular}
\begin{tabular}{ccccccccc}
Quark content & $\mathrm{d\overline{d^{\prime}}}$ & $nL$ & $S$ & $J^{P}$ & $M_{\mathrm{T}}$ & $M_{\mathrm{thr}}$ & $\Delta$ & Meson pair
\\
\hline
\multirow[c]{9}{*}{$\mathrm{cc\overline{qq^{\prime}}}$} & \multirow[c]{6}{*}{$\mathrm{A\overline{A}}$} & \multirow[c]{3}{*}{$1S$} & \multirow[c]{3}{*}{2} & \multirow[c]{3}{*}{$2^{+}$} & \multirow[c]{3}{*}{$4074$} & $4014$ & $60$ & $D^{*}(2007)^{0} \ D^{*}(2007)^{0}$
\\
\cline{7-9}
 &  &  &  &  &  & $4017$ & $57$ & $D^{*}(2007)^{0} \ D^{*}(2010)^{+}$
\\
\cline{7-9}
 &  &  &  &  &  & $4021$ & $53$ & $D^{*}(2010)^{+} \ D^{*}(2010)^{+}$
\\
\cline{3-9}
 &  & \multirow[c]{3}{*}{$1P$} & \multirow[c]{3}{*}{2} & \multirow[c]{3}{*}{$3^{-}$} & \multirow[c]{3}{*}{$4473$} & \multirow[c]{2}{*}{$4470$} & \multirow[c]{2}{*}{$3$} & $D^{*}(2007)^{0} \ D^{*}_{2}(2460)^{0}$
\\
 &  &  &  &  &  &  &  & $D^{*}(2007)^{0} \ D^{*}_{2}(2460)^{+}$
\\
\cline{7-9}
 &  &  &  &  &  & $4474$ & $-1$ & $D^{*}(2010)^{+} \ D^{*}_{2}(2460)^{+}$
\\
\cline{2-9}
 & \multirow[c]{3}{*}{$\mathrm{A\overline{S}}$} & $1S$ & \multirow[c]{3}{*}{1} & $1^{+}$ & $3869$ & $3875$ & $-6$ & $D^{0} \ D^{*}(2010)^{+}$
\\
\cline{3-3}
\cline{5-9}
 &  & \multirow[c]{2}{*}{$1P$} &  & $1^{-}$ & $4299$ & $4282$ & $17$ & $D^{+} \ D_{1}(2430)^{0}$
\\
\cline{5-9}
 &  &  &  & $2^{-}$ & $4316$ & $4328$ & $-12$ & $D^{0} \ D^{*}_{2}(2460)^{+}$
\\
\hline
\multirow[c]{2}{*}{$\mathrm{cc\overline{sq}}$} & \multirow[c]{2}{*}{$\mathrm{A\overline{A}}$} & \multirow[c]{2}{*}{$1P$} & \multirow[c]{2}{*}{2} & \multirow[c]{2}{*}{$3^{-}$} & \multirow[c]{2}{*}{$4616$} & $4573$ & $43$ & $D^{*}_{2}(2460)^{0} \ D_{\mathrm{s}}^{*+}$
\\
 &  &  &  &  &  & $4576$ & $40$ & $D^{*}_{2}(2460)^{+} \ D_{\mathrm{s}}^{*+}$
\\
\hline
$\mathrm{cc\overline{ss}}$ & $\mathrm{A\overline{A}}$ & $1P$ & 2 & $3^{-}$ & $4738$ & $4681$ & $57$ & $D_{\mathrm{s}}^{*+} \ D_{\mathrm{s}2}^{*}(2573)^{+}$
\\
\end{tabular}
\end{ruledtabular}
\end{table*}

\begin{table*}
\caption{Charmed--bottom tetraquarks ($\mathrm{cb\overline{qq^{\prime}}}$, $\mathrm{cb\overline{sq}}$, $\mathrm{cb\overline{ss}}$) that lie below or slightly above the lowest allowed strong fall--apart decay thresholds into a pair of mesons with $\Delta \le 60$ MeV.}
\label{tab:comp.cb}
\centering
\begin{ruledtabular}
\begin{tabular}{ccccccccc}
Quark content & $\mathrm{d\overline{d^{\prime}}}$ & $nL$ & $S$ & $J^{P}$ & $M_{\mathrm{T}}$ & $M_{\mathrm{thr}}$ & $\Delta$ & Meson pair
\\
\hline
\multirow[c]{25}{*}{$\mathrm{cb\overline{qq}}$} & \multirow[c]{15}{*}{$\mathrm{A\overline{A}}$} & \multirow[c]{3}{*}{$1S$} & \multirow[c]{3}{*}{2} & \multirow[c]{3}{*}{$2^{+}$} & \multirow[c]{3}{*}{$7381$} & \multirow[c]{2}{*}{$7332$} & \multirow[c]{2}{*}{$49$} & $D^{*}(2007)^{0} \ B^{*-}$
\\
 &  &  &  &  &  &  &  & $D^{*}(2007)^{0} \ \overline{B}^{*0}$
\\
\cline{7-9}
 &  &  &  &  &  & $7335$ & $46$ & $D^{*}(2010)^{+} \ \overline{B}^{*0}$
\\
\cline{3-9}
 &  & \multirow[c]{3}{*}{$1P$} & \multirow[c]{3}{*}{2} & \multirow[c]{3}{*}{$3^{-}$} & \multirow[c]{3}{*}{$7718$} & $7744$ & $-26$ & $D^{*}(2007)^{0} \ B^{*}_{2}(5747)^{-}$
\\
\cline{7-9}
 &  &  &  &  &  & $7746$ & $-28$ & $D^{*}(2007)^{0} \ \overline{B}^{*}_{2}(5747)^{0}$
\\
\cline{7-9}
 &  &  &  &  &  & $7750$ & $-32$ & $D^{*}(2010)^{+} \ \overline{B}^{*}_{2}(5747)^{0}$
\\
\cline{3-9}
 &  & \multirow[c]{9}{*}{$1D$} & \multirow[c]{3}{*}{1} & \multirow[c]{6}{*}{$3^{+}$} & \multirow[c]{3}{*}{$8024$} & \multirow[c]{3}{*}{$8043$} & \multirow[c]{3}{*}{$-19$} & $D^{*}_{3}(2750)^{0} \ B^{-}$
\\
 &  &  &  &  &  &  &  & $D^{*}_{3}(2750)^{0} \ \overline{B}^{0}$
\\
 &  &  &  &  &  &  &  & $D^{*}_{3}(2750)^{+} \ \overline{B}^{0}$
\\
\cline{6-9}
 &  &  & \multirow[c]{3}{*}{2} &  & \multirow[c]{3}{*}{$8087$} & \multirow[c]{3}{*}{$8043$} & \multirow[c]{3}{*}{$44$} & $D^{*}_{3}(2750)^{0} \ B^{-}$
\\
 &  &  &  &  &  &  &  & $D^{*}_{3}(2750)^{0} \ \overline{B}^{0}$
\\
 &  &  &  &  &  &  &  & $D^{*}_{3}(2750)^{+} \ \overline{B}^{0}$
\\
\cline{4-9}
 &  &  & \multirow[c]{3}{*}{2} & \multirow[c]{3}{*}{$4^{+}$} & \multirow[c]{3}{*}{$8034$} & \multirow[c]{3}{*}{$8088$} & \multirow[c]{3}{*}{$-54$} & $D^{*}_{3}(2750)^{0} \ B^{*-}$
\\
 &  &  &  &  &  &  &  & $D^{*}_{3}(2750)^{0} \ \overline{B}^{*0}$
\\
 &  &  &  &  &  &  &  & $D^{*}_{3}(2750)^{+} \ \overline{B}^{*0}$
\\
\cline{2-9}
 & \multirow[c]{5}{*}{$\mathrm{A\overline{S}}$} & $1S$ & \multirow[c]{5}{*}{1} & $1^{+}$ & $7198$ & $7190$ & $8$ & $D^{0} \ \overline{B}^{*0}$
\\
\cline{3-3}
\cline{5-9}
 &  & \multirow[c]{3}{*}{$1P$} &  & $0^{-}$ & $7552$ & $7622$ & $-70$ & $D^{*}_{0}(2300)^{+} \ B^{-}$
\\
\cline{5-9}
 &  &  &  & $1^{-}$ & $7557$ & $7591$ & $-34$ & $D^{0} \ \overline{B}_{1}(5721)^{0}$
\\
\cline{5-9}
 &  &  &  & $2^{-}$ & $7568$ & $7604$ & $-36$ & $D^{0} \ \overline{B}^{*}_{2}(5747)^{0}$
\\
\cline{3-3}
\cline{5-9}
 &  & $1D$ &  & $3^{+}$ & $7921$ & $8043$ & $-122$ & $D^{*}_{3}(2750)^{0} \ \overline{B}^{0}$
\\
\cline{2-9}
 & \multirow[c]{3}{*}{$\mathrm{S\overline{A}}$} & \multirow[c]{3}{*}{$1D$} & \multirow[c]{3}{*}{1} & \multirow[c]{3}{*}{$3^{+}$} & \multirow[c]{3}{*}{$8020$} & \multirow[c]{3}{*}{$8043$} & \multirow[c]{3}{*}{$-23$} & $D^{*}_{3}(2750)^{0} \ B^{-}$
\\
 &  &  &  &  &  &  &  & $D^{*}_{3}(2750)^{0} \ \overline{B}^{0}$
\\
 &  &  &  &  &  &  &  & $D^{*}_{3}(2750)^{+} \ \overline{B}^{0}$
\\
\cline{2-9}
 & \multirow[c]{2}{*}{$\mathrm{S\overline{S}}$} & $1S$ & \multirow[c]{2}{*}{0} & $0^{+}$ & $7191$ & $7145$ & $46$ & $D^{0} \ \overline{B}^{0}$
\\
\cline{3-3}
\cline{5-9}
 &  & $1P$ &  & $1^{-}$ & $7556$ & $7591$ & $-35$ & $D^{0} \ \overline{B}_{1}(5721)^{0}$
\\
\hline
\multirow[c]{12}{*}{$\mathrm{cb\overline{sq}}$} & \multirow[c]{6}{*}{$\mathrm{A\overline{A}}$} & \multirow[c]{2}{*}{$1P$} & \multirow[c]{2}{*}{2} & \multirow[c]{2}{*}{$3^{-}$} & \multirow[c]{2}{*}{$7861$} & $7847$ & $14$ & $D^{*}(2007)^{0} \ \overline{B}^{*}_{\mathrm{s}2}(5840)^{0}$
\\
\cline{7-9}
 &  &  &  &  &  & $7850$ & $11$ & $D^{*}(2010)^{+} \ \overline{B}^{*}_{\mathrm{s}2}(5840)^{0}$
\\
\cline{3-9}
 &  & \multirow[c]{4}{*}{$1D$} & \multirow[c]{2}{*}{1} & \multirow[c]{2}{*}{$3^{+}$} & \multirow[c]{2}{*}{$8160$} & \multirow[c]{2}{*}{$8130$} & \multirow[c]{2}{*}{$30$} & $D^{*}_{3}(2750)^{0} \ \overline{B}_{\mathrm{s}}^{0}$
\\
 &  &  &  &  &  &  &  & $D^{*}_{3}(2750)^{+} \ \overline{B}_{\mathrm{s}}^{0}$
\\
\cline{4-9}
 &  &  & \multirow[c]{2}{*}{2} & \multirow[c]{2}{*}{$4^{+}$} & \multirow[c]{2}{*}{$8168$} & \multirow[c]{2}{*}{$8179$} & \multirow[c]{2}{*}{$-11$} & $D^{*}_{3}(2750)^{0} \ \overline{B}_{\mathrm{s}}^{*0}$
\\
 &  &  &  &  &  &  &  & $D^{*}_{3}(2750)^{+} \ \overline{B}_{\mathrm{s}}^{*0}$
\\
\cline{2-9}
 & \multirow[c]{4}{*}{$\mathrm{A\overline{S}}$} & \multirow[c]{2}{*}{$1P$} & \multirow[c]{4}{*}{1} & \multirow[c]{2}{*}{$2^{-}$} & \multirow[c]{2}{*}{$7765$} & $7705$ & $60$ & $D^{0} \ \overline{B}^{*}_{\mathrm{s}2}(5840)^{0}$
\\
\cline{7-9}
 &  &  &  &  &  & $7708$ & $57$ & $D_{\mathrm{s}}^{+} \ \overline{B}^{*}_{2}(5747)^{0}$
\\
\cline{3-3}
\cline{5-9}
 &  & \multirow[c]{2}{*}{$1D$} &  & \multirow[c]{2}{*}{$3^{+}$} & \multirow[c]{2}{*}{$8100$} & \multirow[c]{2}{*}{$8130$} & \multirow[c]{2}{*}{$-30$} & $D^{*}_{3}(2750)^{0} \ \overline{B}_{\mathrm{s}}^{0}$
\\
 &  &  &  &  &  &  &  & $D^{*}_{3}(2750)^{+} \ \overline{B}_{\mathrm{s}}^{0}$
\\
\cline{2-9}
 & \multirow[c]{2}{*}{$\mathrm{S\overline{A}}$} & \multirow[c]{2}{*}{$1D$} & \multirow[c]{2}{*}{1} & \multirow[c]{2}{*}{$3^{+}$} & \multirow[c]{2}{*}{$8155$} & \multirow[c]{2}{*}{$8130$} & \multirow[c]{2}{*}{$25$} & $D^{*}_{3}(2750)^{0} \ \overline{B}_{\mathrm{s}}^{0}$
\\
 &  &  &  &  &  &  &  & $D^{*}_{3}(2750)^{+} \ \overline{B}_{\mathrm{s}}^{0}$
\\
\hline
\multirow[c]{4}{*}{$\mathrm{cb\overline{ss}}$} & \multirow[c]{3}{*}{$\mathrm{A\overline{A}}$} & $1P$ & 2 & $3^{-}$ & $7982$ & $7952$ & $30$ & $D_{\mathrm{s}}^{*+} \ \overline{B}^{*}_{\mathrm{s}2}(5840)^{0}$
\\
\cline{3-9}
 &  & \multirow[c]{2}{*}{$1D$} & 1 & $3^{+}$ & $8274$ & $8227$ & $47$ & $D_{\mathrm{s}3}^{*}(2860)^{+} \ \overline{B}_{\mathrm{s}}^{0}$
\\
\cline{4-9}
 &  &  & 2 & $4^{+}$ & $8282$ & $8276$ & $6$ & $D_{\mathrm{s}3}^{*}(2860)^{+} \ \overline{B}_{\mathrm{s}}^{*0}$
\\
\cline{2-9}
 & $\mathrm{S\overline{A}}$ & $1D$ & 1 & $3^{+}$ & $8269$ & $8227$ & $42$ & $D_{\mathrm{s}3}^{*}(2860)^{+} \ \overline{B}_{\mathrm{s}}^{0}$
\\
\end{tabular}
\end{ruledtabular}
\end{table*}

\begin{table*}
\caption{Doubly bottom tetraquarks ($\mathrm{bb\overline{qq^{\prime}}}$) that lie below or slightly above the lowest allowed strong fall--apart decay thresholds into a pair of mesons with $\Delta \le 60$ MeV.}
\label{tab:comp.bb.qq}
\centering
\begin{ruledtabular}
\begin{tabular}{ccccccccc}
Quark content & $\mathrm{d\overline{d^{\prime}}}$ & $nL$ & $S$ & $J^{P}$ & $M_{\mathrm{T}}$ & $M_{\mathrm{thr}}$ & $\Delta$ & Meson pair
\\
\hline
\multirow[c]{44}{*}{$\mathrm{bb\overline{qq}}$} & \multirow[c]{39}{*}{$\mathrm{A\overline{A}}$} & \multirow[c]{9}{*}{$1S$} & \multirow[c]{3}{*}{0} & \multirow[c]{3}{*}{$0^{+}$} & \multirow[c]{3}{*}{$10606$} & \multirow[c]{3}{*}{$10559$} & \multirow[c]{3}{*}{$47$} & $B^{-} \ B^{-}$
\\
 &  &  &  &  &  &  &  & $B^{-} \ \overline{B}^{0}$
\\
 &  &  &  &  &  &  &  & $\overline{B}^{0} \ \overline{B}^{0}$
\\
\cline{4-9}
 &  &  & \multirow[c]{3}{*}{1} & \multirow[c]{3}{*}{$1^{+}$} & \multirow[c]{3}{*}{$10617$} & \multirow[c]{3}{*}{$10604$} & \multirow[c]{3}{*}{$13$} & $B^{-} \ B^{*-}$
\\
 &  &  &  &  &  &  &  & $\overline{B}^{0} \ B^{*-}$
\\
 &  &  &  &  &  &  &  & $\overline{B}^{0} \ \overline{B}^{*0}$
\\
\cline{4-9}
 &  &  & \multirow[c]{3}{*}{2} & \multirow[c]{3}{*}{$2^{+}$} & \multirow[c]{3}{*}{$10638$} & \multirow[c]{3}{*}{$10650$} & \multirow[c]{3}{*}{$-12$} & $B^{*-} \ B^{*-}$
\\
 &  &  &  &  &  &  &  & $B^{*-} \ \overline{B}^{*0}$
\\
 &  &  &  &  &  &  &  & $\overline{B}^{*0} \ \overline{B}^{*0}$
\\
\cline{3-9}
 &  & \multirow[c]{21}{*}{$1P$} & \multirow[c]{3}{*}{1} & \multirow[c]{3}{*}{$0^{-}$} & \multirow[c]{3}{*}{$10969$} & \multirow[c]{3}{*}{$11051$} & \multirow[c]{3}{*}{$-82$} & $B^{*-} \ B_{1}(5721)^{-}$
\\
 &  &  &  &  &  &  &  & $B^{*-} \ \overline{B}_{1}(5721)^{0}$
\\
 &  &  &  &  &  &  &  & $\overline{B}^{*0} \ \overline{B}_{1}(5721)^{0}$
\\
\cline{4-9}
 &  &  & \multirow[c]{3}{*}{0} & \multirow[c]{9}{*}{$1^{-}$} & \multirow[c]{3}{*}{$10940$} & $11005$ & $-65$ & $B^{-} \ B_{1}(5721)^{-}$
\\
\cline{7-9}
 &  &  &  &  &  & \multirow[c]{2}{*}{$11006$} & \multirow[c]{2}{*}{$-66$} & $B^{-} \ \overline{B}_{1}(5721)^{0}$
\\
 &  &  &  &  &  &  &  & $\overline{B}^{0} \ \overline{B}_{1}(5721)^{0}$
\\
\cline{6-9}
 &  &  & \multirow[c]{3}{*}{1} &  & \multirow[c]{3}{*}{$10976$} & $11005$ & $-29$ & $B^{-} \ B_{1}(5721)^{-}$
\\
\cline{7-9}
 &  &  &  &  &  & \multirow[c]{2}{*}{$11006$} & \multirow[c]{2}{*}{$-30$} & $B^{-} \ \overline{B}_{1}(5721)^{0}$
\\
 &  &  &  &  &  &  &  & $\overline{B}^{0} \ \overline{B}_{1}(5721)^{0}$
\\
\cline{6-9}
 &  &  & \multirow[c]{3}{*}{2} &  & \multirow[c]{3}{*}{$10998$} & $11005$ & $-7$ & $B^{-} \ B_{1}(5721)^{-}$
\\
\cline{7-9}
 &  &  &  &  &  & \multirow[c]{2}{*}{$11006$} & \multirow[c]{2}{*}{$-8$} & $B^{-} \ \overline{B}_{1}(5721)^{0}$
\\
 &  &  &  &  &  &  &  & $\overline{B}^{0} \ \overline{B}_{1}(5721)^{0}$
\\
\cline{4-9}
 &  &  & \multirow[c]{3}{*}{1} & \multirow[c]{6}{*}{$2^{-}$} & \multirow[c]{3}{*}{$10947$} & \multirow[c]{2}{*}{$11017$} & \multirow[c]{2}{*}{$-70$} & $B^{-} \ B^{*}_{2}(5747)^{-}$
\\
 &  &  &  &  &  &  &  & $\overline{B}^{0} \ B^{*}_{2}(5747)^{-}$
\\
\cline{7-9}
 &  &  &  &  &  & $11019$ & $-72$ & $\overline{B}^{0} \ \overline{B}^{*}_{2}(5747)^{0}$
\\
\cline{6-9}
 &  &  & \multirow[c]{3}{*}{2} &  & \multirow[c]{3}{*}{$10989$} & \multirow[c]{2}{*}{$11017$} & \multirow[c]{2}{*}{$-28$} & $B^{-} \ B^{*}_{2}(5747)^{-}$
\\
 &  &  &  &  &  &  &  & $\overline{B}^{0} \ B^{*}_{2}(5747)^{-}$
\\
\cline{7-9}
 &  &  &  &  &  & $11019$ & $-30$ & $\overline{B}^{0} \ \overline{B}^{*}_{2}(5747)^{0}$
\\
\cline{4-9}
 &  &  & \multirow[c]{3}{*}{2} & \multirow[c]{3}{*}{$3^{-}$} & \multirow[c]{3}{*}{$10958$} & \multirow[c]{2}{*}{$11062$} & \multirow[c]{2}{*}{$-104$} & $B^{*-} \ B^{*}_{2}(5747)^{-}$
\\
 &  &  &  &  &  &  &  & $\overline{B}^{*0} \ B^{*}_{2}(5747)^{-}$
\\
\cline{7-9}
 &  &  &  &  &  & $11064$ & $-106$ & $\overline{B}^{*0} \ \overline{B}^{*}_{2}(5747)^{0}$
\\
\cline{3-9}
 &  & \multirow[c]{9}{*}{$1D$} & \multirow[c]{3}{*}{1} & \multirow[c]{6}{*}{$3^{+}$} & \multirow[c]{3}{*}{$11243$} & \multirow[c]{2}{*}{$11463$} & \multirow[c]{2}{*}{$-220$} & $B_{1}(5721)^{-} \ B^{*}_{2}(5747)^{-}$
\\
 &  &  &  &  &  &  &  & $\overline{B}_{1}(5721)^{0} \ B^{*}_{2}(5747)^{-}$
\\
\cline{7-9}
 &  &  &  &  &  & $11466$ & $-223$ & $\overline{B}_{1}(5721)^{0} \ \overline{B}^{*}_{2}(5747)^{0}$
\\
\cline{6-9}
 &  &  & \multirow[c]{3}{*}{2} &  & \multirow[c]{3}{*}{$11305$} & \multirow[c]{2}{*}{$11463$} & \multirow[c]{2}{*}{$-158$} & $B_{1}(5721)^{-} \ B^{*}_{2}(5747)^{-}$
\\
 &  &  &  &  &  &  &  & $\overline{B}_{1}(5721)^{0} \ B^{*}_{2}(5747)^{-}$
\\
\cline{7-9}
 &  &  &  &  &  & $11466$ & $-161$ & $\overline{B}_{1}(5721)^{0} \ \overline{B}^{*}_{2}(5747)^{0}$
\\
\cline{4-9}
 &  &  & \multirow[c]{3}{*}{2} & \multirow[c]{3}{*}{$4^{+}$} & \multirow[c]{3}{*}{$11250$} & $11475$ & $-225$ & $B^{*}_{2}(5747)^{-} \ B^{*}_{2}(5747)^{-}$
\\
\cline{7-9}
 &  &  &  &  &  & $11477$ & $-227$ & $B^{*}_{2}(5747)^{-} \ \overline{B}^{*}_{2}(5747)^{0}$
\\
\cline{7-9}
 &  &  &  &  &  & $11479$ & $-229$ & $\overline{B}^{*}_{2}(5747)^{0} \ \overline{B}^{*}_{2}(5747)^{0}$
\\
\cline{2-9}
 & \multirow[c]{5}{*}{$\mathrm{A\overline{S}}$} & $1S$ & \multirow[c]{5}{*}{1} & $1^{+}$ & $10461$ & $10604$ & $-143$ & $\overline{B}^{0} \ B^{*-}$
\\
\cline{3-3}
\cline{5-9}
 &  & \multirow[c]{3}{*}{$1P$} &  & $0^{-}$ & $10799$ & $11051$ & $-252$ & $B^{*-} \ \overline{B}_{1}(5721)^{0}$
\\
\cline{5-9}
 &  &  &  & $1^{-}$ & $10803$ & $11006$ & $-203$ & $B^{-} \ \overline{B}_{1}(5721)^{0}$
\\
\cline{5-9}
 &  &  &  & $2^{-}$ & $10812$ & $11017$ & $-205$ & $\overline{B}^{0} \ B^{*}_{2}(5747)^{-}$
\\
\cline{3-3}
\cline{5-9}
 &  & $1D$ &  & $3^{+}$ & $11138$ & $11463$ & $-325$ & $\overline{B}_{1}(5721)^{0} \ B^{*}_{2}(5747)^{-}$
\\
\end{tabular}
\end{ruledtabular}
\end{table*}

\begin{table*}
\caption{Doubly bottom tetraquarks ($\mathrm{bb\overline{sq}}$, $\mathrm{bb\overline{ss}}$) that lie below or slightly above the lowest allowed strong fall--apart decay thresholds into a pair of mesons with $\Delta \le 60$ MeV.}
\label{tab:comp.bb.sqss}
\centering
\begin{ruledtabular}
\begin{tabular}{ccccccccc}
Quark content & $\mathrm{d\overline{d^{\prime}}}$ & $nL$ & $S$ & $J^{P}$ & $M_{\mathrm{T}}$ & $M_{\mathrm{thr}}$ & $\Delta$ & Meson pair
\\
\hline
\multirow[c]{32}{*}{$\mathrm{bb\overline{sq}}$} & \multirow[c]{22}{*}{$\mathrm{A\overline{A}}$} & \multirow[c]{2}{*}{$1S$} & \multirow[c]{2}{*}{2} & \multirow[c]{2}{*}{$2^{+}$} & \multirow[c]{2}{*}{$10790$} & \multirow[c]{2}{*}{$10740$} & \multirow[c]{2}{*}{$50$} & $B^{*-} \ \overline{B}_{\mathrm{s}}^{*0}$
\\
 &  &  &  &  &  &  &  & $\overline{B}^{*0} \ \overline{B}_{\mathrm{s}}^{*0}$
\\
\cline{3-9}
 &  & \multirow[c]{14}{*}{$1P$} & \multirow[c]{2}{*}{1} & \multirow[c]{2}{*}{$0^{-}$} & \multirow[c]{2}{*}{$11108$} & $11141$ & $-33$ & $B_{1}(5721)^{-} \ \overline{B}_{\mathrm{s}}^{*0}$
\\
\cline{7-9}
 &  &  &  &  &  & $11142$ & $-34$ & $\overline{B}_{1}(5721)^{0} \ \overline{B}_{\mathrm{s}}^{*0}$
\\
\cline{4-9}
 &  &  & \multirow[c]{2}{*}{0} & \multirow[c]{6}{*}{$1^{-}$} & \multirow[c]{2}{*}{$11084$} & \multirow[c]{2}{*}{$11093$} & \multirow[c]{2}{*}{$-9$} & $B_{1}(5721)^{-} \ \overline{B}_{\mathrm{s}}^{0}$
\\
 &  &  &  &  &  &  &  & $\overline{B}_{1}(5721)^{0} \ \overline{B}_{\mathrm{s}}^{0}$
\\
\cline{6-9}
 &  &  & \multirow[c]{2}{*}{1} &  & \multirow[c]{2}{*}{$11113$} & \multirow[c]{2}{*}{$11093$} & \multirow[c]{2}{*}{$20$} & $B_{1}(5721)^{-} \ \overline{B}_{\mathrm{s}}^{0}$
\\
 &  &  &  &  &  &  &  & $\overline{B}_{1}(5721)^{0} \ \overline{B}_{\mathrm{s}}^{0}$
\\
\cline{6-9}
 &  &  & \multirow[c]{2}{*}{2} &  & \multirow[c]{2}{*}{$11131$} & \multirow[c]{2}{*}{$11093$} & \multirow[c]{2}{*}{$38$} & $B_{1}(5721)^{-} \ \overline{B}_{\mathrm{s}}^{0}$
\\
 &  &  &  &  &  &  &  & $\overline{B}_{1}(5721)^{0} \ \overline{B}_{\mathrm{s}}^{0}$
\\
\cline{4-9}
 &  &  & \multirow[c]{2}{*}{1} & \multirow[c]{4}{*}{$2^{-}$} & \multirow[c]{2}{*}{$11091$} & $11104$ & $-13$ & $B^{*}_{2}(5747)^{-} \ \overline{B}_{\mathrm{s}}^{0}$
\\
\cline{7-9}
 &  &  &  &  &  & $11107$ & $-16$ & $\overline{B}^{*}_{2}(5747)^{0} \ \overline{B}_{\mathrm{s}}^{0}$
\\
\cline{6-9}
 &  &  & \multirow[c]{2}{*}{2} &  & \multirow[c]{2}{*}{$11124$} & $11104$ & $20$ & $B^{*}_{2}(5747)^{-} \ \overline{B}_{\mathrm{s}}^{0}$
\\
\cline{7-9}
 &  &  &  &  &  & $11107$ & $17$ & $\overline{B}^{*}_{2}(5747)^{0} \ \overline{B}_{\mathrm{s}}^{0}$
\\
\cline{4-9}
 &  &  & \multirow[c]{2}{*}{2} & \multirow[c]{2}{*}{$3^{-}$} & \multirow[c]{2}{*}{$11100$} & $11153$ & $-53$ & $B^{*}_{2}(5747)^{-} \ \overline{B}_{\mathrm{s}}^{*0}$
\\
\cline{7-9}
 &  &  &  &  &  & $11155$ & $-55$ & $\overline{B}^{*}_{2}(5747)^{0} \ \overline{B}_{\mathrm{s}}^{*0}$
\\
\cline{3-9}
 &  & \multirow[c]{6}{*}{$1D$} & \multirow[c]{2}{*}{1} & \multirow[c]{4}{*}{$3^{+}$} & \multirow[c]{2}{*}{$11379$} & \multirow[c]{2}{*}{$11566$} & \multirow[c]{2}{*}{$-187$} & $B^{*}_{2}(5747)^{-} \ \overline{B}_{\mathrm{s}1}(5830)^{0}$
\\
 &  &  &  &  &  &  &  & $\overline{B}_{1}(5721)^{0} \ \overline{B}^{*}_{\mathrm{s}2}(5840)^{0}$
\\
\cline{6-9}
 &  &  & \multirow[c]{2}{*}{2} &  & \multirow[c]{2}{*}{$11426$} & \multirow[c]{2}{*}{$11566$} & \multirow[c]{2}{*}{$-140$} & $B^{*}_{2}(5747)^{-} \ \overline{B}_{\mathrm{s}1}(5830)^{0}$
\\
 &  &  &  &  &  &  &  & $\overline{B}_{1}(5721)^{0} \ \overline{B}^{*}_{\mathrm{s}2}(5840)^{0}$
\\
\cline{4-9}
 &  &  & \multirow[c]{2}{*}{2} & \multirow[c]{2}{*}{$4^{+}$} & \multirow[c]{2}{*}{$11385$} & $11577$ & $-192$ & $B^{*}_{2}(5747)^{-} \ \overline{B}^{*}_{\mathrm{s}2}(5840)^{0}$
\\
\cline{7-9}
 &  &  &  &  &  & $11579$ & $-194$ & $\overline{B}^{*}_{2}(5747)^{0} \ \overline{B}^{*}_{\mathrm{s}2}(5840)^{0}$
\\
\cline{2-9}
 & \multirow[c]{10}{*}{$\mathrm{A\overline{S}}$} & \multirow[c]{2}{*}{$1S$} & \multirow[c]{10}{*}{1} & \multirow[c]{2}{*}{$1^{+}$} & \multirow[c]{2}{*}{$10670$} & \multirow[c]{2}{*}{$10692$} & \multirow[c]{2}{*}{$-22$} & $B^{*-} \ \overline{B}_{\mathrm{s}}^{0}$
\\
 &  &  &  &  &  &  &  & $\overline{B}^{*0} \ \overline{B}_{\mathrm{s}}^{0}$
\\
\cline{3-3}
\cline{5-9}
 &  & \multirow[c]{6}{*}{$1P$} &  & \multirow[c]{2}{*}{$0^{-}$} & \multirow[c]{2}{*}{$10996$} & $11141$ & $-145$ & $B_{1}(5721)^{-} \ \overline{B}_{\mathrm{s}}^{*0}$
\\
\cline{7-9}
 &  &  &  &  &  & $11142$ & $-146$ & $\overline{B}_{1}(5721)^{0} \ \overline{B}_{\mathrm{s}}^{*0}$
\\
\cline{5-9}
 &  &  &  & \multirow[c]{2}{*}{$1^{-}$} & \multirow[c]{2}{*}{$11000$} & \multirow[c]{2}{*}{$11093$} & \multirow[c]{2}{*}{$-93$} & $B_{1}(5721)^{-} \ \overline{B}_{\mathrm{s}}^{0}$
\\
 &  &  &  &  &  &  &  & $\overline{B}_{1}(5721)^{0} \ \overline{B}_{\mathrm{s}}^{0}$
\\
\cline{5-9}
 &  &  &  & \multirow[c]{2}{*}{$2^{-}$} & \multirow[c]{2}{*}{$11007$} & $11104$ & $-97$ & $B^{*}_{2}(5747)^{-} \ \overline{B}_{\mathrm{s}}^{0}$
\\
\cline{7-9}
 &  &  &  &  &  & $11107$ & $-100$ & $\overline{B}^{*}_{2}(5747)^{0} \ \overline{B}_{\mathrm{s}}^{0}$
\\
\cline{3-3}
\cline{5-9}
 &  & \multirow[c]{2}{*}{$1D$} &  & \multirow[c]{2}{*}{$3^{+}$} & \multirow[c]{2}{*}{$11317$} & \multirow[c]{2}{*}{$11566$} & \multirow[c]{2}{*}{$-249$} & $B^{*}_{2}(5747)^{-} \ \overline{B}_{\mathrm{s}1}(5830)^{0}$
\\
 &  &  &  &  &  &  &  & $\overline{B}_{1}(5721)^{0} \ \overline{B}^{*}_{\mathrm{s}2}(5840)^{0}$
\\
\hline
\multirow[c]{10}{*}{$\mathrm{bb\overline{ss}}$} & \multirow[c]{10}{*}{$\mathrm{A\overline{A}}$} & \multirow[c]{7}{*}{$1P$} & 1 & $0^{-}$ & $11228$ & $11244$ & $-16$ & $\overline{B}_{\mathrm{s}}^{*0} \ \overline{B}_{\mathrm{s}1}(5830)^{0}$
\\
\cline{4-9}
 &  &  & 0 & \multirow[c]{3}{*}{$1^{-}$} & $11207$ & \multirow[c]{3}{*}{$11196$} & $11$ & \multirow[c]{3}{*}{$\overline{B}_{\mathrm{s}}^{0} \ \overline{B}_{\mathrm{s}1}(5830)^{0}$}
\\
\cline{6-6}
\cline{8-8}
 &  &  & 1 &  & $11232$ &  & $36$ & 
\\
\cline{6-6}
\cline{8-8}
 &  &  & 2 &  & $11247$ &  & $51$ & 
\\
\cline{4-9}
 &  &  & 1 & \multirow[c]{2}{*}{$2^{-}$} & $11212$ & \multirow[c]{2}{*}{$11207$} & $5$ & \multirow[c]{2}{*}{$\overline{B}_{\mathrm{s}}^{0} \ \overline{B}^{*}_{\mathrm{s}2}(5840)^{0}$}
\\
\cline{6-6}
\cline{8-8}
 &  &  & 2 &  & $11241$ &  & $34$ & 
\\
\cline{4-9}
 &  &  & 2 & $3^{-}$ & $11221$ & $11255$ & $-34$ & $\overline{B}_{\mathrm{s}}^{*0} \ \overline{B}^{*}_{\mathrm{s}2}(5840)^{0}$
\\
\cline{3-9}
 &  & \multirow[c]{3}{*}{$1D$} & 1 & \multirow[c]{2}{*}{$3^{+}$} & $11493$ & \multirow[c]{2}{*}{$11669$} & $-176$ & \multirow[c]{2}{*}{$\overline{B}_{\mathrm{s}1}(5830)^{0} \ \overline{B}^{*}_{\mathrm{s}2}(5840)^{0}$}
\\
\cline{6-6}
\cline{8-8}
 &  &  & 2 &  & $11532$ &  & $-137$ & 
\\
\cline{4-9}
 &  &  & 2 & $4^{+}$ & $11499$ & $11680$ & $-181$ & $\overline{B}^{*}_{\mathrm{s}2}(5840)^{0} \ \overline{B}^{*}_{\mathrm{s}2}(5840)^{0}$
\\
\end{tabular}
\end{ruledtabular}
\end{table*}

\par
Most of the calculated tetraquark masses exceed their lowest allowed $S$--wave thresholds. However, the positions of these thresholds depend substantially on $J^{P}$. Pairs of ground state heavy--light mesons with $J^{P}=0^{-}$ or $1^{-}$ can form only positive parity $S$--wave states with $J \leq 2$. For negative parity tetraquarks and states with $J > 2$, at least one excited meson is therefore required in an $S$--wave final state. This can place the lowest allowed threshold above the mass of an orbitally excited tetraquark, even when the latter is heavier than the corresponding ground state.
\begin{itemize}
	
	\item For the doubly charmed tetraquarks (Table~\ref{tab:comp.cc}), only a limited number of states satisfy the adopted threshold criterion. Among the orbital excitations, these states predominantly have $J^{P}=3^{-}$, for which the lowest allowed $S$--wave channels contain orbitally excited $D$ mesons. The nonstrange $\mathrm{A\overline{S}}$ ground state with $J^{P}=1^{+}$ also lies slightly below its $DD^{*}$ threshold. Its relatively low mass compared with the corresponding $\mathrm{A\overline{A}}$ states is associated with the lighter scalar antidiquark.
	
	\item The charmed--bottom tetraquarks (Table~\ref{tab:comp.cb}) exhibit a richer spectrum due to the possible scalar and axialvector configurations of the heavy $\mathrm{cb}$ diquark. Several ground and orbitally excited states lie close to or below their corresponding $S$--wave thresholds. In the singly strange sector, the two possible quark rearrangements lead to distinct meson--pair channels, such as $D B_{\mathrm{s}}$ and $D_{\mathrm{s}} B$, with different thresholds. Thus, the threshold structure depends on both the internal spin configuration and the flavor composition of the final mesons.
	
	\item The doubly bottom tetraquarks (Tables~\ref{tab:comp.bb.qq}--\ref{tab:comp.bb.sqss}) have the largest set of states lying below their lowest allowed $S$--wave thresholds. This is particularly evident for the $\mathrm{A\overline{S}}$ configurations containing a scalar light antidiquark, although several $\mathrm{A\overline{A}}$ states also satisfy the selection criterion. The increasing number of such states from the doubly charmed to the doubly bottom sector reflects the combined behavior of the tetraquark masses and the heavy--light meson thresholds, rather than the tetraquark binding alone.
\end{itemize}
\par
The threshold comparison therefore identifies states for which the lowest $S$--wave strong fall--apart channels are kinematically closed or have limited phase space. This does not imply stability against strong decays, since higher--partial--wave channels may remain open. Quantitative predictions of the decay widths require the corresponding transition amplitudes, which are not calculated in the present work.

\section{Other models}
\label{sec:othermodels}

\par
Doubly heavy tetraquarks have been investigated within a variety of theoretical approaches which use different effective degrees of freedom and different treatments of the four--quark dynamics. The methods relevant to our mass comparisons are summarized below (general reviews of the theoretical approaches can be found in Refs.~\cite{Lebed2017,Multiquark2017}).

\subsection{Theoretical approaches}
\label{subsec:theor.app}

\par
Diquark--antidiquark models approximate the four--quark dynamics by an effective interaction between two color clusters. The calculations used for comparison employ a nonrelativistic constituent potential~\cite{ccqq.Lin2025}, relativized and nonrelativistic quark potentials~\cite{ccqq.Dong2025}, Regge mass relations~\cite{ccqq.cbqq.bbqq.Song2023}, and a dynamical diquark model based on the Born--Oppenheimer approximation~\cite{ccqq.ccsq.ccss.Mutuk2024DD}.
\par
Four--body constituent quark models retain the quarks and antiquarks as separate constituents and allow different color and spatial configurations to contribute. Examples include early potential model calculations~\cite{Zouzou1986}, variational calculations with different quark interactions~\cite{cbqq.Carames2019}, and calculations using the real--scaling method to identify resonance candidates~\cite{bbqq.Meng2022}. Other approaches employ hyperspherical harmonics~\cite{Hyper2006}, Gaussian expansions combined with complex scaling~\cite{ccqq.cbqq.bbqq.Wu2024}, or diffusion Monte Carlo~\cite{ccqq.cbqq.bbqq.Mutuk2024DMC}. In the three--body chiral--diquark model of Ref.~\cite{ccqq.cbqq.bbqq.Kim2022}, the two heavy quarks interact with an effective light antidiquark.
\par
Hadronic molecular approaches describe tetraquarks in terms of interacting heavy--light mesons rather than compact color clusters. Long range pion exchange was considered in an early molecular model~\cite{Molecule1991}, while meson exchange, contact, and coupled channel descriptions are reviewed in Ref.~\cite{Molecule2018}. In Ref.~\cite{QQqq.molecule.2021}, short range contact interactions estimated through resonance saturation are used to predict doubly heavy molecular states. Such states, typically associated with meson--pair thresholds, cannot be directly identified with radial or orbital excitations of compact diquark--antidiquark systems, even when their flavor content and $J^{P}$ coincide.
\par
Lattice QCD provides nonperturbative information about doubly heavy systems from correlation functions computed on the lattice. Some calculations determine energy levels relative to meson--pair thresholds~\cite{bbud.lattice.2019}, while scattering analyses can extract pole positions from finite volume spectra~\cite{cbqq.Alexandrou2024}. Alternatively, the HAL QCD method determines hadronic interaction potentials from correlation functions, which can then be used to study scattering amplitudes and their poles~\cite{ccud.lattice.2023}. For mixed bottom--charm systems, lattice studies have investigated the $J^{P}=1^{+}$ channel using both meson--meson and diquark--antidiquark interpolating operators~\cite{cbqq.Padmanath2024}, and the $J^{P}=0^{+}$ channel through meson--meson scattering~\cite{cbqq.Radhakrishnan2024}. Doubly bottom systems, including strange channels, have been studied using lattice NRQCD~\cite{bbqq.bbsq.Hudspith2023} and enlarged correlation matrix bases~\cite{bbqq.bbsq.Alexandrou2024}. Another approach uses lattice QCD static potentials as input to subsequent bound state calculations~\cite{bbqq.Bicudo2017Spin} and resonance searches~\cite{bbqq.Bicudo2017Res}. A broader survey of lattice studies is given in Ref.~\cite{Lattice2023}.
\par
QCD sum rules relate correlators of interpolating currents to a hadronic spectral representation through the operator product expansion and a treatment of the continuum. Subtleties in applying QCD sum rules to tetraquark currents are discussed in Ref.~\cite{Melikhov2019}, while explicit mass estimates for doubly heavy tetraquarks are given in Ref.~\cite{ccqq.ccsq.bbqq.bbsq.Tang2020}. The extracted masses depend on the choice of interpolating current and sum rule parameters, and different currents may couple to the same physical state.
\par
Complementary mass predictions are obtained using heavy--quark--diquark symmetry and effective field theory mass relations~\cite{ccqq.ccsq.ccss.cbqq.bbqq.Braaten2021} or heavy antiquark--diquark symmetry combined with hadron mass relations~\cite{ccqq.ccsq.ccss.cbqq.bbqq.Zhang2026}. Chromomagnetic approaches include an extended chromomagnetic model~\cite{ccqq.cbqq.bbqq.Weng2022} and a treatment incorporating heavy--quark binding, flux tube corrections, and chromomagnetic mixing~\cite{ccqq.cbqq.bbqq.Liu2023}.

\subsection{Comparison conventions}
\label{subsec:comp.conv}

\par
Tables~\ref{tab:other.ccqq.S}--\ref{tab:other.bbss.P} compare the predicted masses in MeV for the same quark compositions and, as far as available, the same isospin $I$ and spin--parity $J^{P}$. For nonstrange systems, subscripts on masses denote $I=0$ or $I=1$. For singly and doubly strange systems the isospin is fixed at $I=\frac{1}{2}$ and $I=0$, respectively. No isospin is assigned when it is not specified in the original calculation. The symbols $\mathrm{S}$ and $\mathrm{A}$ denote scalar and axialvector (anti)diquarks, and the corresponding configurations are distinguished where necessary. Multiple masses in one cell denote distinct levels in the cited calculation, not necessarily states in one--to--one correspondence with our spectrum.
\par
In our model, the $1S$, $1P$, $2S$, $1D$, $2P$, and $3S$ labels refer to the relative motion of ground--state diquarks and antidiquarks. Other approaches may use different excitation coordinates, effective constituents, or assignments of radial quantum numbers. In particular, the $P$--wave $\lambda$--mode of the three--body calculation~\cite{ccqq.cbqq.bbqq.Kim2022}, the Born--Oppenheimer label $P_{\mathrm{BO}}$~\cite{bbqq.Bicudo2017Res}, and molecular configurations containing internally excited mesons~\cite{QQqq.molecule.2021} are not identified with our $1P$ excitation. A direct $1D$ comparison is included only for $\mathrm{cc\overline{qq^{\prime}}}$, for which suitable predictions were identified.
\par
When several parameter choices are reported, the choice used in the tables is specified. Both Scheme~$\RNumb{1}$ and Scheme~$\RNumb{2}$ of Ref.~\cite{ccqq.cbqq.bbqq.Weng2022} are shown because they can give noticeably different spectra. For Ref.~\cite{ccqq.Lin2025} we use Set~$\RNumb{1}$: its parameters include an input from the observed $\mathrm{T}_{\mathrm{cc}}^{+}$ mass, so the resulting agreement with that mass is not an independent prediction. For Ref.~\cite{ccqq.Dong2025} the GI potential is used, and for Ref.~\cite{ccqq.ccsq.ccss.cbqq.bbqq.Zhang2026} we quote the HADS--based masses, including the eigenvalues of mixing matrices where mixing is present in that scheme. The nonstrange $\mathrm{cb\overline{qq^{\prime}}}$ entries remain unmixed in the HADS approximation. In the four--body calculation~\cite{cbqq.Carames2019}, the near--threshold $2^{+}$ state found with the AL1 potential is not bound with the alternative CQC interaction. Likewise, estimates based on different QCD sum rule currents are not necessarily different physical levels.
\par
The superscripts $\mathrm{mol}$ and $\mathrm{virt}$ distinguish molecular bound states and virtual poles. The symbol $\dagger$ denotes the real part of a resonance pole, $\mathrm{RSM}$ a resonance candidate identified by the real--scaling method rather than by a pole calculation, and $\mathrm{FV}$ a finite volume lattice energy level. Values from different analyses of the same lattice ensembles or from different channel treatments may also be correlated: in particular, the single and coupled channel results of Ref.~\cite{bbqq.Aoki2023} concern the same physical channel.
\par
When a source reports an energy $\Delta E$ relative to a meson--pair threshold rather than an absolute tetraquark mass, we quote a reference value
\begin{equation}\label{eq:Mref}
M_{\mathrm{ref}} = M_{\mathrm{thr}} + \Delta E
\end{equation}
\noindent
with negative $\Delta E$ indicating a position below the chosen reference threshold, without necessarily implying a bound state. The reference thresholds used in the comparisons are $DD^{*}$ ($3875$~MeV), $D^{*}D_{s}$ ($3975$~MeV), $DD_{s}^{*}$ ($3977$~MeV), $BD$ (about $7145$~MeV), $B^{*}D$ (about $7190$~MeV), $BB^{*}$ ($10604$~MeV), $BB_{s}^{*}$ (about $10695$~MeV), and $B_{s}B^{*}$ (about $10692$~MeV). The two strange bottom thresholds are kept distinct and the value appropriate to each source is used. The superscript $\mathrm{ref}$ indicates a mass reconstructed on this common reference scale, not a direct lattice determination of an absolute mass at the physical point. The uncertainties of these reconstructed masses are inherited from the source values of $\Delta E$. Uncertainties in the reference threshold masses are not included.

\FloatBarrier

\subsection{Doubly charmed tetraquarks}
\label{sec:othermodels.cc}

\subsubsection{Nonstrange: $\mathrm{cc\overline{qq^{\prime}}}$}
\label{sec:othermodels.ccqq}

\par
For the lowest isoscalar $\mathrm{cc\overline{qq^{\prime}}}$ state with $J^{P}=1^{+}$, Scheme~$\RNumb{2}$ of Ref.~\cite{ccqq.cbqq.bbqq.Weng2022} gives $3869$~MeV, equal to our value of the mass, whereas Scheme~$\RNumb{1}$ gives $3750$~MeV. Close values for this state are also obtained from chromomagnetic calculations~\cite{ccqq.cbqq.bbqq.Liu2023,ccqq.ccsq.ccss.cbqq.bbqq.Zhang2025} and a heavy antiquark--diquark symmetry estimate~\cite{ccqq.ccsq.ccss.cbqq.bbqq.Zhang2026}. The nearby $3864$~MeV level of Ref.~\cite{ccqq.cbqq.bbqq.Wu2024} is interpreted there as a $DD^{*}$ molecule.


\setlength{\jot}{0pt}
\begin{longtblr}[caption={Comparison of the predicted masses (in MeV) of the $\mathrm{cc\overline{qq^{\prime}}}$ tetraquarks in the $1S$, $2S$, and $3S$ states with other theoretical approaches. Mass subscripts denote isospin $I=0,1$ where assigned.},label={tab:other.ccqq.S},theme=prdwidetable]{
	width=\textwidth,
	rowhead=4,
	colspec={Q[l,m,wd=230pt] Q[c,m,wd=30pt] Q[c,m] Q[c,m] Q[c,m]},
	stretch=0.07,
	rowsep=0.3pt,
	colsep=4pt
	}
\hline\hline
Approach and reference & $nL$ & $J^{P}=0^{+}$ & $J^{P}=1^{+}$ & $J^{P}=2^{+}$
\\
\hline
\SetCell[r=3]{m}\shortstack{\textbf{This work} \\ \textbf{(relativistic diquark--antidiquark)}} & $1S$ & $3983_{1}$ & $\begin{gathered}3869_{0} \\ 4017_{1}\end{gathered}$ & $4074_{1}$
\\
\cline{2-5}
 & $2S$ & $4760_{1}$ & $\begin{gathered}4622_{0} \\ 4764_{1}\end{gathered}$ & $4768_{1}$
\\
\cline{2-5}
 & $3S$ & $5456_{1}$ & $\begin{gathered}5332_{0} \\ 5454_{1}\end{gathered}$ & $5445_{1}$
\\
\hline
\SetCell[c=5]{l}\textit{Diquark--antidiquark models} &  &  &  &
\\
\nopagebreak
\cline{1-5}
Regge mass relations~\cite{ccqq.cbqq.bbqq.Song2023} & $1S$ & $4163_{1}$ & $\begin{gathered}3997_{0} \\ 4185_{1}\end{gathered}$ & $4229_{1}$
\\
\cline{1-5}
Dynamical diquarks, Born--Oppenheimer~\cite{ccqq.ccsq.ccss.Mutuk2024DD} & $1S$ & $(3904 \pm 76)_{1}$ & $(3952 \pm 80)_{1}$ & $(4001 \pm 82)_{1}$
\\
\cline{1-5}
\SetCell[r=2]{m}\shortstack[l]{Nonrelativistic potential \\ (Set $\RNumb{1}$)~\cite{ccqq.Lin2025}} & $1S$ & $3834_{1}$ & $\begin{gathered}3875_{0} \\ 3908_{1}\end{gathered}$ & $4055_{1}$
\\
\cline{2-5}
 & $2S$ & $4527_{1}$ & $\begin{gathered}4511_{0} \\ 4558_{1}\end{gathered}$ & $4622_{1}$
\\
\cline{1-5}
\SetCell[r=2]{m}Relativized GI potential~\cite{ccqq.Dong2025} & $2S$ & $4554_{1}$ & $\begin{gathered}4534_{0} \\ 4595_{1}\end{gathered}$ & $4674_{1}$
\\
\cline{2-5}
 & $3S$ & $4999_{1}$ & $\begin{gathered}4956_{0} \\ 5026_{1}\end{gathered}$ & $5082_{1}$
\\
\hline
\SetCell[c=5]{l}\textit{Three--body heavy--quark--antidiquark models} &  &  &  &
\\
\nopagebreak
\cline{1-5}
\SetCell[r=2]{m}Chiral light antidiquark~\cite{ccqq.cbqq.bbqq.Kim2022} & $1S$ & $4132_{1}$ & $\begin{gathered}3961_{0} \\ 4151_{1}\end{gathered}$ & $4185_{1}$
\\
\cline{2-5}
 & $2S$ & $4546_{1}$ & $\begin{gathered}4363_{0} \\ 4560_{1}\end{gathered}$ & $4585_{1}$
\\
\hline
\SetCell[c=5]{l}\textit{Heavy--quark symmetry and hadron--mass relations} &  &  &  &
\\
\nopagebreak
\cline{1-5}
Constituent-mass estimate~\cite{ccqq.cbqq.bbqq.Karliner2017} & $1S$ &  & $(3882 \pm 12)_{0}$ & 
\\
\cline{1-5}
Heavy--quark symmetry~\cite{ccqq.cbqq.bbqq.Eichten2017} & $1S$ & $4146_{1}$ & $\begin{gathered}3978_{0} \\ 4167_{1}\end{gathered}$ & $4210_{1}$
\\
\cline{1-5}
Heavy--quark--diquark symmetry~\cite{ccqq.ccsq.ccss.cbqq.bbqq.Braaten2021} & $1S$ & $(4111 \pm 11)_{1}$ & $\begin{gathered}(3947 \pm 11)_{0} \\ (4133 \pm 11)_{1}\end{gathered}$ & $(4177 \pm 11)_{1}$
\\
\cline{1-5}
HADS mass formula~\cite{ccqq.ccsq.ccss.cbqq.bbqq.Zhang2026} & $1S$ & $4035_{1}$ & $\begin{gathered}3876_{0} \\ 4058_{1}\end{gathered}$ & $4103_{1}$
\\
\hline
\SetCell[c=5]{l}\textit{Chromomagnetic interaction models} &  &  &  &
\\
\nopagebreak
\cline{1-5}
\shortstack[l]{Extended chromomagnetic \\ (Scheme $\RNumb{1}$)~\cite{ccqq.cbqq.bbqq.Weng2022}} & $1S$ & $\begin{gathered}3833_{1} \\ 4127_{1}\end{gathered}$ & $\begin{gathered}3750_{0} \\ 3976_{0} \\ 3946_{1}\end{gathered}$ & $4017_{1}$
\\
\cline{1-5}
\shortstack[l]{Extended chromomagnetic \\ (Scheme $\RNumb{2}$)~\cite{ccqq.cbqq.bbqq.Weng2022}} & $1S$ & $\begin{gathered}3969_{1} \\ 4365_{1}\end{gathered}$ & $\begin{gathered}3869_{0} \\ 4231_{0} \\ 4053_{1}\end{gathered}$ & $4124_{1}$
\\
\cline{1-5}
Chromomagnetic + flux tube~\cite{ccqq.cbqq.bbqq.Liu2023} & $1S$ & $\begin{gathered}3975_{1} \\ 4414_{1}\end{gathered}$ & $\begin{gathered}3879_{0} \\ 4288_{0} \\ 4054_{1}\end{gathered}$ & $4125_{1}$
\\
\cline{1-5}
Flux--tube + chromomagnetic~\cite{ccqq.ccsq.ccss.cbqq.bbqq.Zhang2025} & $1S$ & $\begin{gathered}(3981 \pm 14)_{1} \\ (4374 \pm 14)_{1}\end{gathered}$ & $\begin{gathered}(3890 \pm 13)_{0} \\ (4244 \pm 15)_{0} \\ (4066 \pm 13)_{1}\end{gathered}$ & $(4137 \pm 13)_{1}$
\\
\hline
\SetCell[c=5]{l}\textit{Four--body constituent quark models} &  &  &  &
\\
\nopagebreak
\cline{1-5}
\shortstack[l]{Nonrelativistic diffusion \\ Monte Carlo~\cite{ccqq.cbqq.bbqq.Mutuk2024DMC}} & $1S$ & $\begin{gathered}4062_{1} \\ 4283_{1}\end{gathered}$ & $\begin{gathered}3892_{0} \\ 4130_{0} \\ 4104_{1}\end{gathered}$ & $4207_{1}$
\\
\cline{1-5}
Gaussian expansion + complex scaling~\cite{ccqq.cbqq.bbqq.Wu2024} & $1S$ &  & $3864_{0}^{\mathrm{mol}}$ & $4673_{1}^{\dagger}$
\\
\hline
\SetCell[c=5]{l}\textit{Hadronic molecules} &  &  &  &
\\
\nopagebreak
\cline{1-5}
\shortstack[l]{Meson--meson contact interaction \\ ($\Lambda=0.5$~GeV)~\cite{QQqq.molecule.2021}} &   &  & $\begin{gathered}3872.42_{0}^{\mathrm{virt}} \\ 4014.32_{0}^{\mathrm{virt}}\end{gathered}$ &
\\
\cline{2-5}
\shortstack[l]{Meson--meson contact interaction \\ ($\Lambda=1.0$~GeV)~\cite{QQqq.molecule.2021}} &   &  & $\begin{gathered}3870.04_{0}^{\mathrm{mol}} \\ 4009.93_{0}^{\mathrm{mol}}\end{gathered}$ &
\\
\hline
\SetCell[c=5]{l}\textit{QCD sum rules} &  &  &  &
\\
\nopagebreak
\cline{1-5}
Axialvector--diquark--antidiquark currents~\cite{ccqq.ccsq.ccss.Wang2018} & $1S$ & $(3870 \pm 90)_{1}$ & $(3900 \pm 90)_{1}$ & $(3950 \pm 90)_{1}$
\\
\cline{1-5}
Octet--octet currents~\cite{ccqq.ccsq.bbqq.bbsq.Tang2020} & $1S$ &  & $4720 \pm 130$ & 
\\
\cline{1-5}
Singlet--singlet currents~\cite{ccqq.ccsq.bbqq.bbsq.Tang2020} & $1S$ &  & $(4040 \pm 130)_{0}$ & 
\\
\hline
\SetCell[c=5]{l}\textit{Lattice QCD (finite volume levels and scattering poles)} &  &  &  &
\\
\nopagebreak
\cline{1-5}
Lattice finite-volume level~\cite{ccqq.ccsq.ccss.bbqq.bbsq.bbss.Junnarkar2019} &   & & $(3852 \pm 11)_{0}^{\mathrm{FV,ref}}$ &
\\
\cline{1-5}
Lattice pole, $m_{\pi} \simeq 280$~MeV~\cite{ccqq.Padmanath2022} &   & & $(3865.1^{+7.1}_{-3.6})_{0}^{\mathrm{virt,ref}}$ &
\\
\cline{1-5}
HAL QCD, $m_{\pi} = 146$~MeV~\cite{ccud.lattice.2023} &   & & $\big( \simeq 3875\big)_{0}^{\mathrm{virt,ref}}$ &
\\
\hline\hline
\end{longtblr}


\begin{table*}
\caption{Comparison of the predicted masses (in MeV) of the $\mathrm{cc\overline{qq^{\prime}}}$ tetraquarks in the $1P$ and $2P$ states with other theoretical approaches. Mass subscripts denote isospin $I=0,1$ where assigned.}
\label{tab:other.ccqq.P}
\centering
\setlength{\jot}{0pt}
\begin{ruledtabular}
\begin{tblr}{
	width=\textwidth,
	colspec={Q[l,m,wd=200pt] Q[c,m,wd=30pt] Q[c,m] Q[c,m] Q[c,m] Q[c,m]},
	rowsep=1pt,
	colsep=7pt
	}
\hline\hline
Approach and reference & $nL$ & $J^{P}=0^{-}$ & $J^{P}=1^{-}$ & $J^{P}=2^{-}$ & $J^{P}=3^{-}$
\\
\hline
\SetCell[r=2]{m}\shortstack{\textbf{This work} \\ \textbf{(relativistic diquark--antidiquark)}} & $1P$ & $\begin{gathered}4291_{0} \\ 4456_{1}\end{gathered}$ & $\begin{gathered}4299_{0} \\ 4423_{1} \\ 4471_{1} \\ 4500_{1}\end{gathered}$ & $\begin{gathered}4316_{0} \\ 4443_{1} \\ 4500_{1}\end{gathered}$ & $4473_{1}$
\\
\cline{2-6}
 & $2P$ & $\begin{gathered}5000_{0} \\ 5150_{1}\end{gathered}$ & $\begin{gathered}5000_{0} \\ 5084_{1} \\ 5149_{1} \\ 5188_{1}\end{gathered}$ & $\begin{gathered}5000_{0} \\ 5095_{1} \\ 5162_{1}\end{gathered}$ & $5112_{1}$
\\
\hline
\SetCell[c=6]{l}\textit{Diquark--antidiquark models} &  &  &  &  &
\\
\cline{1-6}
Regge relations~\cite{ccqq.cbqq.bbqq.Song2023} & $1P$ & $\begin{gathered}4253_{0} \\ 4429_{1}\end{gathered}$ & $\begin{gathered}4268_{0} \\ 4446_{1} \\ 4447_{1} \\ 4487_{1}\end{gathered}$ & $\begin{gathered}4298_{0} \\ 4484_{1} \\ 4499_{1}\end{gathered}$ & $4517_{1}$
\\
\cline{1-6}
\SetCell[r=2]{m}\shortstack[l]{Nonrelativistic potential \\ (Set $\RNumb{1}$)~\cite{ccqq.Lin2025}} & $1P$ & $\begin{gathered}4275_{0} \\ 4351_{1}\end{gathered}$ & $\begin{gathered}4285_{0} \\ 4335_{1} \\ 4380_{1} \\ 4417_{1}\end{gathered}$ & $\begin{gathered}4305_{0} \\ 4359_{1} \\ 4432_{1}\end{gathered}$ & $4400_{1}$
\\
\cline{2-6}
 & $2P$ & $\begin{gathered}4792_{0} \\ 4836_{1}\end{gathered}$ & $\begin{gathered}4802_{0} \\ 4837_{1} \\ 4863_{1} \\ 4889_{1}\end{gathered}$ & $\begin{gathered}4822_{0} \\ 4859_{1} \\ 4911_{1}\end{gathered}$ & $4896_{1}$
\\
\cline{1-6}
\SetCell[r=2]{m}Relativized GI potential~\cite{ccqq.Dong2025} & $1P$ & $\begin{gathered}4349_{0} \\ 4404_{1}\end{gathered}$ & $\begin{gathered}4370_{0} \\ 4410_{1} \\ 4477_{1} \\ 4480_{1}\end{gathered}$ & $\begin{gathered}4408_{0} \\ 4501_{1} \\ 4503_{1}\end{gathered}$ & $4529_{1}$
\\
\cline{2-6}
 & $2P$ & $\begin{gathered}4796_{0} \\ 4872_{1}\end{gathered}$ & $\begin{gathered}4810_{0} \\ 4881_{1} \\ 4908_{1} \\ 4916_{1}\end{gathered}$ & $\begin{gathered}4837_{0} \\ 4926_{1} \\ 4934_{1}\end{gathered}$ & $4950_{1}$
\\
\hline
\SetCell[c=6]{l}\textit{Three--body quark--antidiquark model} &  &  &  &  &
\\
\cline{1-6}
Chiral three--body model~\cite{ccqq.cbqq.bbqq.Kim2022} & $1P$ & $\begin{gathered}4253_{0} \\ 4430_{1}\end{gathered}$ & $\begin{gathered}4253_{0} \\ 4423_{1} \\ 4430_{1} \\ 4442_{1}\end{gathered}$ & $\begin{gathered}4253_{0} \\ 4430_{1} \\ 4442_{1}\end{gathered}$ & $4442_{1}$
\\
\hline
\SetCell[c=6]{l}\textit{Hadronic molecules (negative--parity meson--meson channels)} &  &  &  &  &
\\
\cline{1-6}
\shortstack[l]{Meson--meson contact interaction \\ ($\Lambda=0.5$~GeV)~\cite{QQqq.molecule.2021}} &   & $4429.65_{0}^{\mathrm{virt}}$ & $\begin{gathered}4286.52_{0}^{\mathrm{virt}} \\ 4429.65_{0}^{\mathrm{virt}} \\ 4471.00_{0}^{\mathrm{virt}}\end{gathered}$ & $\begin{gathered}4328.35_{0}^{\mathrm{virt}} \\ 4429.65_{0}^{\mathrm{virt}} \\ 4471.00_{0}^{\mathrm{virt}}\end{gathered}$ & $4471.00_{0}^{\mathrm{virt}}$
\\
\cline{2-6}
\shortstack[l]{Meson--meson contact interaction \\ ($\Lambda=1.0$~GeV)~\cite{QQqq.molecule.2021}} &   & $4421.88_{0}^{\mathrm{mol}}$ & $\begin{gathered}4282.06_{0}^{\mathrm{mol}} \\ 4421.88_{0}^{\mathrm{mol}} \\ 4461.90_{0}^{\mathrm{mol}}\end{gathered}$ & $\begin{gathered}4321.31_{0}^{\mathrm{mol}} \\ 4421.88_{0}^{\mathrm{mol}} \\ 4461.90_{0}^{\mathrm{mol}}\end{gathered}$ & $4461.90_{0}^{\mathrm{mol}}$
\\
\hline\hline
\end{tblr}
\end{ruledtabular}
\end{table*}

\begin{table*}
\caption{Comparison of the predicted masses (in MeV) of the $\mathrm{cc\overline{qq^{\prime}}}$ tetraquarks in the $1D$ states with other theoretical approaches. Mass subscripts denote isospin $I=0,1$ where assigned.}
\label{tab:other.ccqq.D}
\centering
\setlength{\jot}{0pt}
\begin{ruledtabular}
\begin{tblr}{
	width=\textwidth,
	colspec={Q[l,m,wd=200pt] Q[c,m] Q[c,m] Q[c,m] Q[c,m] Q[c,m]},
	rowsep=1pt,
	colsep=5.5pt
	}
\hline\hline
Approach and reference & $J^{P}=0^{+}$ & $J^{P}=1^{+}$ & $J^{P}=2^{+}$ & $J^{P}=3^{+}$ & $J^{P}=4^{+}$
\\
\hline
\shortstack{\textbf{This work} \\ \textbf{(relativistic diquark--antidiquark)}} & $4910_{1}$ & $\begin{gathered}4721_{0} \\ 4878_{1} \\ 4921_{1}\end{gathered}$ & $\begin{gathered}4733_{0} \\ 4834_{1} \\ 4892_{1} \\ 4941_{1}\end{gathered}$ & $\begin{gathered}4750_{0} \\ 4846_{1} \\ 4912_{1}\end{gathered}$ & $4863_{1}$
\\
\hline
\SetCell[c=6]{l}\textit{Diquark--antidiquark models} &  &  &  &  &
\\
\cline{1-6}
Relativized GI potential~\cite{ccqq.Dong2025} & $4776_{1}$ & $\begin{gathered}4689_{0} \\ 4783_{1} \\ 4793_{1}\end{gathered}$ & $\begin{gathered}4701_{0} \\ 4792_{1} \\ 4799_{1} \\ 4822_{1}\end{gathered}$ & $\begin{gathered}4718_{0} \\ 4802_{1} \\ 4821_{1}\end{gathered}$ & $4813_{1}$
\\
\hline\hline
\end{tblr}
\end{ruledtabular}
\end{table*}

\subsubsection{Singly strange: $\mathrm{cc\overline{sq}}$}
\label{sec:othermodels.ccsq}

\par
For $\mathrm{cc\overline{sq}}$, our lowest $1^{+}$ mass of $4093$~MeV is close to the mass $4106 \pm 12$~MeV estimated in the constituent quark model~\cite{ccsq.Karliner2022}. The heavy antiquark--diquark symmetry predictions~\cite{ccqq.ccsq.ccss.cbqq.bbqq.Zhang2026} differ from our $0^{+}$ and higher $1^{+}$ masses by approximately $30$--$35$~MeV. The flux--tube chromomagnetic calculation~\cite{ccqq.ccsq.ccss.cbqq.bbqq.Zhang2025} gives $4232 \pm 22$~MeV for the $2^{+}$ state, compared with our result of $4233$~MeV.

\begin{table*}
\caption{Comparison of the predicted masses (in MeV) of the $\mathrm{cc\overline{sq}}$ tetraquarks in the $1S$, $2S$, and $3S$ states with other theoretical approaches. All states have isospin $I=\frac{1}{2}$.}
\label{tab:other.ccsq.S}
\centering
\setlength{\jot}{0pt}
\begin{ruledtabular}
\begin{tblr}{
	width=\textwidth,
	colspec={Q[l,m,wd=200pt] Q[c,m,wd=30pt] Q[c,m] Q[c,m] Q[c,m]},
	rowsep=1pt,
	colsep=7pt
	}
\hline\hline
Approach and reference & $nL$ & $J^{P}=0^{+}$ & $J^{P}=1^{+}$ & $J^{P}=2^{+}$
\\
\hline
\SetCell[r=3]{m}\shortstack{\textbf{This work} \\ \textbf{(relativistic diquark--antidiquark)}} & $1S$ & $4161$ & $\begin{gathered}4093 \, (\mathrm{A\overline{S}}) \\ 4188 \, (\mathrm{A\overline{A}})\end{gathered}$ & $4233$
\\
\cline{2-5}
 & $2S$ & $4897$ & $\begin{gathered}4801 \, (\mathrm{A\overline{S}}) \\ 4899 \, (\mathrm{A\overline{A}})\end{gathered}$ & $4901$
\\
\cline{2-5}
 & $3S$ & $5567$ & $\begin{gathered}5479 \, (\mathrm{A\overline{S}}) \\ 5565 \, (\mathrm{A\overline{A}})\end{gathered}$ & $5556$
\\
\hline
\SetCell[c=5]{l}\textit{Diquark--antidiquark models} &  &  &  & 
\\
\cline{1-5}
Regge mass relations~\cite{ccqq.cbqq.bbqq.Song2023} & $1S$ & $4297$ & $\begin{gathered}4180 \\ 4315\end{gathered}$ & $4352$
\\
\hline
\SetCell[c=5]{l}\textit{Heavy--quark symmetry and hadron--mass relations} &  &  &  & 
\\
\cline{1-5}
Heavy--quark--diquark symmetry~\cite{ccqq.ccsq.ccss.cbqq.bbqq.Braaten2021} & $1S$ & $4232 \pm 11$ & $\begin{gathered}4124 \pm 12 \\ 4254 \pm 11\end{gathered}$ & $4298 \pm 11$
\\
\cline{1-5}
Constituent-mass estimate~\cite{ccsq.Karliner2022} & $1S$ &  & $4106 \pm 12$ & 
\\
\cline{1-5}
Superflavor symmetry~\cite{ccsq.Tanaka2024} & $1S$ &  & $4047 \pm 11$ & 
\\
\cline{1-5}
HADS mass formula~\cite{ccqq.ccsq.ccss.cbqq.bbqq.Zhang2026} & $1S$ & $4127$ & $\begin{gathered}4014 \\ 4158\end{gathered}$ & $4184$
\\
\hline
\SetCell[c=5]{l}\textit{Three--body heavy--quark--antidiquark models} &  &  &  & 
\\
\cline{1-5}
\SetCell[r=2]{m}Chiral light antidiquark~\cite{ccqq.cbqq.bbqq.Kim2022} & $1S$ & $4260$ & $\begin{gathered}4141 \\ 4277\end{gathered}$ & $4309$
\\
\cline{2-5}
 & $2S$ & $4673$ & $\begin{gathered}4545 \\ 4685\end{gathered}$ & $4709$
\\
\hline
\SetCell[c=5]{l}\textit{Chromomagnetic interaction models} &  &  &  & 
\\
\cline{1-5}
Color--magnetic mixing (threshold scheme)~\cite{ccqq.ccsq.ccss.cbqq.bbqq.Luo2017} & $1S$ & $\begin{gathered}3933 \\ 4210\end{gathered}$ & $\begin{gathered}3921 \\ 4060 \\ 4096\end{gathered}$ & $4131$
\\
\cline{1-5}
\shortstack[l]{Extended chromomagnetic \\ (Scheme $\RNumb{1}$)~\cite{ccqq.cbqq.bbqq.Weng2022}} & $1S$ & $\begin{gathered}3938 \\ 4209\end{gathered}$ & $\begin{gathered}3919 \\ 4073 \\ 4087\end{gathered}$ & $4144$
\\
\cline{1-5}
\shortstack[l]{Extended chromomagnetic \\ (Scheme $\RNumb{2}$)~\cite{ccqq.cbqq.bbqq.Weng2022}} & $1S$ & $\begin{gathered}4086 \\ 4436\end{gathered}$ & $\begin{gathered}4052 \\ 4180 \\ 4329\end{gathered}$ & $4252$
\\
\cline{1-5}
\shortstack[l]{Flux--tube + \\ chromomagnetic~\cite{ccqq.ccsq.ccss.cbqq.bbqq.Zhang2025}} & $1S$ & $\begin{gathered}4054 \pm 23 \\ 4368 \pm 26\end{gathered}$ & $\begin{gathered}4034 \pm 31 \\ 4160 \pm 4 \\ 4260 \pm 14\end{gathered}$ & $4232 \pm 22$
\\
\hline
\SetCell[c=5]{l}\textit{Four--body constituent quark models} &  &  &  & 
\\
\cline{1-5}
Chiral quark model, coupled channels~\cite{ccsq.ccss.bbsq.bbss.Chen2023} & $1S$ &  & $3923.8$ & 
\\
\cline{1-5}
Delocalization + coupled channels~\cite{ccsq.Liu2024} & $1S$ & $\big( 4113\text{--}4114 \big)^{\dagger}$ & $3971$--$3975$ & 
\\
\cline{1-5}
\shortstack[l]{Nonrelativistic diffusion \\ Monte Carlo~\cite{ccqq.cbqq.bbqq.Mutuk2024DMC}} & $1S$ & $\begin{gathered}4064 \\ 4349\end{gathered}$ & $\begin{gathered}4056 \\ 4128 \\ 4201\end{gathered}$ & $4314$
\\
\cline{1-5}
Gaussian expansion + complex scaling~\cite{ccqq.cbqq.bbqq.Wu2024} & $1S$ &  & $4578^{\dagger}$ & $4723^{\dagger}$
\\
\hline
\SetCell[c=5]{l}\textit{QCD sum rules} &  &  &  & 
\\
\cline{1-5}
Axialvector--diquark--antidiquark currents~\cite{ccqq.ccsq.ccss.Wang2018} & $1S$ & $3940 \pm 100$ & $3960 \pm 80$ & $4010 \pm 90$
\\
\cline{1-5}
Octet--octet currents~\cite{ccqq.ccsq.bbqq.bbsq.Tang2020} & $1S$ &  & $\begin{gathered}4760 \pm 140 \\ 4780 \pm 140\end{gathered}$ & 
\\
\cline{1-5}
Singlet--singlet currents~\cite{ccqq.ccsq.bbqq.bbsq.Tang2020} & $1S$ &  & $4180 \pm 130$ & 
\\
\cline{1-5}
Diquark QCD sum rules~\cite{ccsq.Agaev2023} & $1S$ & $\begin{gathered}4035 \pm 145 \\ 4128 \pm 142\end{gathered}$ & $3995 \pm 143$ & 
\\
\hline
\SetCell[c=5]{l}\textit{Lattice QCD (threshold-referenced level)} & & & &
\\
\cline{1-5}
Finite-volume level~\cite{ccqq.ccsq.ccss.bbqq.bbsq.bbss.Junnarkar2019} &   & & $(3969 \pm 8)^{\mathrm{FV,ref}}$ &
\\
\hline\hline
\end{tblr}
\end{ruledtabular}
\end{table*}

\begin{table*}
\caption{Comparison of the predicted masses (in MeV) of the $\mathrm{cc\overline{sq}}$ tetraquarks in the $1P$ and $2P$ states with other theoretical approaches. All states have isospin $I=\frac{1}{2}$.}
\label{tab:other.ccsq.P}
\centering
\setlength{\jot}{0pt}
\begin{ruledtabular}
\begin{tblr}{
	width=\textwidth,
	colspec={Q[l,m,wd=200pt] Q[c,m,wd=30pt] Q[c,m] Q[c,m] Q[c,m] Q[c,m]},
	rowsep=1pt,
	colsep=7pt
	}
\hline\hline
Approach and reference & $nL$ & $J^{P}=0^{-}$ & $J^{P}=1^{-}$ & $J^{P}=2^{-}$ & $J^{P}=3^{-}$
\\
\hline
\SetCell[r=2]{m}\shortstack{\textbf{This work} \\ \textbf{(relativistic diquark--antidiquark)}} & $1P$ & $\begin{gathered}4493 \, (\mathrm{A\overline{S}}) \\ 4600 \, (\mathrm{A\overline{A}})\end{gathered}$ & $\begin{gathered}4500 \, (\mathrm{A\overline{S}}) \\ 4575 \, (\mathrm{A\overline{A}}) \\ 4612 \, (\mathrm{A\overline{A}}) \\ 4635 \, (\mathrm{A\overline{A}})\end{gathered}$ & $\begin{gathered}4514 \, (\mathrm{A\overline{S}}) \\ 4591 \, (\mathrm{A\overline{A}}) \\ 4636 \, (\mathrm{A\overline{A}})\end{gathered}$ & $4616$
\\
\cline{2-6}
 & $2P$ & $\begin{gathered}5162 \, (\mathrm{A\overline{S}}) \\ 5266 \, (\mathrm{A\overline{A}})\end{gathered}$ & $\begin{gathered}5162 \, (\mathrm{A\overline{S}}) \\ 5215 \, (\mathrm{A\overline{A}}) \\ 5265 \, (\mathrm{A\overline{A}}) \\ 5297 \, (\mathrm{A\overline{A}})\end{gathered}$ & $\begin{gathered}5162 \, (\mathrm{A\overline{S}}) \\ 5223 \, (\mathrm{A\overline{A}}) \\ 5276 \, (\mathrm{A\overline{A}})\end{gathered}$ & $5237$
\\
\hline
\SetCell[c=6]{l}\textit{Diquark--antidiquark models} &  &  &  &  & 
\\
\cline{1-6}
Regge mass relations~\cite{ccqq.cbqq.bbqq.Song2023} & $1P$ & $\begin{gathered}4458 \\ 4563\end{gathered}$ & $\begin{gathered}4472 \\ 4579 \\ 4579 \\ 4613\end{gathered}$ & $\begin{gathered}4499 \\ 4612 \\ 4624\end{gathered}$ & $4641$
\\
\hline
\SetCell[c=6]{l}\textit{Three--body heavy--quark--antidiquark models} &  &  &  &  & 
\\
\cline{1-6}
Chiral light antidiquark~\cite{ccqq.cbqq.bbqq.Kim2022} & $1P$ & $\begin{gathered}4410 \\ 4544 \\ 4551\end{gathered}$ & $\begin{gathered}4410 \\ 4538 \\ 4544 \\ 4551 \\ 4556\end{gathered}$ & $\begin{gathered}4410 \\ 4544 \\ 4551 \\ 4556\end{gathered}$ & $4556$
\\
\hline\hline
\end{tblr}
\end{ruledtabular}
\end{table*}

\subsubsection{Doubly strange: $\mathrm{cc\overline{ss}}$}
\label{sec:othermodels.ccss}

\par
Our ground--state $\mathrm{cc\overline{ss}}$ masses differ from the heavy--quark--diquark symmetry estimates of Ref.~\cite{ccqq.ccsq.ccss.cbqq.bbqq.Braaten2021} by approximately $45$~MeV, with similar level ordering and spin splittings. Close predictions for the $1^{+}$ and $2^{+}$ states are also obtained in the four--body diffusion Monte Carlo calculations~\cite{ccqq.cbqq.bbqq.Mutuk2024DMC}.

\begin{table*}
\caption{Comparison of the predicted masses (in MeV) of the $\mathrm{cc\overline{ss}}$ tetraquarks in the $1S$, $2S$, and $3S$ states with other theoretical approaches. All states have isospin $I=0$.}
\label{tab:other.ccss.S}
\centering
\setlength{\jot}{0pt}
\begin{ruledtabular}
\begin{tblr}{
	width=\textwidth,
	colspec={Q[l,m,wd=200pt] Q[c,m,wd=30pt] Q[c,m] Q[c,m] Q[c,m]},
	rowsep=1pt,
	colsep=7pt
	}
\hline\hline
Approach and reference & $nL$ & $J^{P}=0^{+}$ & $J^{P}=1^{+}$ & $J^{P}=2^{+}$
\\
\hline
\SetCell[r=3]{m}\shortstack{\textbf{This work} \\ \textbf{(relativistic diquark--antidiquark)}} & $1S$ & $4308$ & $4330$ & $4368$
\\
\cline{2-5}
 & $2S$ & $5013$ & $5015$ & $5016$
\\
\cline{2-5}
 & $3S$ & $5663$ & $5660$ & $5653$
\\
\hline
\SetCell[c=5]{l}\textit{Diquark--antidiquark models} &  &  &  & 
\\
\cline{1-5}
Regge mass relations~\cite{ccqq.cbqq.bbqq.Song2023} & $1S$ & $4420$ & $4436$ & $4470$
\\
\hline
\SetCell[c=5]{l}\textit{Heavy--quark symmetry and hadron--mass relations} &  &  &  & 
\\
\cline{1-5}
Heavy--quark--diquark symmetry~\cite{ccqq.ccsq.ccss.cbqq.bbqq.Braaten2021} & $1S$ & $4352 \pm 12$ & $4374 \pm 12$ & $4418 \pm 12$
\\
\cline{1-5}
HADS mass formula~\cite{ccqq.ccsq.ccss.cbqq.bbqq.Zhang2026} & $1S$ & $4224$ & $4239$ & $4270$
\\
\hline
\SetCell[c=5]{l}\textit{Three--body heavy--quark--antidiquark models} &  &  &  & 
\\
\cline{1-5}
\SetCell[r=2]{m}Chiral light antidiquark~\cite{ccqq.cbqq.bbqq.Kim2022} & $1S$ & $4376$ & $4392$ & $4421$
\\
\cline{2-5}
 & $2S$ & $4786$ & $4798$ & $4820$
\\
\hline
\SetCell[c=5]{l}\textit{Chromomagnetic interaction models} &  &  &  & 
\\
\cline{1-5}
Color--magnetic mixing (threshold scheme)~\cite{ccqq.ccsq.ccss.cbqq.bbqq.Luo2017} & $1S$ & $\begin{gathered}4016 \\ 4293\end{gathered}$ & $4146$ & $4218$
\\
\cline{1-5}
\shortstack[l]{Extended chromomagnetic \\ (Scheme $\RNumb{1}$)~\cite{ccqq.cbqq.bbqq.Weng2022}} & $1S$ & $\begin{gathered}4044 \\ 4311\end{gathered}$ & $4193$ & $4265$
\\
\cline{1-5}
\shortstack[l]{Extended chromomagnetic \\ (Scheme $\RNumb{2}$)~\cite{ccqq.cbqq.bbqq.Weng2022}} & $1S$ & $\begin{gathered}4199 \\ 4532\end{gathered}$ & $4300$ & $4372$
\\
\cline{1-5}
\shortstack[l]{Flux--tube + \\ chromomagnetic~\cite{ccqq.ccsq.ccss.cbqq.bbqq.Zhang2025}} & $1S$ & $\begin{gathered}4122 \pm 22 \\ 4392 \pm 23\end{gathered}$ & $4265 \pm 20$ & $4337 \pm 20$
\\
\hline
\SetCell[c=5]{l}\textit{Four--body constituent quark models} &  &  &  & 
\\
\cline{1-5}
\shortstack[l]{Nonrelativistic diffusion \\ Monte Carlo~\cite{ccqq.cbqq.bbqq.Mutuk2024DMC}} & $1S$ & $\begin{gathered}4205 \\ 4474\end{gathered}$ & $4323$ & $4381$
\\
\cline{1-5}
Gaussian expansion + complex scaling~\cite{ccqq.cbqq.bbqq.Wu2024} & $1S$ &  &  & $4808^{\dagger}$
\\
\cline{1-5}
Nonrelativistic Gaussian expansion~\cite{ccss.Liu2025} & $1S$ & $\begin{gathered}4386 \\ 4510\end{gathered}$ & $4449$ & $4481$
\\
\hline
\SetCell[c=5]{l}\textit{QCD sum rules} &  &  &  & 
\\
\cline{1-5}
Axialvector--diquark--antidiquark currents~\cite{ccqq.ccsq.ccss.Wang2018} & $1S$ & $3990 \pm 100$ & $4020 \pm 90$ & $4060 \pm 90$
\\
\hline\hline
\end{tblr}
\end{ruledtabular}
\end{table*}

\begin{table*}
\caption{Comparison of the predicted masses (in MeV) of the $\mathrm{cc\overline{ss}}$ tetraquarks in the $1P$ and $2P$ states with other theoretical approaches. All states have isospin $I=0$.}
\label{tab:other.ccss.P}
\centering
\setlength{\jot}{0pt}
\begin{ruledtabular}
\begin{tblr}{
	width=\textwidth,
	colspec={Q[l,m,wd=200pt] Q[c,m,wd=30pt] Q[c,m] Q[c,m] Q[c,m] Q[c,m]},
	rowsep=1pt,
	colsep=7pt
	}
\hline\hline
Approach and reference & $nL$ & $J^{P}=0^{-}$ & $J^{P}=1^{-}$ & $J^{P}=2^{-}$ & $J^{P}=3^{-}$
\\
\hline
\SetCell[r=2]{m}\shortstack{\textbf{This work} \\ \textbf{(relativistic diquark--antidiquark)}} & $1P$ & $4724$ & $\begin{gathered}4702 \\ 4733 \\ 4753\end{gathered}$ & $\begin{gathered}4716 \\ 4754\end{gathered}$ & $4738$
\\
\cline{2-6}
 & $2P$ & $5368$ & $\begin{gathered}5325 \\ 5366 \\ 5395\end{gathered}$ & $\begin{gathered}5332 \\ 5376\end{gathered}$ & $5343$
\\
\hline
\SetCell[c=6]{l}\textit{Diquark--antidiquark models} &  &  &  &  & 
\\
\cline{1-6}
Regge mass relations~\cite{ccqq.cbqq.bbqq.Song2023} & $1P$ & $4696$ & $\begin{gathered}4710 \\ 4711 \\ 4739\end{gathered}$ & $\begin{gathered}4741 \\ 4751\end{gathered}$ & $4766$
\\
\hline
\SetCell[c=6]{l}\textit{Three--body heavy--quark--antidiquark models} &  &  &  &  & 
\\
\cline{1-6}
Chiral light antidiquark~\cite{ccqq.cbqq.bbqq.Kim2022} & $1P$ & $\begin{gathered}4650 \\ 4661\end{gathered}$ & $\begin{gathered}4644 \\ 4650 \\ 4661 \\ 4665\end{gathered}$ & $\begin{gathered}4650 \\ 4661 \\ 4665\end{gathered}$ & $4665$
\\
\hline\hline
\end{tblr}
\end{ruledtabular}
\end{table*}

\FloatBarrier

\subsection{Charmed--bottom tetraquarks}
\label{sec:othermodels.cb}

\subsubsection{Nonstrange: $\mathrm{cb\overline{qq^{\prime}}}$}
\label{sec:othermodels.cbqq}

\par
For $\mathrm{cb\overline{qq^{\prime}}}$, our lowest isoscalar $0^{+}$ and $1^{+}$ masses, $7191$ and $7198$~MeV, differ by approximately $20$--$50$~MeV from the corresponding heavy antiquark--diquark symmetry estimates~\cite{ccqq.ccsq.ccss.cbqq.bbqq.Zhang2026} and flux--tube chromomagnetic predictions~\cite{ccqq.ccsq.ccss.cbqq.bbqq.Zhang2025}. The extended chromomagnetic model~\cite{ccqq.cbqq.bbqq.Weng2022} produces a wider range of masses depending on the parameter scheme. Comparisons of excited states remain more limited.


\setlength{\jot}{0pt}
\begin{longtblr}[caption={Comparison of the predicted masses (in MeV) of the $\mathrm{cb\overline{qq^{\prime}}}$ tetraquarks in the $1S$, $2S$, and $3S$ states with other theoretical approaches. Mass subscripts denote isospin $I=0,1$ where assigned.},label={tab:other.cbqq.S},theme=prdwidetable]{
	width=\textwidth,
	rowhead=4,
	colspec={Q[l,m,wd=200pt] Q[c,m,wd=30pt] Q[c,m] Q[c,m] Q[c,m]},
	rowsep=0.5pt,
	colsep=7pt
	}
\hline\hline
Approach and reference & $nL$ & $J^{P}=0^{+}$ & $J^{P}=1^{+}$ & $J^{P}=2^{+}$
\\
\hline
\SetCell[r=3]{m}\shortstack{\textbf{This work} \\ \textbf{(relativistic diquark--antidiquark)}} & $1S$ & $\begin{gathered}7337_{1} \\ 7191_{0}\end{gathered}$ & $\begin{gathered}7353_{1} \, (\mathrm{A\overline{A}}) \\ 7198_{0} \, (\mathrm{A\overline{S}}) \\ 7361_{1} \, (\mathrm{S\overline{A}})\end{gathered}$ & $7381_{1}$
\\
\cline{2-5}
 & $2S$ & $\begin{gathered}7964_{1} \\ 7815_{0}\end{gathered}$ & $\begin{gathered}7967_{1} \, (\mathrm{A\overline{A}}) \\ 7822_{0} \, (\mathrm{A\overline{S}}) \\ 7963_{1} \, (\mathrm{S\overline{A}})\end{gathered}$ & $7972_{1}$
\\
\cline{2-5}
 & $3S$ & $\begin{gathered}8517_{1} \\ 8383_{0}\end{gathered}$ & $\begin{gathered}8516_{1} \, (\mathrm{A\overline{A}}) \\ 8389_{0} \, (\mathrm{A\overline{S}}) \\ 8508_{1} \, (\mathrm{S\overline{A}})\end{gathered}$ & $8513_{1}$
\\
\hline
\SetCell[c=5]{l}\textit{Diquark--antidiquark models} &  &  &  & 
\\
\cline{1-5}
Regge mass relations~\cite{ccqq.cbqq.bbqq.Song2023} & $1S$ & $\begin{gathered}7268_{0} \\ 7458_{1}\end{gathered}$ & $\begin{gathered}7269_{0} \\ 7469_{1} \\ 7478_{1}\end{gathered}$ & $7490_{1}$
\\
\hline
\SetCell[c=5]{l}\textit{Three--body heavy--quark--antidiquark models} &  &  &  & 
\\
\cline{1-5}
\SetCell[r=2]{m}Chiral light antidiquark~\cite{ccqq.cbqq.bbqq.Kim2022} & $1S$ & $\begin{gathered}7225_{0} \\ 7431_{1}\end{gathered}$ & $\begin{gathered}7246_{0} \\ 7429_{1} \\ 7441_{1}\end{gathered}$ & $7459_{1}$
\\
\cline{2-5}
 & $2S$ & $\begin{gathered}7601_{0} \\ 7800_{1}\end{gathered}$ & $\begin{gathered}7610_{0} \\ 7805_{1} \\ 7808_{1}\end{gathered}$ & $7823_{1}$
\\
\hline
\SetCell[c=5]{l}\textit{Heavy--quark symmetry and hadron--mass relations} &  &  &  & 
\\
\cline{1-5}
Constituent--quark mass relations~\cite{ccqq.cbqq.bbqq.Karliner2017} & $1S$ & $(7134 \pm 13)_{0}$ &  & 
\\
\cline{1-5}
Heavy--quark symmetry~\cite{ccqq.cbqq.bbqq.Eichten2017} & $1S$ & $\begin{gathered}7229_{0} \\ 7461_{1}\end{gathered}$ & $\begin{gathered}7272_{0} \\ 7439_{1} \\ 7472_{1}\end{gathered}$ & $7493_{1}$
\\
\cline{1-5}
\shortstack[l]{Heavy--quark EFT \\ mass relations~\cite{ccqq.ccsq.ccss.cbqq.bbqq.Braaten2021}} & $1S$ & $\begin{gathered}(7248 \pm 37)_{0} \\ (7460 \pm 27)_{1}\end{gathered}$ & $\begin{gathered}(7282 \pm 28)_{0} \\ (7455 \pm 37)_{1} \\ (7474 \pm 27)_{1}\end{gathered}$ & $(7503 \pm 27)_{1}$
\\
\cline{1-5}
\shortstack[l]{Heavy antiquark--diquark \\ symmetry (unmixed)~\cite{ccqq.ccsq.ccss.cbqq.bbqq.Zhang2026}} & $1S$ & $\begin{gathered}7142_{0} \\ 7358_{1}\end{gathered}$ & $\begin{gathered}7176_{0} \\ 7347_{1} \\ 7369_{1}\end{gathered}$ & $7392_{1}$
\\
\hline
\SetCell[c=5]{l}\textit{Chromomagnetic interaction models} &  &  &  & 
\\
\cline{1-5}
\shortstack[l]{Extended chromomagnetic \\ (Scheme $\RNumb{1}$)~\cite{ccqq.cbqq.bbqq.Weng2022}} & $1S$ & $\begin{gathered}7003_{0}, \; 7220_{0} \\ 7190_{1}, \; 7441_{1}\end{gathered}$ & $\begin{gathered}7046_{0}, \; 7233_{0} \\ 7329_{0}, \; 7211_{1} \\ 7264_{1}, \; 7417_{1}\end{gathered}$ & $\begin{gathered}7353_{0} \\ 7293_{1}\end{gathered}$
\\
\cline{1-5}
\shortstack[l]{Extended chromomagnetic \\ (Scheme $\RNumb{2}$)~\cite{ccqq.cbqq.bbqq.Weng2022}} & $1S$ & $\begin{gathered}7125_{0}, \; 7459_{0} \\ 7306_{1}, \; 7685_{1}\end{gathered}$ & $\begin{gathered}7158_{0}, \; 7482_{0} \\ 7585_{0}, \; 7323_{1} \\ 7367_{1}, \; 7665_{1}\end{gathered}$ & $\begin{gathered}7610_{0} \\ 7396_{1}\end{gathered}$
\\
\cline{1-5}
Chromomagnetic + flux tube~\cite{ccqq.cbqq.bbqq.Liu2023} & $1S$ & $\begin{gathered}7128_{0}, \; 7523_{0} \\ 7313_{1}, \; 7733_{1}\end{gathered}$ & $\begin{gathered}7159_{0}, \; 7655_{0} \\ 7537_{0}, \; 7323_{1} \\ 7717_{1}, \; 7412_{1}\end{gathered}$ & $\begin{gathered}7657_{0} \\ 7386_{1}\end{gathered}$
\\
\cline{1-5}
\shortstack[l]{Flux--tube + \\ chromomagnetic~\cite{ccqq.ccsq.ccss.cbqq.bbqq.Zhang2025}} & $1S$ & $\begin{gathered}(7142 \pm 13)_{0} \\ (7471 \pm 14)_{0} \\ (7312 \pm 16)_{1} \\ (7691 \pm 17)_{1}\end{gathered}$ & $\begin{gathered}(7173 \pm 16)_{0} \\ (7494 \pm 15)_{0} \\ (7596 \pm 16)_{0} \\ (7330 \pm 14)_{1} \\ (7375 \pm 14)_{1} \\ (7672 \pm 14)_{1}\end{gathered}$ & $\begin{gathered}(7622 \pm 14)_{0} \\ (7402 \pm 16)_{1}\end{gathered}$
\\
\hline
\SetCell[c=5]{l}\textit{Four--body constituent quark models} &  &  &  & 
\\
\cline{1-5}
Four--body AL1 potential~\cite{cbqq.Carames2019} & $1S$ & $(7132)_{0}$ & $(7189)_{0}$ & $(7363)_{0}$
\\
\cline{1-5}
\shortstack[l]{Nonrelativistic diffusion \\ Monte Carlo~\cite{ccqq.cbqq.bbqq.Mutuk2024DMC}} & $1S$ & $\begin{gathered}7108_{0} \\ 7201_{0} \\ 7396_{1} \\ 7620_{1}\end{gathered}$ & $\begin{gathered}7084_{0} \\ 7204_{0} \\ 7324_{0} \\ 7510_{1} \\ 7448_{1} \\ 7542_{1}\end{gathered}$ & $\begin{gathered}7360_{0} \\ 7568_{1}\end{gathered}$
\\
\cline{1-5}
Gaussian expansion + complex scaling~\cite{ccqq.cbqq.bbqq.Wu2024} & $1S$ & $7129_{0}$ & $7185_{0}$ & $7363_{0}^{\mathrm{mol}}$
\\
\hline
\SetCell[c=5]{l}\textit{Lattice QCD (threshold--referenced results)} & & & &
\\
\cline{1-5}
NRQCD~\cite{cbqq.bbqq.bbsq.Francis2019} & $1S$ & & $(7129\text{--}7175)_{0}^{\mathrm{ref}}$ &
\\
\cline{1-5}
Lattice scattering poles~\cite{cbqq.Alexandrou2024} & & $( \simeq 7145^{+0.4}_{-1.5})_{0}^{\mathrm{ref}}$ & $( \simeq 7188^{+2.0}_{-0.7})_{0}^{\mathrm{ref}}$ & 
\\
\cline{1-5}
Lattice $D\bar B^{*}$ scattering~\cite{cbqq.Padmanath2024} & $1S$ & & $( \simeq 7147^{+6+14}_{-7-24})_{0}^{\mathrm{ref}}$ & 
\\
\cline{1-5}
Lattice $D\bar B$ scattering~\cite{cbqq.Radhakrishnan2024} & $1S$ & $( \simeq 7106^{+4+8}_{-6-18})_{0}^{\mathrm{ref}}$ & & 
\\
\hline\hline
\end{longtblr}


\begin{table*}
\caption{Comparison of the predicted masses (in MeV) of the $\mathrm{cb\overline{qq^{\prime}}}$ tetraquarks in the $1P$ and $2P$ states with other theoretical approaches. Mass subscripts denote isospin $I=0,1$ where assigned.}
\label{tab:other.cbqq.P}
\centering
\setlength{\jot}{0pt}
\begin{ruledtabular}
\begin{tblr}{
	width=\textwidth,
	colspec={Q[l,m,wd=200pt] Q[c,m,wd=30pt] Q[c,m] Q[c,m] Q[c,m] Q[c,m]},
	rowsep=1pt,
	colsep=7pt
	}
\hline\hline
Approach and reference & $nL$ & $J^{P}=0^{-}$ & $J^{P}=1^{-}$ & $J^{P}=2^{-}$ & $J^{P}=3^{-}$
\\
\hline
\SetCell[r=2]{m}\shortstack{\textbf{This work} \\ \textbf{(relativistic diquark--antidiquark)}} & $1P$ & $\begin{gathered}7722_{1} \, (\mathrm{A\overline{A}}) \\ 7552_{0} \, (\mathrm{A\overline{S}}) \\ 7749_{1} \, (\mathrm{S\overline{A}})\end{gathered}$ & $\begin{gathered}7691_{1} \, (\mathrm{A\overline{A}}) \\ 7731_{1} \, (\mathrm{A\overline{A}}) \\ 7756_{1} \, (\mathrm{A\overline{A}}) \\ 7557_{0} \, (\mathrm{A\overline{S}}) \\ 7733_{1} \, (\mathrm{S\overline{A}}) \\ 7556_{0} \, (\mathrm{S\overline{S}})\end{gathered}$ & $\begin{gathered}7702_{1} \, (\mathrm{A\overline{A}}) \\ 7748_{1} \, (\mathrm{A\overline{A}}) \\ 7568_{0} \, (\mathrm{A\overline{S}}) \\ 7700_{1} \, (\mathrm{S\overline{A}})\end{gathered}$ & $7718_{1}$
\\
\cline{2-6}
 & $2P$ & $\begin{gathered}8281_{1} \, (\mathrm{A\overline{A}}) \\ 8123_{0} \, (\mathrm{A\overline{S}}) \\ 8306_{1} \, (\mathrm{S\overline{A}})\end{gathered}$ & $\begin{gathered}8225_{1} \, (\mathrm{A\overline{A}}) \\ 8284_{1} \, (\mathrm{A\overline{A}}) \\ 8314_{1} \, (\mathrm{A\overline{A}}) \\ 8125_{0} \, (\mathrm{A\overline{S}}) \\ 8281_{1} \, (\mathrm{S\overline{A}}) \\ 8120_{0} \, (\mathrm{S\overline{S}})\end{gathered}$ & $\begin{gathered}8232_{1} \, (\mathrm{A\overline{A}}) \\ 8292_{1} \, (\mathrm{A\overline{A}}) \\ 8129_{0} \, (\mathrm{A\overline{S}}) \\ 8229_{1} \, (\mathrm{S\overline{A}})\end{gathered}$ & $8244_{1}$
\\
\hline
\SetCell[c=6]{l}\textit{Diquark--antidiquark models} &  &  &  &  & 
\\
\cline{1-6}
Regge mass relations~\cite{ccqq.cbqq.bbqq.Song2023} & $1P$ & $\begin{gathered}7520_{0} \\ 7687_{1}\end{gathered}$ & $\begin{gathered}7528_{0} \\ 7695_{1} \\ 7696_{1} \\ 7715_{1}\end{gathered}$ & $\begin{gathered}7542_{0} \\ 7714_{1} \\ 7721_{1}\end{gathered}$ & $7730_{1}$
\\
\hline\hline
\end{tblr}
\end{ruledtabular}
\end{table*}

\subsubsection{Singly strange: $\mathrm{cb\overline{sq}}$}
\label{sec:othermodels.cbsq}

\par
For $\mathrm{cb\overline{sq}}$, our lowest $0^{+}$ mass of $7403$~MeV almost coincides with the three--body prediction of $7402$~MeV~\cite{ccqq.cbqq.bbqq.Kim2022}. For the lowest $1^{+}$ state, our result of $7410$~MeV is also close to the flux--tube chromomagnetic estimate of $7409\pm22$~MeV~\cite{ccqq.ccsq.ccss.cbqq.bbqq.Zhang2025}. The QCD sum--rule masses~\cite{cbsq.Wang2020} are generally lower by about $0.2$~GeV.

\begin{table*}
\caption{Comparison of the predicted masses (in MeV) of the $\mathrm{cb\overline{sq}}$ tetraquarks in the $1S$, $2S$, and $3S$ states with other theoretical approaches. All states have isospin $I=\frac{1}{2}$.}
\label{tab:other.cbsq.S}
\centering
\setlength{\jot}{0pt}
\begin{ruledtabular}
\begin{tblr}{
	width=\textwidth,
	colspec={Q[l,m,wd=200pt] Q[c,m,wd=30pt] Q[c,m] Q[c,m] Q[c,m]},
	rowsep=1pt,
	colsep=7pt
	}
\hline\hline
Approach and reference & $nL$ & $J^{P}=0^{+}$ & $J^{P}=1^{+}$ & $J^{P}=2^{+}$
\\
\hline
\SetCell[r=3]{m}\shortstack{\textbf{This work} \\ \textbf{(relativistic diquark--antidiquark)}} & $1S$ & $\begin{gathered}7499 \, (\mathrm{A\overline{A}}) \\ 7403 \, (\mathrm{S\overline{S}})\end{gathered}$ & $\begin{gathered}7511 \, (\mathrm{A\overline{A}}) \\ 7410 \, (\mathrm{A\overline{S}}) \\ 7516 \, (\mathrm{S\overline{A}})\end{gathered}$ & $7535$
\\
\cline{2-5}
 & $2S$ & $\begin{gathered}8101 \, (\mathrm{A\overline{A}}) \\ 7998 \, (\mathrm{S\overline{S}})\end{gathered}$ & $\begin{gathered}8103 \, (\mathrm{A\overline{A}}) \\ 8004 \, (\mathrm{A\overline{S}}) \\ 8098 \, (\mathrm{S\overline{A}})\end{gathered}$ & $8106$
\\
\cline{2-5}
 & $3S$ & $\begin{gathered}8633 \, (\mathrm{A\overline{A}}) \\ 8538 \, (\mathrm{S\overline{S}})\end{gathered}$ & $\begin{gathered}8632 \, (\mathrm{A\overline{A}}) \\ 8545 \, (\mathrm{A\overline{S}}) \\ 8624 \, (\mathrm{S\overline{A}})\end{gathered}$ & $8629$
\\
\hline
\SetCell[c=5]{l}\textit{Diquark--antidiquark models} &  &  &  & 
\\
\cline{1-5}
Regge mass relations~\cite{ccqq.cbqq.bbqq.Song2023} & $1S$ & $\begin{gathered}7451 \\ 7558\end{gathered}$ & $\begin{gathered}7452 \\ 7579 \\ 7605\end{gathered}$ & $7616$
\\
\hline
\SetCell[c=5]{l}\textit{Three--body heavy--quark--antidiquark models} &  &  &  & 
\\
\cline{1-5}
\SetCell[r=2]{m}Chiral light antidiquark~\cite{ccqq.cbqq.bbqq.Kim2022} & $1S$ & $\begin{gathered}7402 \\ 7557\end{gathered}$ & $\begin{gathered}7423 \\ 7552 \\ 7565\end{gathered}$ & $7582$
\\
\cline{2-5}
 & $2S$ & $\begin{gathered}7779 \\ 7924\end{gathered}$ & $\begin{gathered}7788 \\ 7929 \\ 7931\end{gathered}$ & $7945$
\\
\hline
\SetCell[c=5]{l}\textit{Heavy--quark symmetry and hadron--mass relations} &  &  &  & 
\\
\cline{1-5}
Heavy--quark symmetry~\cite{ccqq.cbqq.bbqq.Eichten2017} & $1S$ & $7406$ & $7445$ & 
\\
\cline{1-5}
\shortstack[l]{Heavy--quark EFT \\ mass relations~\cite{ccqq.ccsq.ccss.cbqq.bbqq.Braaten2021}} & $1S$ & $\begin{gathered}7422 \pm 38 \\ 7578 \pm 28\end{gathered}$ & $\begin{gathered}7456 \pm 28 \\ 7573 \pm 37 \\ 7592 \pm 28\end{gathered}$ & $7621 \pm 28$
\\
\cline{1-5}
\shortstack[l]{Heavy \\ antiquark--diquark symmetry~\cite{ccqq.ccsq.ccss.cbqq.bbqq.Zhang2026}} & $1S$ & $\begin{gathered}7286 \\ 7440\end{gathered}$ & $\begin{gathered}7317 \\ 7425 \\ 7452\end{gathered}$ & $7468$
\\
\hline
\SetCell[c=5]{l}\textit{Chromomagnetic interaction models} &  &  &  & 
\\
\cline{1-5}
\shortstack[l]{Extended chromomagnetic \\ (Scheme $\RNumb{1}$)~\cite{ccqq.cbqq.bbqq.Weng2022}} & $1S$ & $\begin{gathered}7157, \; 7299 \\ 7333, \; 7507\end{gathered}$ & $\begin{gathered}7213, \; 7324 \\ 7331, \; 7386 \\ 7416, \; 7480\end{gathered}$ & $\begin{gathered}7415 \\ 7439\end{gathered}$
\\
\cline{1-5}
\shortstack[l]{Extended chromomagnetic \\ (Scheme $\RNumb{2}$)~\cite{ccqq.cbqq.bbqq.Weng2022}} & $1S$ & $\begin{gathered}7296, \; 7421 \\ 7548, \; 7739\end{gathered}$ & $\begin{gathered}7336, \; 7441 \\ 7488, \; 7565 \\ 7665, \; 7717\end{gathered}$ & $\begin{gathered}7518 \\ 7691\end{gathered}$
\\
\cline{1-5}
\shortstack[l]{Flux--tube + \\ chromomagnetic~\cite{ccqq.ccsq.ccss.cbqq.bbqq.Zhang2025}} & $1S$ & $\begin{gathered}7266 \pm 8 \\ 7387 \pm 24 \\ 7477 \pm 8 \\ 7657 \pm 28\end{gathered}$ & $\begin{gathered}7310 \pm 7 \\ 7409 \pm 22 \\ 7457 \pm 20 \\ 7490 \pm 11 \\ 7584 \pm 13 \\ 7635 \pm 26\end{gathered}$ & $\begin{gathered}7488 \pm 24 \\ 7610 \pm 12\end{gathered}$
\\
\hline
\SetCell[c=5]{l}\textit{QCD sum rules} &  &  &  & 
\\
\cline{1-5}
Diquark--antidiquark currents~\cite{cbsq.Wang2020} & $1S$ & $\begin{gathered}7240 \pm 130 \\ 7140 \pm 120 \\ 7180 \pm 120 \\ 7200 \pm 140\end{gathered}$ & $\begin{gathered}7230 \pm 130 \\ 7140 \pm 120 \\ 7170 \pm 140 \\ 7230 \pm 130\end{gathered}$ & 
\\
\hline\hline
\end{tblr}
\end{ruledtabular}
\end{table*}

\begin{table*}
\caption{Comparison of the predicted masses (in MeV) of the $\mathrm{cb\overline{sq}}$ tetraquarks in the $1P$ and $2P$ states with other theoretical approaches. All states have isospin $I=\frac{1}{2}$.}
\label{tab:other.cbsq.P}
\centering
\setlength{\jot}{0pt}
\begin{ruledtabular}
\begin{tblr}{
	width=\textwidth,
	colspec={Q[l,m,wd=200pt] Q[c,m,wd=30pt] Q[c,m] Q[c,m] Q[c,m] Q[c,m]},
	rowsep=1pt,
	colsep=7pt
	}
\hline\hline
Approach and reference & $nL$ & $J^{P}=0^{-}$ & $J^{P}=1^{-}$ & $J^{P}=2^{-}$ & $J^{P}=3^{-}$
\\
\hline
\SetCell[r=2]{m}\shortstack{\textbf{This work} \\ \textbf{(relativistic diquark--antidiquark)}} & $1P$ & $\begin{gathered}7863 \, (\mathrm{A\overline{A}}) \\ 7751 \, (\mathrm{A\overline{S}}) \\ 7883 \, (\mathrm{S\overline{A}})\end{gathered}$ & $\begin{gathered}7837 \, (\mathrm{A\overline{A}}) \\ 7870 \, (\mathrm{A\overline{A}}) \\ 7889 \, (\mathrm{A\overline{A}}) \\ 7756 \, (\mathrm{A\overline{S}}) \\ 7870 \, (\mathrm{S\overline{A}}) \\ 7753 \, (\mathrm{S\overline{S}})\end{gathered}$ & $\begin{gathered}7847 \, (\mathrm{A\overline{A}}) \\ 7884 \, (\mathrm{A\overline{A}}) \\ 7765 \, (\mathrm{A\overline{S}}) \\ 7845 \, (\mathrm{S\overline{A}})\end{gathered}$ & $7861$
\\
\cline{2-6}
 & $2P$ & $\begin{gathered}8401 \, (\mathrm{A\overline{A}}) \\ 8292 \, (\mathrm{A\overline{S}}) \\ 8419 \, (\mathrm{S\overline{A}})\end{gathered}$ & $\begin{gathered}8356 \, (\mathrm{A\overline{A}}) \\ 8402 \, (\mathrm{A\overline{A}}) \\ 8426 \, (\mathrm{A\overline{A}}) \\ 8294 \, (\mathrm{A\overline{S}}) \\ 8399 \, (\mathrm{S\overline{A}}) \\ 8288 \, (\mathrm{S\overline{S}})\end{gathered}$ & $\begin{gathered}8362 \, (\mathrm{A\overline{A}}) \\ 8409 \, (\mathrm{A\overline{A}}) \\ 8297 \, (\mathrm{A\overline{S}}) \\ 8359 \, (\mathrm{S\overline{A}})\end{gathered}$ & $8372$
\\
\hline
\SetCell[c=6]{l}\textit{Diquark--antidiquark models} &  &  &  &  & 
\\
\cline{1-6}
Regge mass relations~\cite{ccqq.cbqq.bbqq.Song2023} & $1P$ & $\begin{gathered}7735 \\ 7855\end{gathered}$ & $\begin{gathered}7741 \\ 7863 \\ 7863 \\ 7879\end{gathered}$ & $\begin{gathered}7754 \\ 7879 \\ 7884\end{gathered}$ & $7893$
\\
\hline\hline
\end{tblr}
\end{ruledtabular}
\end{table*}

\subsubsection{Doubly strange: $\mathrm{cb\overline{ss}}$}
\label{sec:othermodels.cbss}

\par
For $\mathrm{cb\overline{ss}}$, the three--body calculation~\cite{ccqq.cbqq.bbqq.Kim2022} gives ground--state masses of $7670$~MeV for $0^{+}$, $7665$ and $7678$~MeV for $1^{+}$, and $7693$~MeV for $2^{+}$. These values are $20$--$36$~MeV above our corresponding predictions.

\begin{table*}
\caption{Comparison of the predicted masses (in MeV) of the $\mathrm{cb\overline{ss}}$ tetraquarks in the $1S$, $2S$, and $3S$ states with other theoretical approaches. All states have isospin $I=0$.}
\label{tab:other.cbss.S}
\centering
\setlength{\jot}{0pt}
\begin{ruledtabular}
\begin{tblr}{
	width=\textwidth,
	colspec={Q[l,m,wd=200pt] Q[c,m,wd=30pt] Q[c,m] Q[c,m] Q[c,m]},
	rowsep=1pt,
	colsep=7pt
	}
\hline\hline
Approach and reference & $nL$ & $J^{P}=0^{+}$ & $J^{P}=1^{+}$ & $J^{P}=2^{+}$
\\
\hline
\SetCell[r=3]{m}\shortstack{\textbf{This work} \\ \textbf{(relativistic diquark--antidiquark)}} & $1S$ & $7634$ & $\begin{gathered}7645 \, (\mathrm{A\overline{A}}) \\ 7649 \, (\mathrm{S\overline{A}})\end{gathered}$ & $7665$
\\
\cline{2-5}
 & $2S$ & $8217$ & $\begin{gathered}8218 \, (\mathrm{A\overline{A}}) \\ 8213 \, (\mathrm{S\overline{A}})\end{gathered}$ & $8221$
\\
\cline{2-5}
 & $3S$ & $8733$ & $\begin{gathered}8732 \, (\mathrm{A\overline{A}}) \\ 8724 \, (\mathrm{S\overline{A}})\end{gathered}$ & $8729$
\\
\hline
\SetCell[c=5]{l}\textit{Diquark--antidiquark models} &  &  &  & 
\\
\cline{1-5}
Regge mass relations~\cite{ccqq.cbqq.bbqq.Song2023} & $1S$ & $7717$ & $\begin{gathered}7710 \\ 7724 \\ 7733\end{gathered}$ & 
\\
\hline
\SetCell[c=5]{l}\textit{Three--body heavy--quark--antidiquark models} &  &  &  & 
\\
\cline{1-5}
\SetCell[r=2]{m}Chiral light antidiquark~\cite{ccqq.cbqq.bbqq.Kim2022} & $1S$ & $7670$ & $\begin{gathered}7665 \\ 7678\end{gathered}$ & $7693$
\\
\cline{2-5}
 & $2S$ & $8035$ & $\begin{gathered}8038 \\ 8042\end{gathered}$ & $8055$
\\
\hline
\SetCell[c=5]{l}\textit{Heavy--quark symmetry and hadron--mass relations} &  &  &  & 
\\
\cline{1-5}
\shortstack[l]{Heavy--quark EFT \\ mass relations~\cite{ccqq.ccsq.ccss.cbqq.bbqq.Braaten2021}} & $1S$ & $7696 \pm 28$ & $\begin{gathered}7691 \pm 38 \\ 7710 \pm 28\end{gathered}$ & $7739 \pm 28$
\\
\cline{1-5}
\shortstack[l]{Heavy \\ antiquark--diquark symmetry~\cite{ccqq.ccsq.ccss.cbqq.bbqq.Zhang2026}} & $1S$ & $7527$ & $\begin{gathered}7508 \\ 7535\end{gathered}$ & $7550$
\\
\hline
\SetCell[c=5]{l}\textit{Chromomagnetic interaction models} &  &  &  & 
\\
\cline{1-5}
\shortstack[l]{Extended chromomagnetic \\ (Scheme $\RNumb{1}$)~\cite{ccqq.cbqq.bbqq.Weng2022}} & $1S$ & $\begin{gathered}7404 \\ 7597\end{gathered}$ & $\begin{gathered}7432 \\ 7503 \\ 7569\end{gathered}$ & $7534$
\\
\cline{1-5}
\shortstack[l]{Extended chromomagnetic \\ (Scheme $\RNumb{2}$)~\cite{ccqq.cbqq.bbqq.Weng2022}} & $1S$ & $\begin{gathered}7531 \\ 7819\end{gathered}$ & $\begin{gathered}7554 \\ 7604 \\ 7795\end{gathered}$ & $7634$
\\
\cline{1-5}
\shortstack[l]{Flux--tube + \\ chromomagnetic~\cite{ccqq.ccsq.ccss.cbqq.bbqq.Zhang2025}} & $1S$ & $\begin{gathered}7457 \pm 24 \\ 7651 \pm 26\end{gathered}$ & $\begin{gathered}7486 \pm 21 \\ 7557 \pm 21 \\ 7624 \pm 23\end{gathered}$ & $7587 \pm 23$
\\
\hline
\SetCell[c=5]{l}\textit{Four--body constituent quark models} &  &  &  & 
\\
\cline{1-5}
\shortstack[l]{Nonrelativistic diffusion \\ Monte Carlo~\cite{ccqq.cbqq.bbqq.Mutuk2024DMC}} & $1S$ & $\begin{gathered}7653 \\ 7833\end{gathered}$ & $\begin{gathered}7711 \\ 7764 \\ 7816\end{gathered}$ & $7862$
\\
\hline\hline
\end{tblr}
\end{ruledtabular}
\end{table*}

\begin{table*}
\caption{Comparison of the predicted masses (in MeV) of the $\mathrm{cb\overline{ss}}$ tetraquarks in the $1P$ and $2P$ states with other theoretical approaches. All states have isospin $I=0$.}
\label{tab:other.cbss.P}
\centering
\setlength{\jot}{0pt}
\begin{ruledtabular}
\begin{tblr}{
	width=\textwidth,
	colspec={Q[l,m,wd=200pt] Q[c,m,wd=30pt] Q[c,m] Q[c,m] Q[c,m] Q[c,m]},
	rowsep=1pt,
	colsep=7pt
	}
\hline\hline
Approach and reference & $nL$ & $J^{P}=0^{-}$ & $J^{P}=1^{-}$ & $J^{P}=2^{-}$ & $J^{P}=3^{-}$
\\
\hline
\SetCell[r=2]{m}\shortstack{\textbf{This work} \\ \textbf{(relativistic diquark--antidiquark)}} & $1P$ & $\begin{gathered}7983 \, (\mathrm{A\overline{A}}) \\ 7999 \, (\mathrm{S\overline{A}})\end{gathered}$ & $\begin{gathered}7961 \, (\mathrm{A\overline{A}}) \\ 7989 \, (\mathrm{A\overline{A}}) \\ 8006 \, (\mathrm{A\overline{A}}) \\ 7989 \, (\mathrm{S\overline{A}})\end{gathered}$ & $\begin{gathered}7969 \, (\mathrm{A\overline{A}}) \\ 8002 \, (\mathrm{A\overline{A}}) \\ 7967 \, (\mathrm{S\overline{A}})\end{gathered}$ & $7982$
\\
\cline{2-6}
 & $2P$ & $\begin{gathered}8505 \, (\mathrm{A\overline{A}}) \\ 8518 \, (\mathrm{S\overline{A}})\end{gathered}$ & $\begin{gathered}8468 \, (\mathrm{A\overline{A}}) \\ 8506 \, (\mathrm{A\overline{A}}) \\ 8526 \, (\mathrm{A\overline{A}}) \\ 8502 \, (\mathrm{S\overline{A}})\end{gathered}$ & $\begin{gathered}8473 \, (\mathrm{A\overline{A}}) \\ 8512 \, (\mathrm{A\overline{A}}) \\ 8469 \, (\mathrm{S\overline{A}})\end{gathered}$ & $8481$
\\
\hline
\SetCell[c=6]{l}\textit{Diquark--antidiquark models} &  &  &  &  & 
\\
\cline{1-6}
Regge mass relations~\cite{ccqq.cbqq.bbqq.Song2023} & $1P$ & $7959$ & $\begin{gathered}7966 \\ 7966 \\ 7980\end{gathered}$ & $\begin{gathered}7981 \\ 7985\end{gathered}$ & $7993$
\\
\hline\hline
\end{tblr}
\end{ruledtabular}
\end{table*}

\FloatBarrier

\subsection{Doubly bottom tetraquarks}
\label{sec:othermodels.bb}

\subsubsection{Nonstrange: $\mathrm{bb\overline{qq^{\prime}}}$}
\label{sec:othermodels.bbqq}

\par
For $\mathrm{bb\overline{qq^{\prime}}}$, our lowest isoscalar $1^{+}$ mass of $10461$~MeV differs by only $10$~MeV from the heavy--quark EFT estimate~\cite{ccqq.ccsq.ccss.cbqq.bbqq.Braaten2021}. For the isovector $0^{+}$ and $2^{+}$ states, our masses of $10606$ and $10638$~MeV differ by $30$ and $25$~MeV, respectively, from the flux--tube chromomagnetic predictions~\cite{ccqq.ccsq.ccss.cbqq.bbqq.Zhang2025}. The comparison with lattice QCD results for the lowest $1^{+}$ state is discussed below.


\begin{longtblr}[caption={Comparison of the predicted masses (in MeV) of the $\mathrm{bb\overline{qq^{\prime}}}$ tetraquarks in the $1S$, $2S$, and $3S$ states with other theoretical approaches. Mass subscripts denote isospin $I=0,1$ where assigned.},label={tab:other.bbqq.S},theme=prdwidetable]{
	width=\textwidth,
	rowhead=4,
	colspec={Q[l,m,wd=200pt] Q[c,m,wd=30pt] Q[c,m] Q[c,m] Q[c,m]},
	rowsep=1pt,
	colsep=7pt
	}
\hline\hline
Approach and reference & $nL$ & $J^{P}=0^{+}$ & $J^{P}=1^{+}$ & $J^{P}=2^{+}$
\\
\hline
\SetCell[r=3]{m}\shortstack{\textbf{This work} \\ \textbf{(relativistic diquark--antidiquark)}} & $1S$ & $10606_{1}$ & $\begin{gathered}10617_{1} \\ 10461_{0}\end{gathered}$ & $10638_{1}$
\\
\cline{2-5}
 & $2S$ & $11188_{1}$ & $\begin{gathered}11191_{1} \\ 11045_{0}\end{gathered}$ & $11196_{1}$
\\
\cline{2-5}
 & $3S$ & $11691_{1}$ & $\begin{gathered}11691_{1} \\ 11563_{0}\end{gathered}$ & $11690_{1}$
\\
\hline
\SetCell[c=5]{l}\textit{Diquark--antidiquark models} &  &  &  & 
\\
\nopagebreak
\cline{1-5}
Regge mass relations~\cite{ccqq.cbqq.bbqq.Song2023} & $1S$ & $10726_{1}$ & $\begin{gathered}10530_{0} \\ 10733_{1}\end{gathered}$ & $10747_{1}$
\\
\hline
\SetCell[c=5]{l}\textit{Three--body heavy--quark--antidiquark models} &  &  &  & 
\\
\nopagebreak
\cline{1-5}
\SetCell[r=2]{m}Chiral light antidiquark~\cite{ccqq.cbqq.bbqq.Kim2022} & $1S$ & $10679_{1}$ & $\begin{gathered}10489_{0} \\ 10685_{1}\end{gathered}$ & $10696_{1}$
\\
\cline{2-5}
 & $2S$ & $11011_{1}$ & $\begin{gathered}10815_{0} \\ 11016_{1}\end{gathered}$ & $11025_{1}$
\\
\hline
\SetCell[c=5]{l}\textit{Heavy--quark symmetry and hadron--mass relations} &  &  &  & 
\\
\nopagebreak
\cline{1-5}
Constituent--quark mass relations~\cite{ccqq.cbqq.bbqq.Karliner2017} & $1S$ &  & $(10389 \pm 12)_{0}$ & 
\\
\cline{1-5}
Heavy--quark symmetry~\cite{ccqq.cbqq.bbqq.Eichten2017} & $1S$ & $10674_{1}$ & $\begin{gathered}10482_{0} \\ 10681_{1}\end{gathered}$ & $10695_{1}$
\\
\cline{1-5}
\shortstack[l]{Heavy--quark EFT \\ mass relations~\cite{ccqq.ccsq.ccss.cbqq.bbqq.Braaten2021}} & $1S$ & $(10664 \pm 25)_{1}$ & $\begin{gathered}(10471 \pm 25)_{0} \\ (10671 \pm 25)_{1}\end{gathered}$ & $(10692 \pm 25)_{1}$
\\
\cline{1-5}
\shortstack[l]{Heavy \\ antiquark--diquark symmetry~\cite{ccqq.ccsq.ccss.cbqq.bbqq.Zhang2026}} & $1S$ & $10586_{1}$ & $\begin{gathered}10396_{0} \\ 10593_{1}\end{gathered}$ & $10608_{1}$
\\
\hline
\SetCell[c=5]{l}\textit{Chromomagnetic interaction models} &  &  &  & 
\\
\nopagebreak
\cline{1-5}
\shortstack[l]{Extended chromomagnetic \\ (Scheme $\RNumb{1}$)~\cite{ccqq.cbqq.bbqq.Weng2022}} & $1S$ & $\begin{gathered}10469_{1} \\ 10809_{1}\end{gathered}$ & $\begin{gathered}10292_{0} \\ 10703_{0} \\ 10485_{1}\end{gathered}$ & $10508_{1}$
\\
\cline{1-5}
\shortstack[l]{Extended chromomagnetic \\ (Scheme $\RNumb{2}$)~\cite{ccqq.cbqq.bbqq.Weng2022}} & $1S$ & $\begin{gathered}10569_{1} \\ 11055_{1}\end{gathered}$ & $\begin{gathered}10391_{0} \\ 10950_{0} \\ 10584_{1}\end{gathered}$ & $10607_{1}$
\\
\cline{1-5}
Chromomagnetic + flux tube~\cite{ccqq.cbqq.bbqq.Liu2023} & $1S$ & $\begin{gathered}10486_{1} \\ 11151_{1}\end{gathered}$ & $\begin{gathered}10298_{0} \\ 11053_{0} \\ 10490_{1}\end{gathered}$ & $10498_{1}$
\\
\cline{1-5}
\shortstack[l]{Flux--tube + \\ chromomagnetic~\cite{ccqq.ccsq.ccss.cbqq.bbqq.Zhang2025}} & $1S$ & $\begin{gathered}(10576 \pm 17)_{1} \\ (11057 \pm 18)_{1}\end{gathered}$ & $\begin{gathered}(10406 \pm 17)_{0} \\ (10957 \pm 18)_{0} \\ (10590 \pm 17)_{1}\end{gathered}$ & $(10613 \pm 17)_{1}$
\\
\hline
\SetCell[c=5]{l}\textit{Four--body constituent quark models} &  &  &  & 
\\
\nopagebreak
\cline{1-5}
Variational four--body (bound)~\cite{bbqq.Meng2022} & & & $\begin{gathered}(10444)_{0} \\ (10613)_{0}^{\mathrm{mol}}\end{gathered}$ & 
\\
\cline{1-5}
Variational four--body (RSM)~\cite{bbqq.Meng2022} & & & $(10667)_{0}^{\mathrm{RSM}}$ & 
\\
\cline{1-5}
\shortstack[l]{Nonrelativistic diffusion \\ Monte Carlo~\cite{ccqq.cbqq.bbqq.Mutuk2024DMC}} & $1S$ & $\begin{gathered}10624_{1} \\ 10748_{1}\end{gathered}$ & $\begin{gathered}10338_{0} \\ 10624_{0} \\ 10680_{1}\end{gathered}$ & $10702_{1}$
\\
\cline{1-5}
Gaussian expansion + complex scaling~\cite{ccqq.cbqq.bbqq.Wu2024} & $1S$ &  & $\begin{gathered}10491_{0} \\ 10642_{0}^{\mathrm{mol}}\end{gathered}$ & 
\\
\hline
\SetCell[c=5]{l}\textit{QCD sum rules} &  &  &  &
\\
\nopagebreak
\cline{1-5}
Octet--octet currents~\cite{ccqq.ccsq.bbqq.bbsq.Tang2020} & $1S$ &  & $11280 \pm 150$ & 
\\
\cline{1-5}
Singlet--singlet currents~\cite{ccqq.ccsq.bbqq.bbsq.Tang2020} & $1S$ &  & $(10360 \pm 150)_{0}$ & 
\\
\hline
\SetCell[c=5]{l}\textit{Lattice QCD and lattice--derived potential models} &  &  &  & 
\\
\nopagebreak
\cline{1-5}
Lattice-QCD potential~\cite{bbqq.Bicudo2017Spin} & $1S$ & & $(10545^{+38}_{-30})_{0}$ &
\\
\cline{1-5}
NRQCD~\cite{bbqq.bbsq.Francis2017} & $1S$ & & $(10415 \pm 10 \pm 3)_{0}$ &
\\
\cline{1-5}
Lattice QCD, derived~\cite{ccqq.ccsq.ccss.bbqq.bbsq.bbss.Junnarkar2019} & $1S$ & & $(10461 \pm 34)_{0}^{\mathrm{ref}}$ &
\\
\cline{1-5}
Lattice QCD (NRQCD)~\cite{bbud.lattice.2019} & $1S$ &  & $(10476 \pm 24 \pm 10)_{0}$ &
\\
\cline{1-5}
NRQCD--HISQ~\cite{bbqq.Mohanta2020} & $1S$ & & $(10417 \pm 9)_{0}$ & 
\\
\cline{1-5}
Lattice scattering (prelim.), derived~\cite{cbqq.bbqq.bbsq.Pflaumer2023} & $1S$ & & $(10501 \pm 8)_{0}^{\mathrm{ref}}$ &
\\
\cline{1-5}
Tuned lattice NRQCD~\cite{bbqq.bbsq.Hudspith2023} & $1S$ & & $(10492 \pm 13)_{0}^{\mathrm{ref}}$ & 
\\
\cline{1-5}
HAL QCD (coupled), derived~\cite{bbqq.Aoki2023} & $1S$ & & $(10521 \pm 10 \pm 20)_{0}^{\mathrm{ref}}$ &
\\
\cline{1-5}
HAL QCD (single), derived~\cite{bbqq.Aoki2023} & $1S$ & & $(10449 \pm 17 \pm 20)_{0}^{\mathrm{ref}}$ &
\\
\cline{1-5}
Lattice NRQCD (GEVP)~\cite{bbqq.bbsq.Alexandrou2024} & $1S$ & & $(10504 \pm 10^{+36}_{-51})_{0}^{\mathrm{ref}}$ & 
\\
\cline{1-5}
\shortstack[l]{Lattice QCD \\ derived mass~\cite{bbqq.bbsq.Colquhoun2024}} & $1S$ &  & $(10489 \pm 17)_{0}^{\mathrm{ref}}$ & 
\\
\cline{1-5}
\shortstack[l]{Lattice QCD (RHQ) \\ derived mass~\cite{bbqq.Hoffmann2026}} & $1S$ &  & $(10528 \pm 23)_{0}^{\mathrm{ref}}$ & 
\\
\hline\hline
\end{longtblr}


\begin{table*}
\caption{Comparison of the predicted masses (in MeV) of the $\mathrm{bb\overline{qq^{\prime}}}$ tetraquarks in the $1P$ and $2P$ states with other theoretical approaches. Mass subscripts denote isospin $I=0,1$ where assigned.}
\label{tab:other.bbqq.P}
\centering
\setlength{\jot}{0pt}
\begin{ruledtabular}
\begin{tblr}{
	width=\textwidth,
	colspec={Q[l,m,wd=200pt] Q[c,m,wd=30pt] Q[c,m] Q[c,m] Q[c,m] Q[c,m]},
	rowsep=1pt,
	colsep=7pt
	}
\hline\hline
Approach and reference & $nL$ & $J^{P}=0^{-}$ & $J^{P}=1^{-}$ & $J^{P}=2^{-}$ & $J^{P}=3^{-}$
\\
\hline
\SetCell[r=2]{m}\shortstack{\textbf{This work} \\ \textbf{(relativistic diquark--antidiquark)}} & $1P$ & $\begin{gathered}10969_{1} \\ 10799_{0}\end{gathered}$ & $\begin{gathered}10940_{1} \\ 10976_{1} \\ 10998_{1} \\ 10803_{0}\end{gathered}$ & $\begin{gathered}10947_{1} \\ 10989_{1} \\ 10812_{0}\end{gathered}$ & $10958_{1}$
\\
\cline{2-6}
 & $2P$ & $\begin{gathered}11481_{1} \\ 11323_{0}\end{gathered}$ & $\begin{gathered}11430_{1} \\ 11484_{1} \\ 11511_{1} \\ 11325_{0}\end{gathered}$ & $\begin{gathered}11436_{1} \\ 11490_{1} \\ 11329_{0}\end{gathered}$ & $11445_{1}$
\\
\hline
\SetCell[c=6]{l}\textit{Diquark--antidiquark models} &  &  &  &  & 
\\
\cline{1-6}
Regge mass relations~\cite{ccqq.cbqq.bbqq.Song2023} & $1P$ & $\begin{gathered}10764_{0} \\ 10941_{1}\end{gathered}$ & $\begin{gathered}10769_{0} \\ 10946_{1} \\ 10946_{1} \\ 10959_{1}\end{gathered}$ & $\begin{gathered}10779_{0} \\ 10958_{1} \\ 10963_{1}\end{gathered}$ & $10969_{1}$
\\
\hline
\SetCell[c=6]{l}\textit{Four--body constituent quark model}
\\
\cline{1-6}
Variational four--body, RSM~\cite{bbqq.Meng2022} & & $(10762)_{0}^{\mathrm{RSM}}$ & $(10762)_{0}^{\mathrm{RSM}}$ & $(10762)_{0}^{\mathrm{RSM}}$ & 
\\
\hline
\SetCell[c=6]{l}\textit{Lattice-QCD potential (resonance pole)} & & & & &
\\
\cline{1-6}
Born--Oppenheimer~\cite{bbqq.Bicudo2017Res} & $P_{\mathrm{BO}}$ & & $(10576 \pm 4)_{0}^{\dagger}$ & &
\\
\hline\hline
\end{tblr}
\end{ruledtabular}
\end{table*}

\subsubsection{Singly strange: $\mathrm{bb\overline{sq}}$}
\label{sec:othermodels.bbsq}

\par
For $\mathrm{bb\overline{sq}}$, all four ground--state masses differ from the heavy--quark EFT predictions~\cite{ccqq.ccsq.ccss.cbqq.bbqq.Braaten2021} by only $16$--$26$~MeV. The three--body calculation~\cite{ccqq.cbqq.bbqq.Kim2022} gives $10664$~MeV for the lowest $1^{+}$ state, compared with our prediction of $10670$~MeV.

\begin{table*}
\caption{Comparison of the predicted masses (in MeV) of the $\mathrm{bb\overline{sq}}$ tetraquarks in the $1S$, $2S$, and $3S$ states with other theoretical approaches. All states have isospin $I=\frac{1}{2}$.}
\label{tab:other.bbsq.S}
\centering
\setlength{\jot}{0pt}
\begin{ruledtabular}
\begin{tblr}{
	width=\textwidth,
	colspec={Q[l,m,wd=200pt] Q[c,m,wd=30pt] Q[c,m] Q[c,m] Q[c,m]},
	rowsep=1pt,
	colsep=7pt
	}
\hline\hline
Approach and reference & $nL$ & $J^{P}=0^{+}$ & $J^{P}=1^{+}$ & $J^{P}=2^{+}$
\\
\hline
\SetCell[r=3]{m}\shortstack{\textbf{This work} \\ \textbf{(relativistic diquark--antidiquark)}} & $1S$ & $10763$ & $\begin{gathered}10772 \, (\mathrm{A\overline{A}}) \\ 10670 \, (\mathrm{A\overline{S}})\end{gathered}$ & $10790$
\\
\cline{2-5}
 & $2S$ & $11324$ & $\begin{gathered}11326 \, (\mathrm{A\overline{A}}) \\ 11228 \, (\mathrm{A\overline{S}})\end{gathered}$ & $11330$
\\
\cline{2-5}
 & $3S$ & $11808$ & $\begin{gathered}11808 \, (\mathrm{A\overline{A}}) \\ 11721 \, (\mathrm{A\overline{S}})\end{gathered}$ & $11807$
\\
\hline
\SetCell[c=5]{l}\textit{Diquark--antidiquark models} &  &  &  & 
\\
\cline{1-5}
Regge mass relations~\cite{ccqq.cbqq.bbqq.Song2023} & $1S$ & $10855$ & $\begin{gathered}10713 \\ 10861\end{gathered}$ & $10873$
\\
\hline
\SetCell[c=5]{l}\textit{Three--body heavy--quark--antidiquark models} &  &  &  & 
\\
\cline{1-5}
\SetCell[r=2]{m}Chiral light antidiquark~\cite{ccqq.cbqq.bbqq.Kim2022} & $1S$ & $10802$ & $\begin{gathered}10664 \\ 10808\end{gathered}$ & $10818$
\\
\cline{2-5}
 & $2S$ & $11136$ & $\begin{gathered}10993 \\ 11140\end{gathered}$ & $11148$
\\
\hline
\SetCell[c=5]{l}\textit{Heavy--quark symmetry and hadron--mass relations} &  &  &  & 
\\
\cline{1-5}
Heavy--quark symmetry~\cite{ccqq.cbqq.bbqq.Eichten2017} & $1S$ &  & $10643$ & 
\\
\cline{1-5}
\shortstack[l]{Heavy--quark EFT \\ mass relations~\cite{ccqq.ccsq.ccss.cbqq.bbqq.Braaten2021}} & $1S$ & $10781 \pm 25$ & $\begin{gathered}10644 \pm 26 \\ 10788 \pm 25\end{gathered}$ & $10809 \pm 25$
\\
\cline{1-5}
\shortstack[l]{Heavy \\ antiquark--diquark symmetry~\cite{ccqq.ccsq.ccss.cbqq.bbqq.Zhang2026}} & $1S$ & $10660$ & $\begin{gathered}10533 \\ 10667\end{gathered}$ & $10679$
\\
\hline
\SetCell[c=5]{l}\textit{Chromomagnetic interaction models} &  &  &  & 
\\
\cline{1-5}
\shortstack[l]{Extended chromomagnetic \\ (Scheme $\RNumb{1}$)~\cite{ccqq.cbqq.bbqq.Weng2022}} & $1S$ & $\begin{gathered}10586 \\ 10855\end{gathered}$ & $\begin{gathered}10473 \\ 10605 \\ 10779\end{gathered}$ & $10629$
\\
\cline{1-5}
\shortstack[l]{Extended chromomagnetic \\ (Scheme $\RNumb{2}$)~\cite{ccqq.cbqq.bbqq.Weng2022}} & $1S$ & $\begin{gathered}10684 \\ 11090\end{gathered}$ & $\begin{gathered}10569 \\ 10701 \\ 11016\end{gathered}$ & $10724$
\\
\cline{1-5}
\shortstack[l]{Flux--tube + \\ chromomagnetic~\cite{ccqq.ccsq.ccss.cbqq.bbqq.Zhang2025}} & $1S$ & $\begin{gathered}10652 \pm 26 \\ 10994 \pm 30\end{gathered}$ & $\begin{gathered}10547 \pm 7 \\ 10669 \pm 26 \\ 10924 \pm 15\end{gathered}$ & $10693 \pm 26$
\\
\hline
\SetCell[c=5]{l}\textit{Four--body constituent quark models} &  &  &  & 
\\
\cline{1-5}
Chiral quark model, coupled channels~\cite{ccsq.ccss.bbsq.bbss.Chen2023} & $1S$ & & $10582.9$ &
\\
\cline{1-5}
\shortstack[l]{Nonrelativistic diffusion \\ Monte Carlo~\cite{ccqq.cbqq.bbqq.Mutuk2024DMC}} & $1S$ & $\begin{gathered}10721 \\ 10852\end{gathered}$ & $\begin{gathered}10684 \\ 10736 \\ 10778\end{gathered}$ & $10747$
\\
\cline{1-5}
Gaussian expansion + complex scaling~\cite{ccqq.cbqq.bbqq.Wu2024} & $1S$ &  & $10647$ & 
\\
\hline
\SetCell[c=5]{l}\textit{QCD sum rules} &  &  &  &
\\
\cline{1-5}
Octet--octet currents~\cite{ccqq.ccsq.bbqq.bbsq.Tang2020} & $1S$ &  & $\begin{gathered}11360 \pm 160 \\ 11390 \pm 160\end{gathered}$ & 
\\
\cline{1-5}
Singlet--singlet currents~\cite{ccqq.ccsq.bbqq.bbsq.Tang2020} & $1S$ &  & $10510 \pm 160$ & 
\\
\hline
\SetCell[c=5]{l}\textit{Lattice QCD} &  &  &  & 
\\
\cline{1-5}
NRQCD~\cite{bbqq.bbsq.Francis2017} & $1S$ & & $10594 \pm 7 \pm 3$ &
\\
\cline{1-5}
Lattice QCD, derived~\cite{ccqq.ccsq.ccss.bbqq.bbsq.bbss.Junnarkar2019} & $1S$ & & $(10608 \pm 32)^{\mathrm{ref}}$ &
\\
\cline{1-5}
Lattice scattering, derived~\cite{cbqq.bbqq.bbsq.Pflaumer2023} & $1S$ & & $(10609 \pm 22 \pm 10)^{\mathrm{ref}}$ &
\\
\cline{1-5}
Nonperturbatively tuned NRQCD~\cite{bbqq.bbsq.Hudspith2023} & $1S$ & & $( \simeq 10645 \pm 12)^{\mathrm{ref}}$ & 
\\
\cline{1-5}
Lattice NRQCD (GEVP)~\cite{bbqq.bbsq.Alexandrou2024} & $1S$ & & $( \simeq 10665 \pm 3^{+11}_{-31})^{\mathrm{ref}}$ & 
\\
\cline{1-5}
\shortstack[l]{Lattice QCD \\ derived mass~\cite{bbqq.bbsq.Colquhoun2024}} & $1S$ &  & $(10645 \pm 8)^{\mathrm{ref}}$ & 
\\
\hline\hline
\end{tblr}
\end{ruledtabular}
\end{table*}

\begin{table*}
\caption{Comparison of the predicted masses (in MeV) of the $\mathrm{bb\overline{sq}}$ tetraquarks in the $1P$ and $2P$ states with other theoretical approaches. All states have isospin $I=\frac{1}{2}$.}
\label{tab:other.bbsq.P}
\centering
\setlength{\jot}{0pt}
\begin{ruledtabular}
\begin{tblr}{
	width=\textwidth,
	colspec={Q[l,m,wd=200pt] Q[c,m,wd=30pt] Q[c,m] Q[c,m] Q[c,m] Q[c,m]},
	rowsep=1pt,
	colsep=7pt
	}
\hline\hline
Approach and reference & $nL$ & $J^{P}=0^{-}$ & $J^{P}=1^{-}$ & $J^{P}=2^{-}$ & $J^{P}=3^{-}$
\\
\hline
\SetCell[r=2]{m}\shortstack{\textbf{This work} \\ \textbf{(relativistic diquark--antidiquark)}} & $1P$ & $\begin{gathered}11108 \, (\mathrm{A\overline{A}}) \\ 10996 \, (\mathrm{A\overline{S}})\end{gathered}$ & $\begin{gathered}11084 \, (\mathrm{A\overline{A}}) \\ 11113 \, (\mathrm{A\overline{A}}) \\ 11131 \, (\mathrm{A\overline{A}}) \\ 11000 \, (\mathrm{A\overline{S}})\end{gathered}$ & $\begin{gathered}11091 \, (\mathrm{A\overline{A}}) \\ 11124 \, (\mathrm{A\overline{A}}) \\ 11007 \, (\mathrm{A\overline{S}})\end{gathered}$ & $11100$
\\
\cline{2-6}
 & $2P$ & $\begin{gathered}11601 \, (\mathrm{A\overline{A}}) \\ 11493 \, (\mathrm{A\overline{S}})\end{gathered}$ & $\begin{gathered}11561 \, (\mathrm{A\overline{A}}) \\ 11603 \, (\mathrm{A\overline{A}}) \\ 11625 \, (\mathrm{A\overline{A}}) \\ 11495 \, (\mathrm{A\overline{S}})\end{gathered}$ & $\begin{gathered}11566 \, (\mathrm{A\overline{A}}) \\ 11608 \, (\mathrm{A\overline{A}}) \\ 11498 \, (\mathrm{A\overline{S}})\end{gathered}$ & $11573$
\\
\hline
\SetCell[c=6]{l}\textit{Diquark--antidiquark models} &  &  &  &  & 
\\
\cline{1-6}
Regge mass relations~\cite{ccqq.cbqq.bbqq.Song2023} & $1P$ & $\begin{gathered}10983 \\ 11110\end{gathered}$ & $\begin{gathered}10988 \\ 11116 \\ 11116 \\ 11127\end{gathered}$ & $\begin{gathered}10996 \\ 11126 \\ 11130\end{gathered}$ & $11136$
\\
\hline\hline
\end{tblr}
\end{ruledtabular}
\end{table*}

\subsubsection{Doubly strange: $\mathrm{bb\overline{ss}}$}
\label{sec:othermodels.bbss}

\par
For $\mathrm{bb\overline{ss}}$, our ground--state masses of $10896$, $10904$, and $10919$~MeV for $J^{P}=0^{+}$, $1^{+}$, and $2^{+}$, respectively, differ by only $1$--$7$~MeV from the heavy--quark EFT predictions~\cite{ccqq.ccsq.ccss.cbqq.bbqq.Braaten2021}. The three--body calculation~\cite{ccqq.cbqq.bbqq.Kim2022} gives masses higher by $10$--$18$~MeV, while the diffusion Monte Carlo results~\cite{ccqq.cbqq.bbqq.Mutuk2024DMC} are within $4$--$7$~MeV for the $1^{+}$ and $2^{+}$ states.

\begin{table*}
\caption{Comparison of the predicted masses (in MeV) of the $\mathrm{bb\overline{ss}}$ tetraquarks in the $1S$, $2S$, and $3S$ states with other theoretical approaches. All states have isospin $I=0$.}
\label{tab:other.bbss.S}
\centering
\setlength{\jot}{0pt}
\begin{ruledtabular}
\begin{tblr}{
	width=\textwidth,
	colspec={Q[l,m,wd=200pt] Q[c,m,wd=30pt] Q[c,m] Q[c,m] Q[c,m]},
	rowsep=1pt,
	colsep=7pt
	}
\hline\hline
Approach and reference & $nL$ & $J^{P}=0^{+}$ & $J^{P}=1^{+}$ & $J^{P}=2^{+}$
\\
\hline
\SetCell[r=3]{m}\shortstack{\textbf{This work} \\ \textbf{(relativistic diquark--antidiquark)}} & $1S$ & $10896$ & $10904$ & $10919$
\\
\cline{2-5}
 & $2S$ & $11440$ & $11442$ & $11444$
\\
\cline{2-5}
 & $3S$ & $11909$ & $11909$ & $11908$
\\
\hline
\SetCell[c=5]{l}\textit{Diquark--antidiquark models} &  &  &  & 
\\
\cline{1-5}
Regge mass relations~\cite{ccqq.cbqq.bbqq.Song2023} & $1S$ & $10976$ & $10981$ & $10991$
\\
\hline
\SetCell[c=5]{l}\textit{Three--body heavy--quark--antidiquark models} &  &  &  & 
\\
\cline{1-5}
\SetCell[r=2]{m}Chiral light antidiquark~\cite{ccqq.cbqq.bbqq.Kim2022} & $1S$ & $10914$ & $10919$ & $10929$
\\
\cline{2-5}
 & $2S$ & $11249$ & $11252$ & $11260$
\\
\hline
\SetCell[c=5]{l}\textit{Heavy--quark symmetry and hadron--mass relations} &  &  &  & 
\\
\cline{1-5}
\shortstack[l]{Heavy--quark EFT \\ mass relations~\cite{ccqq.ccsq.ccss.cbqq.bbqq.Braaten2021}} & $1S$ & $10898 \pm 26$ & $10905 \pm 26$ & $10926 \pm 26$
\\
\cline{1-5}
\shortstack[l]{Heavy \\ antiquark--diquark symmetry~\cite{ccqq.ccsq.ccss.cbqq.bbqq.Zhang2026}} & $1S$ & $10740$ & $10745$ & $10755$
\\
\hline
\SetCell[c=5]{l}\textit{Chromomagnetic interaction models} &  &  &  & 
\\
\cline{1-5}
\shortstack[l]{Extended chromomagnetic \\ (Scheme $\RNumb{1}$)~\cite{ccqq.cbqq.bbqq.Weng2022}} & $1S$ & $\begin{gathered}10697 \\ 10929\end{gathered}$ & $10718$ & $10743$
\\
\cline{1-5}
\shortstack[l]{Extended chromomagnetic \\ (Scheme $\RNumb{2}$)~\cite{ccqq.cbqq.bbqq.Weng2022}} & $1S$ & $\begin{gathered}10792 \\ 11154\end{gathered}$ & $10810$ & $10834$
\\
\cline{1-5}
\shortstack[l]{Flux--tube + \\ chromomagnetic~\cite{ccqq.ccsq.ccss.cbqq.bbqq.Zhang2025}} & $1S$ & $\begin{gathered}10736 \pm 25 \\ 10949 \pm 28\end{gathered}$ & $10758 \pm 24$ & $10783 \pm 24$
\\
\hline
\SetCell[c=5]{l}\textit{Four--body constituent quark models} &  &  &  & 
\\
\cline{1-5}
\shortstack[l]{Nonrelativistic diffusion \\ Monte Carlo~\cite{ccqq.cbqq.bbqq.Mutuk2024DMC}} & $1S$ & $\begin{gathered}10868 \\ 10972\end{gathered}$ & $10908$ & $10926$
\\
\hline\hline
\end{tblr}
\end{ruledtabular}
\end{table*}

\begin{table*}
\caption{Comparison of the predicted masses (in MeV) of the $\mathrm{bb\overline{ss}}$ tetraquarks in the $1P$ and $2P$ states with other theoretical approaches. All states have isospin $I=0$.}
\label{tab:other.bbss.P}
\centering
\setlength{\jot}{0pt}
\begin{ruledtabular}
\begin{tblr}{
	width=\textwidth,
	colspec={Q[l,m,wd=200pt] Q[c,m,wd=30pt] Q[c,m] Q[c,m] Q[c,m] Q[c,m]},
	rowsep=1pt,
	colsep=7pt
	}
\hline\hline
Approach and reference & $nL$ & $J^{P}=0^{-}$ & $J^{P}=1^{-}$ & $J^{P}=2^{-}$ & $J^{P}=3^{-}$
\\
\hline
\SetCell[r=2]{m}\shortstack{\textbf{This work} \\ \textbf{(relativistic diquark--antidiquark)}} & $1P$ & $11228$ & $\begin{gathered}11207 \\ 11232 \\ 11247\end{gathered}$ & $\begin{gathered}11212 \\ 11241\end{gathered}$ & $11221$
\\
\cline{2-6}
 & $2P$ & $11706$ & $\begin{gathered}11672 \\ 11707 \\ 11725\end{gathered}$ & $\begin{gathered}11676 \\ 11712\end{gathered}$ & $11683$
\\
\hline
\SetCell[c=6]{l}\textit{Diquark--antidiquark models} &  &  &  &  & 
\\
\cline{1-6}
Regge mass relations~\cite{ccqq.cbqq.bbqq.Song2023} & $1P$ & $11201$ & $\begin{gathered}11206 \\ 11206 \\ 11215\end{gathered}$ & $\begin{gathered}11216 \\ 11219\end{gathered}$ & $11224$
\\
\hline\hline
\end{tblr}
\end{ruledtabular}
\end{table*}

\subsection{Overall comparison}
\label{subsec:resume}

\par
For all nine flavor compositions, the calculated $1P$ masses show good numerical agreement with the Regge mass relations of Ref.~\cite{ccqq.cbqq.bbqq.Song2023}, generally within several tens of MeV for nearby levels with the same $J^{P}$ and isospin. Comparisons for the higher radial and orbital excitations remain considerably more limited.
\par
Our lowest doubly bottom tetraquark masses are close to several lattice QCD estimates. For the isoscalar $\mathrm{bb\overline{ud}}$ state with $J^{P}=1^{+}$, we obtain $10461$~MeV, compared with $10476 \pm 24 \pm 10$ and $10489 \pm 17$~MeV from Refs.~\cite{bbud.lattice.2019,bbqq.bbsq.Colquhoun2024}, respectively. For the lowest $\mathrm{bb\overline{sq}}$ state with $J^{P}=1^{+}$, our prediction of $10670$~MeV is close to the lattice estimate of $10645 \pm 8$~MeV~\cite{bbqq.bbsq.Colquhoun2024}. Other lattice results exhibit a larger spread.

\section{Experimental data}
\label{sec:experiment}

\par
The experimental information on doubly heavy open--flavor tetraquarks remains limited. The $\mathrm{T}_{\mathrm{cc}}(3875)^{+}$ state, with the minimal quark composition $\mathrm{cc\overline{ud}}$, was observed by the LHCb Collaboration in 2021 in the $D^{0}D^{0}\pi^{+}$ invariant mass spectrum~\cite{ccud:LHCb:2021}. A subsequent analysis~\cite{ccud:LHCb:2022std} found that the state lies very close to the $D^{0}D^{*+}$ threshold and is consistent with $I(J^{P})=0(1^{+})$. No other tetraquark states with the doubly heavy flavor compositions considered here have been experimentally established.
\par
The LHCb Collaboration determined the pole position relative to the $D^{0}D^{*+}$ threshold and its width as~\cite{ccud:LHCb:2022std}
\begin{equation}
\begin{aligned}
\delta m_{\mathrm{pole}} &= -360 \pm 40^{+4}_{-0} \ \mathrm{keV}, 
\\
\Gamma_{\mathrm{pole}} &= 48 \pm 2^{+0}_{-14} \ \mathrm{keV}.
\end{aligned}
\end{equation}
\noindent
These values characterize an extremely narrow near--threshold state and suggest an important role of the $DD^{*}$ component in its structure.
\par
Our lowest isoscalar $\mathrm{cc\overline{ud}}$ state has $J^{P}=1^{+}$, the $\mathrm{A\overline{S}}$ configuration, and a predicted mass of $3869$ MeV (Table~\ref{tab:comp.cc}), about $6$ MeV below the observed mass. However, the theoretical uncertainty of $20$--$50$~MeV does not allow us to determine the very small experimental binding energy relative to the $D^{0}D^{*+}$ threshold with sufficient accuracy. Moreover, our model does not explicitly include coupling to the nearby $DD^{*}$ meson--meson channels, which may affect the mass and decay properties of this near--threshold state.

\section{Conclusion}
\label{sec:conclusion}

\par
We have calculated the masses of the ground and excited doubly heavy open--flavor tetraquarks within the relativistic diquark--antidiquark model. Nine flavor compositions and the $1S$, $1P$, $2S$, $1D$, $2P$, and $3S$ excitations were considered. The interaction includes relativistic spin--independent and spin--dependent terms and the finite sizes of the diquarks. The model parameters were fixed in previous studies of ordinary hadrons and were not readjusted for tetraquarks. All allowed scalar and axialvector configurations were included, with spin mixing taken into account where applicable.
\par
The calculated masses were compared with the lowest allowed $S$--wave strong fall--apart thresholds. Although most states lie well above them, a number of ground and orbitally excited states are below or close to their corresponding thresholds. This occurs in particular for negative--parity states and states with large $J$, for which angular momentum and parity conservation can raise the lowest allowed $S$--wave threshold. The number of such states generally increases from the doubly charmed to the doubly bottom sector.
\par
The calculated ground--state masses and spin splittings are close to several published theoretical estimates. For the lowest doubly bottom tetraquarks, our results also agree within a few tens of MeV with several lattice QCD calculations. The $1P$ spectra are also numerically close to the Regge mass relations for all nine compositions. The comparison is less conclusive for higher excitations, where fewer results are available and the assignments of states can differ between models.
\par
For the lowest isoscalar $\mathrm{cc\overline{ud}}$ state with $J^{P}=1^{+}$ we obtain $M=3869$ MeV, approximately $6$ MeV below the observed $\mathrm{T}_{\mathrm{cc}}(3875)^{+}$ mass. The predicted quantum numbers are also consistent with the experimental analysis. However, the model does not reproduce the extremely small binding energy relative to the $D^{0}D^{*+}$ threshold with the same accuracy. Its description may require explicit coupling to the meson--meson continuum.

\begin{acknowledgments}
The authors are grateful to D. Ebert and A. V. Berezhnoy for useful discussions.
\end{acknowledgments}

\bibliography{GalkinSavchenko_PRD2026}

\end{document}